\documentclass[twocolumn,preprintnumbers,amsmath,amssymb,aps, prx]{revtex4}

\usepackage{graphicx}
\usepackage[latin1]{inputenc}
\usepackage{dcolumn}
\usepackage{bm}
\usepackage{epsfig}
\usepackage{color}
\usepackage{framed}
\usepackage{mathtools}
\usepackage{xcolor}
\usepackage{ulem}		
\usepackage{amsmath}
\usepackage{amssymb}
\usepackage{bbm}

\usepackage{makeidx}

\newcommand{\Mod}[1]{\ (\text{mod}\ #1)}

\newcommand{\rw}{\rightarrow}

\newcommand{\mB}{{\mathcal B}}

\newcommand{\mG}{{\mathcal G}}

\newcommand{\mM}{{\mathcal M}}

\newcommand{\mT}{{\mathcal T}}
\newcommand{\mL}{{\mathcal L}}
\newcommand{\Real}{\mathbb{R}}

\newcommand{\sd}{{\sf d}}
\newcommand{\sT}{{\sf T}}

\newcommand{\mbK}{\mathbb{K}}
\newcommand{\mbM}{\mathbb{M}}
\newcommand{\mbN}{\mathbb{N}}

\newcommand{\lb}{\left(}
\newcommand{\rb}{\right)}

\newcommand{\cred}{}

\newtheorem{teorema}{\bf Theorem}

\newtheorem{definicion}{\bf Definition}

\begin{document}


\title{{Multiplex decomposition of non-Markovian dynamics\\ and the hidden layer reconstruction problem}}

\author{Lucas Lacasa$^{1,*}$, In\'es P. Mari\~no$^{2,3}$, Joaquin Miguez$^{4}$, Vincenzo Nicosia$^1$,\\  \'Edgar Rold\'an$^{5,6}$, Ana Lisica$^7$, Stephan W. Grill$^{8,9}$ and Jes\'us G\'omez-Garde\~nes$^{10,11}$ }

\affiliation{$^1$School of Mathematical Sciences, Queen Mary University of London, Mile End Rd, E14NS London (United Kingdom)\\ $^2$Department of Biology, Geology, Physics $\&$ Inorganic Chemistry, Universidad Rey Juan Carlos, 28933 M\'ostoles, Madrid (Spain)\\
$^3$ Institute for Women's Health, University College London, London WC1E 6BT, London (United Kingdom)\\
$^4$Department of Signal Theory \& Communications, Universidad Carlos III de Madrid. 28911 Legan\'es, Madrid (Spain)\\
  $^5$The Abdus Salam International Center for Theoretical Physics, Strada Costiera 11, 34151 Trieste (Italy)\\
$^6$Max Planck Institute for the Physics of Complex Systems,
  N{\"o}thnitzerstra{\ss}e 38, 01187 Dresden (Germany)\\
  $^7$London Center for Nanotechnology, University College London, London (United Kingdom)\\
$^8$Biotechnology Center, Technical University Dresden, Tatzberg 47/49, 01309 Dresden (Germany)\\
$^9$Max Planck Institute of Molecular Cell Biology and Genetics, Pfotenhauer Str. 108, 01307 Dresden (Germany)\\
 $^{10}$Department of Condensed Matter Physics, University of Zaragoza, 50009 Zaragoza (Spain)\\
 $^{11}$Institute for Biocomputation and Physics of Complex Systems (BIFI), University of Zaragoza, E-50018 Zaragoza (Spain)
}


\begin{abstract}
{Elements composing complex systems usually interact in several different ways and as such the interaction architecture is well modelled by a network with multiple layers --a multiplex network--,} where the system's complex dynamics is often the result of several intertwined processes taking place at different
  levels. However only in a few cases can such multi-layered architecture be empirically observed, as one usually only has experimental access to such structure from an aggregated projection. A fundamental challenge is thus to determine whether 
  the hidden underlying architecture of complex systems is better modelled as a single interaction layer or results from the aggregation and interplay of multiple layers. Here we show that, {assuming that random walkers diffuse Markovianly in each of the hidden layers, then using local information provided by a random walker navigating the aggregated network one can decide in a robust way if the underlying structure is a multiplex or not and, in the former case, to determine the most probable number of hidden layers}. 
  We introduce a method that enables to decipher the underlying multiplex architecture of complex systems by exploiting the non-Markovian signatures on the statistics of a single random walk on the aggregated network. {In fact, the mathematical formalism presented here extends above and beyond detection of physical layers in networked complex systems, as it provides a principled solution for the optimal decomposition and projection of complex, non-Markovian dynamics into a Markov switching combination of diffusive modes.   
  We validate the proposed methodology with numerical simulations of both (i) random walks navigating hidden multiplex networks (thereby reconstructing the true hidden architecture) and (ii) Markovian and non-Markovian continuous stochastic processes (thereby reconstructing an effective multiplex decomposition where each layer accounts for a different diffusive mode). We also state and prove two existence theorems guaranteeing that an exact reconstruction of the dynamics in terms of these hidden jump-Markov models is always possible for arbitrary finite-order Markovian and fully non-Markovian processes. Finally, we} showcase the {applicability of the method} to experimental recordings from (i) the mobility dynamics of human players in an online multiplayer game and (ii) the  dynamics of RNA polymerases at the single-molecule level.
\end{abstract}


\maketitle

 \section{Introduction} \label{sintro}

Network Science has emerged as a powerful unifying framework for
studying the emergence of collective phenomena in real complex systems
from different domains~\cite{newmanbook,barabook}, and has allowed 
to increase the accuracy and predictive power of minimal models of complex
dynamics, including epidemic spreading~\cite{vespignaniRMP},
synchronisation~\cite{arenasphysrep}, or social
dynamics~\cite{castellanoRMP}. 
One of the most fascinating challenges faced in the last few years by
Network Science is the need to incorporate and couple several network
structures in order to correctly capture the inherently multidimensional nature of
interaction patterns in real-world systems. As a result, much effort
has been recently devoted to the definition and study of multilayer
and multiplex networks~\cite{physrep,jcn,battiston2016review}.  The
ubiquity of such structures in social, biological and
technological systems has required the revision of the several canonical dynamical models that were previously studied only on isolated complex
networks, including percolation~\cite{buldy,Gao012,grass,BiD014,radi},
diffusion dynamics~{\cite{gomez13,radi2}},
navigation~\cite{Manlio14,Sole16,Battiston2016RW},
epidemics~\cite{mendiola,granell13,Buono14,Sanz14}, evolutionary games
\cite{GG12,Wang14,Mata15}, synchronization~\cite{DelGenio16}, or
opinion dynamics~\cite{Diakonova2014,Diakonova2016}. Notably the collective behaviour of such complex systems depends strongly on whether they can described by isolated networks or by
coupled networks~\cite{NatPhys16}, highlighting the importance of the
multilayer architecture of real-world systems.
\begin{figure}[h]	
\centering
\includegraphics[width=1.05\columnwidth]{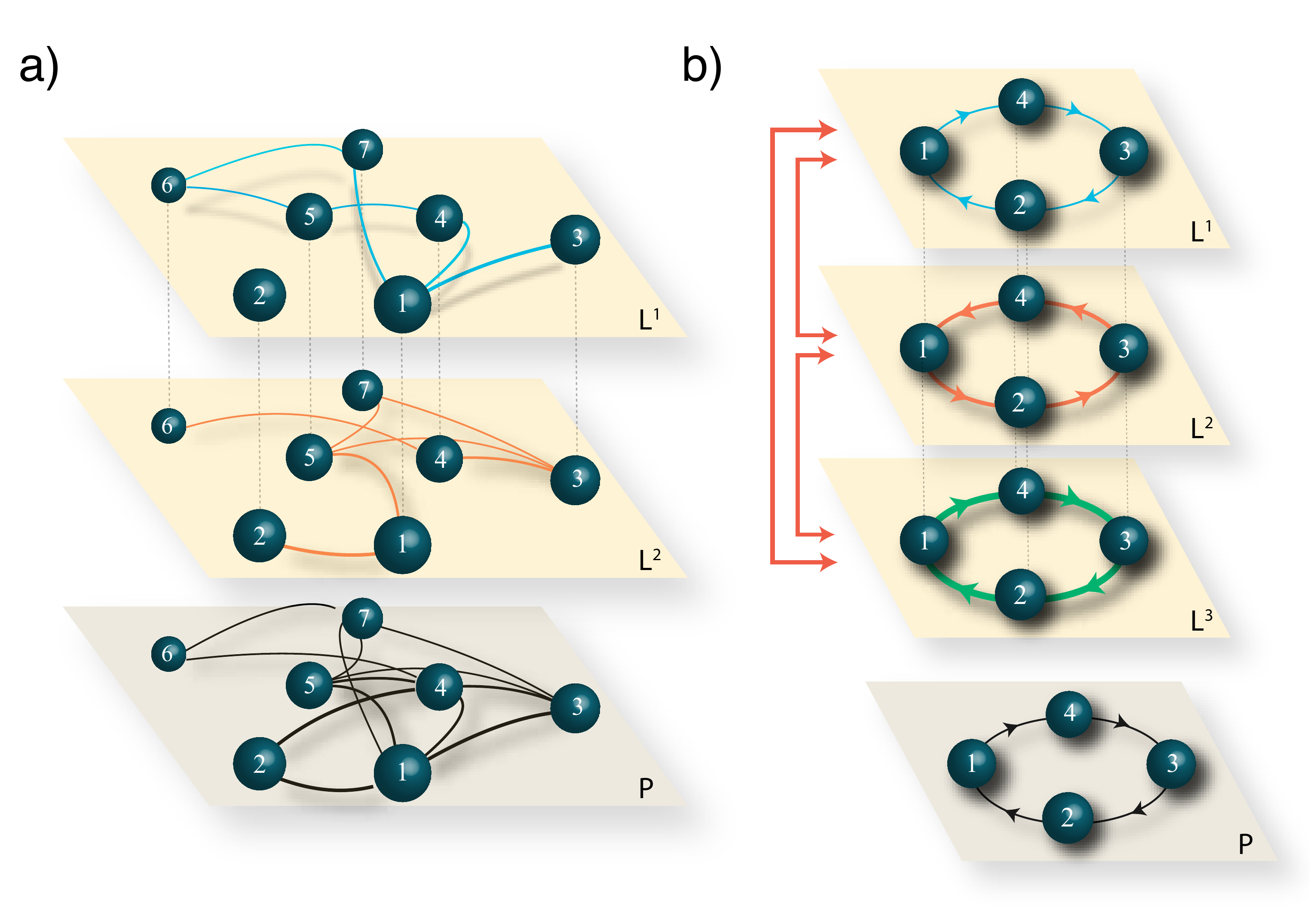}
\caption{(Panel a) A multiplex network with $L=2$ layers ($L^1$ and $L^2$) and $K=7$ nodes.  A
  random walker diffusing over this structure generates a two
  dimensional time series $\{X(t),\ell(t)\}_{t=1}^N$ where
  $X(t)$ and $\ell(t)$ are the vertex and
  layer locations of the walker at time $t$. In many real-world cases the layer indicator $\ell(t)$ is hidden and one has access only to $\{X(t)\}_{t=1}^N$, i.e. to the series of
  states of the walker on the projected network ($P$) shown in the bottom of the figure. For $L>1$ the resulting trajectory is non-Markovian: we rely on this Markovianity-breaking phenomenon property to
  detect multiplexity and to provide an estimate of the number of
  layers in the system by observing only $\{X(t)\}_{t=1}^N$. (Panel b) Simple canonical model where we fix a prior on the topology of each layer: a cycle graph with homogeneous transition rates, with uniform interlayer transition rates. This model serves as a basis for a generic stochastic decomposition of non-Markovian dynamics on the aggregated network.}
\label{fig0}
\end{figure}

Multilayer network models of real-world systems face two fundamental and dual challenges. The first one is the
necessity to assess in a systematic way whether a multilayer network
model is adequate to represent the system, and when such model gives redundant information. This challenge was first addressed in \cite{nicosia},
and constitutes nowadays an intense field of research. 
The dual challenge aims at understanding whether
an empirical network whose multilayer character is not directly observable
is genuinely monolayer or is only an aggregated projection
 of a hidden multilayer network (see Fig.~\ref{fig0}b for an illustration of a multiplex network with $L=2$ layers and its aggregated projection). Such scenario has received much less attention despite being, for instance, central for networks arising in natural systems whose architecture is not directly observable, as in genetic networks or in brain functional networks where pairs of nodes modelling different brain areas can interact according to an a priori unknown range of different biological pathways~\cite{Zanin}.\\

\noindent In this article we provide a method to identify the hidden
multi-layer structure of a complex system from coarse-grained dynamical 
measurements of its state. We show that, by using only
local information extracted from simple random-walk statistics, it is possible to discriminate whether the underlying structure of
the system is actually a single-layer or a multi-layer network, and in
the latter case, to estimate the number of interacting layers in the
system. {Note that methods to infer network topological properties via random walk statistics have been explored previously {\cite{PNASRosvall, tiago3}}.} Notably, our discrimination method  exploits the breaking of Markovianity occuring in a coarse-grained 
multi-layer random walk, while the method to estimate the most probable number of layers is based on a maximum a-posteriori (MAP) probabilistic criterion which can be implemented via numerical integration methods, including conventional grid-based approximations \cite{Davis07} or more sophisticated Monte Carlo algorithms \cite{Chopin12SMC} {which we show increase the computational efficiency}.\\

\noindent {Interestingly, this paper not only deals with a particular problem of Network Science. As a matter of fact, we show that the mathematical formulation can indeed be applied to signals {of} arbitrary origin --not necessarily random walkers navigating a network--, and the multiplexity estimation framework reduces in the general case to a stochastic decomposition of the signal in terms of an effective multiplex network, whose layers play the role of independent dynamical modes. More concretely, non-Markovian dynamics can be thereby decomposed into a stochastic combination of diffusive modes by projecting the dynamics into an appropriate hidden jump-Markov model.}\\

\noindent  {We validate the proposed methodology with numerical simulations of (i) random walks navigating hidden multiplex networks (thereby reconstructing the true architecture of the hidden multiplex networks) and (ii) both Markovian and non-Markovian continuous stochastic processes (thereby reconstructing an effective multiplex decomposition where each layer accounts for a different dynamical mode). 
Furthermore, we state and prove two existence theorems guaranteeing that such multiplex decomposition is always possible for any finite order Markovian and infinite order (fully non-Markovian) processes. {Specifically, we show that} random sequences generated by those processes can be exactly reconstructed as a random walk over an effective multiplex network. Finally, we apply our method  to experimental recordings of two complex systems of different nature, and show that the method can be leveraged to decompose noisy, non-Markovian processes into alternating combinations of simpler dynamics and extract valuable information accordingly}.\\

\noindent The article is structured as follows: in Sec.~II we propose the methods for multiplexity detection and layer estimation, and we explore their performance in a few examples. In Sec.~III we discuss the analogy between multiplexity unfolding and the decomposition of non-Markovian processes as multiple-layer Markovian processes and therefore extend the methodology to continuous time processes. {We also state and discuss the implications of two existence theorems, whose proofs are put in two appendices for readability}. In Sec.~IV we address real-world scenarios where we analyse two sets of experimental recordings, namely human mobility in an online environment and traces of RNA polymerase, that further showcase the applicability of the method. In Sec.~V we provide a discussion of our results. Mathematical details and additional examples can be found in the Appendices.

\section{Description of the methods and illustrative examples} \label{smethods}

Multiplex networks are the most ubiquitous class of multilayer networks. They are a natural model for online social networks~\cite{lamb}, where a given individual can communicate with others via different platforms (e.g. Facebook, Twitter, email, etc) or transportation networks~\cite{barth,Cardillo}, where a set of locations can be connected in a multimodal way (e.g. bus, train, underground, etc).\\
A multiplex network is defined by a set of $L\geq 1$ interaction layers (networks), all of them having the same set of $K$ nodes but different topology (different edge set), with the peculiarity that each node has a replica in each layer  (see Fig.~\ref{fig0}a for an illustration). This structure is thereby fully described by a set of adjacency matrices
$\{{\bf A}^{(\ell)}\}_{\ell=1}^L$, where $A^{(\ell)}_{\alpha\beta}=1$ if there is an edge between nodes $\alpha$ and $\beta$ at layer $\ell$ and zero otherwise.  For simplicity we label the different layers of the multiplex network with Roman letters ($i,j$, etc), and the nodes of each layer with Greek letters ($\alpha,\beta$, etc).\\ 

\noindent We consider a random walker navigating a multiplex~\cite{Manlio14} defined as follows: jumps between layers are governed by a Markov chain with $L \times L$ transition matrix ${\bf R}_L$ ($R_{ij}$ is the probability to jump from layer $i$ to layer $j$) while the dynamics within each individual layer $\ell$ is also Markovian and determined by a $K \times K$ transition matrix ${\bf T}^{(\ell)}$ (where $T^{(\ell)}_{\alpha\beta}$ is the probability to walk from node $\alpha$ to node $\beta$ at layer $\ell$). For simplicity we only consider diffusive dynamics where at each time step the walker at node $\alpha$ on layer $\ell$
 (i) remains in the same layer with probability
$R_{\ell\ell}=1-r$ or instantaneously jumps with uniform probability $R_{\ell\ell'}=r/(L-1)$ to a different layer $\ell'$, and subsequently (ii) diffuses to one of the neighbours of node $\alpha$ in the chosen layer according to the layer internal dynamics (given by ${\bf T}^{(\ell)}$ or ${\bf T}^{(\ell')}$). {Notice that this type of dynamical model can be mathematically formalised in terms of a jump-Markov affine system, as defined in the field of Control Theory (see \cite{Garulli} and references therein for a review)}.\\ 
\noindent When $r\ll 1$, i.e. when
walkers tend to remain in the same layer, this navigation model might mimick for instance human mobility in multilayered transportation
networks~\cite{gallotti1,gallotti2}, where multimodality is minimised
to avoid waiting times related to connections between different
modes.\\


\noindent {In the particular case when layers have a simple cycle-graph topology, the jump-Markov model described above is also reminiscent of
the so-called discrete flashing ratchet model~\cite{ratchet,
  ratchet2},} {better known as a Parrondo game \cite{Parrondogame, Dinis}. This is a paradoxical gambling strategy that allows winning in loosing scenarios, where gamblers can alternate between two different strategies (game A, layer 1; game B, layer 2), each of them having different rules and winning probabilties. Our model for $L=2$ can be seen as a variant of a  Parrondo gambler that plays with probabilities $r$ and $1-r$ with two different biased coins.\\
  More generally, Brownian ratchets are paradigmatic
models used in Nonequilibrium Physics to describe the transport of Brownian
particles embedded in periodic, asymmetric energy potentials, a
paradigm originally proposed by
Smoluchowski~\cite{Smoluchowski} and popularised by
Feynman~\cite{Feynman} in the context of thermodynamic engines, and further
shown to be a minimal model system for molecular motors in
biophysics~\cite{ratchet2}. Again, a Brownian particle subject to a periodic asymmetric potential that is switched on and off stochastically is formally equivalent to our random walker navigating over a multiplex with $L=2$ cycle graphs with different transition matrices~\footnote{In the definition above time is considered discrete, however a similar methodology can be applied for continuous time dynamics by sampling the continous time series.}.}\\

\begin{figure*}[ht]
\centering
\includegraphics[width=0.95\columnwidth]{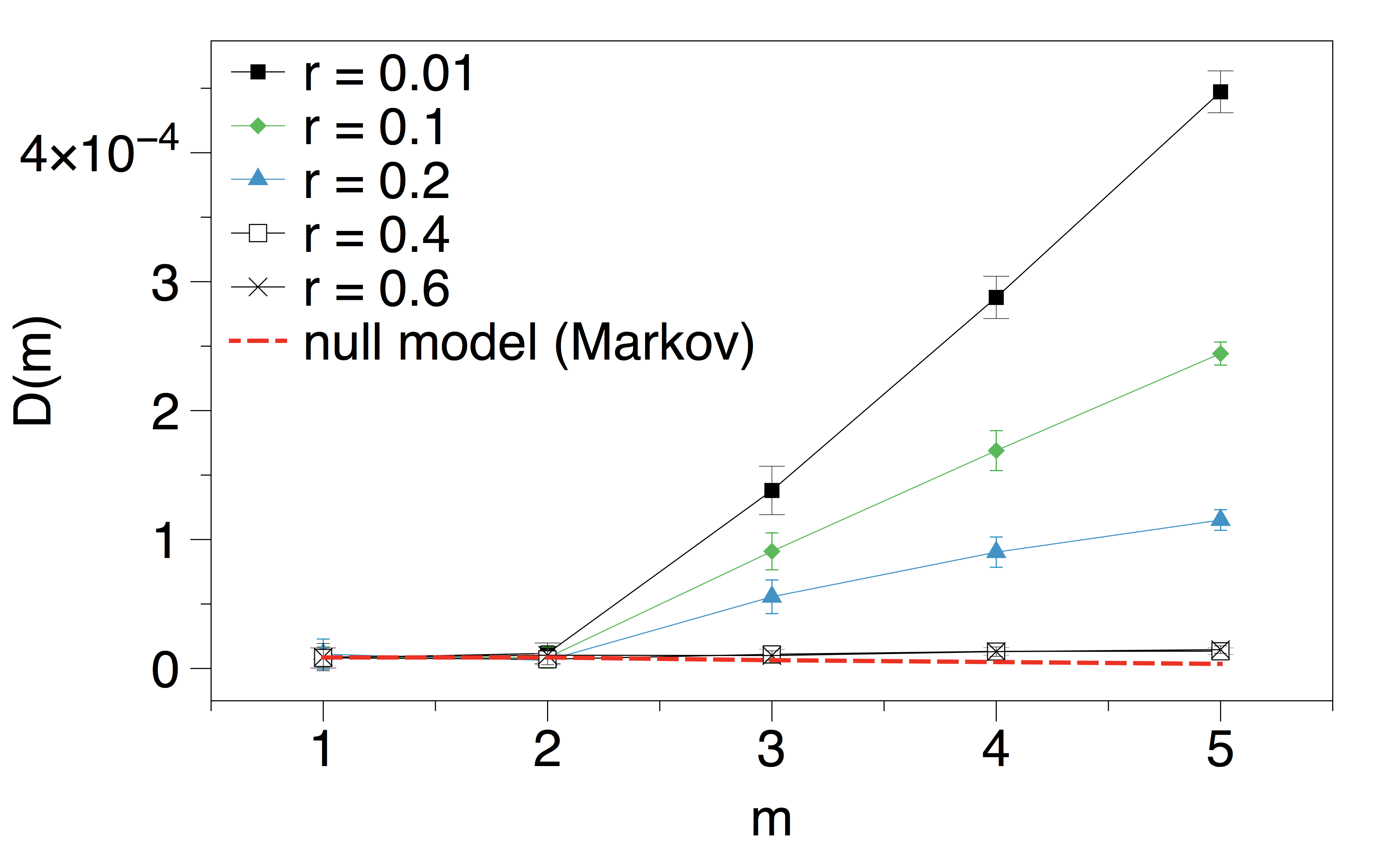}
\includegraphics[width=0.9\columnwidth]{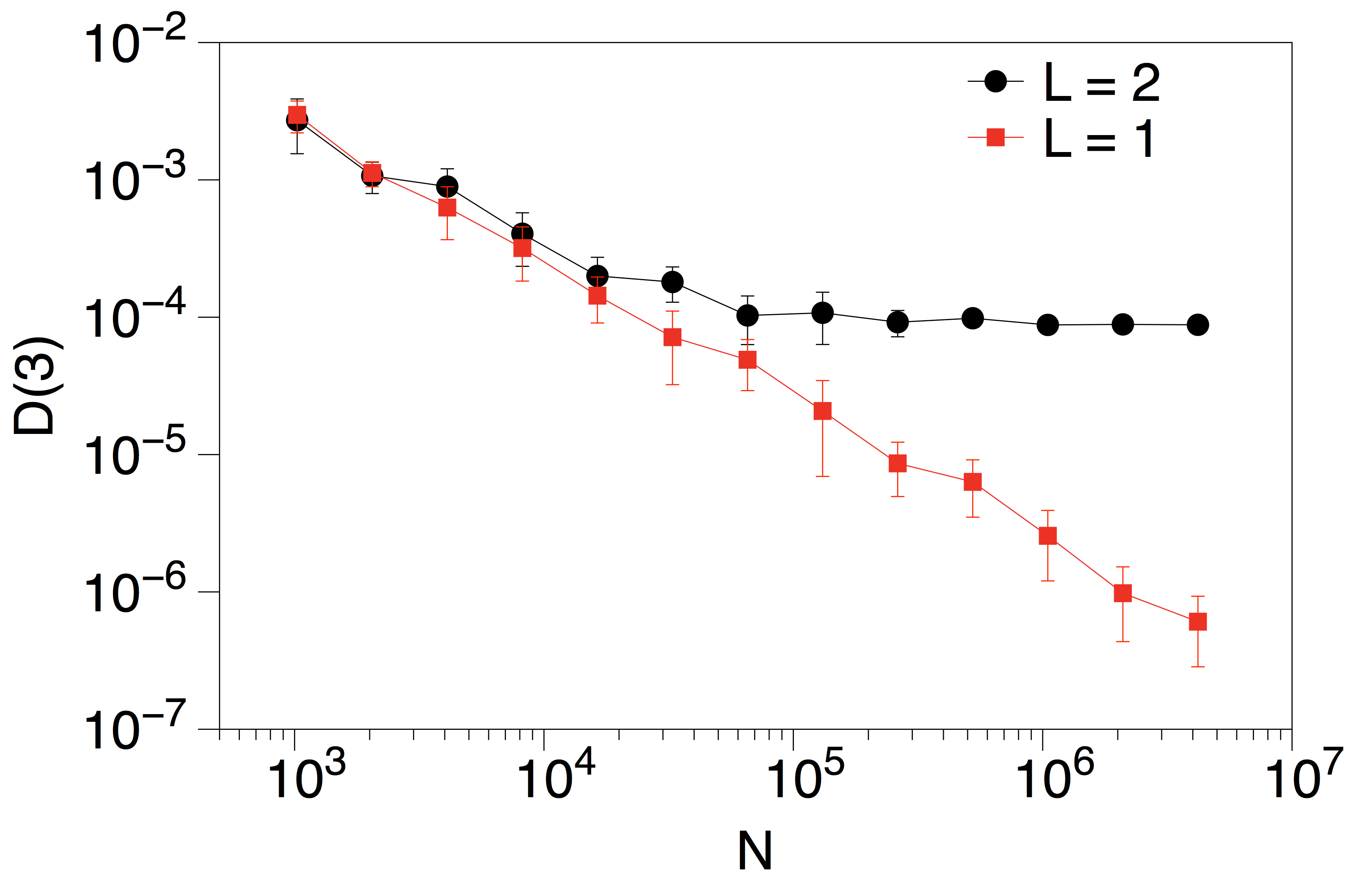}
\caption{(Left panel) Multiplexity detection statistic
  $\mathcal{D}(m)$ (see Eq.\ref{KLD}) for $X(t)$, in a
  case where $X(t)$ shows an induced current, for different values
  of the switching rate $r$. The series $X(t)$ records the position of
  a walker diffusing over two layers, with transition probabilities
  in the layers given by  $T^{(1)}_{\alpha,\alpha+1}=1/3$ and $T^{(2)}_{\alpha,\alpha+1}=1/2$
  respectively, such that the transition probability in the
  Markovian surrogate series is $Q_{\alpha,\alpha+1}=5/12$. We
  correctly find that $X(t)$ is non-Markovian even for large values of
  $r$ as $\mathcal{D}(m)|_{m\geq 3}>0$, which suggests an underlying multiplex structure. (Right panel) Finite-size-scaling analysis where we compute $\mathcal{D}(3)$  as a function of the series size $N$, for the multiplex considered in the left panel ($r=0.1$, green dots) and a null model with equivalent monoplex dynamics (red squares) for which $\mathcal{D}(3)$ should vanish for large values of $N$. In both panels, the symbols represent the mean  and the error bars the  standard deviation calculated over  $10$ different realisations.}
\label{fig1}
\end{figure*}


The general navigation process on a multiplex network discussed above can be expressed as a 
 stochastic process  fully described by an infinite
two-dimensional time series $\{X(t),\ell(t)\}_{t=1}^{\infty}$ where
$X(t)\in\{1,\dots,K\}$ and $\ell(t)\in \{1,\dots,L\}$. As in 
real-world scenarios the multiplex nature of the system is not always empirically accessible, the layer indicator $\ell(t)$ is usually hidden and the only observable is the sequence of node locations $X(1), X(2), \cdots$\,. In such a situation, we only have experimental access to {\it partial} information of
the process, described by a finite sequence of observations of the variable $X$:  ${\cal O}= \{X(t)\}_{t=1}^N$.
 This is formally equivalent to observing a dynamical process on the aggregated (projected) network. Hence the question: is it possible to discern if the system is multiplex and in that case, to estimate the most probable number of layers if we only have access to
$\mathcal{O}$? We now propose and test a novel method to achieve this highly non-trivial task.



\subsection{Method to detect multiplexity} Consider the Markov switching
walker discussed above, navigating on a (multiplex) network from which we only have access
to coarse-grained information given by $\cal O$. Initially assuming $X(t)$ is a Markov process, we can estimate directly from ${\cal O}$ the (monoplex) transition matrix ${\bf Q}$ that would describe such a Markovian dynamics. Accordingly, we can define a Markov chain associated to
${\bf Q}$ and generate a  a Markovian surrogate of the original
process  $\{Y(t)\}_{t=1}^N$. If the underlying network was truly monoplex, then
$X(t)$ would be actually Markov and $X(t)$ and $Y(t)$ would then have asymptotically
equivalent statistics, $P_X (Z_1,\dots,Z_m) = P_Y(Z_1,\dots,Z_m)$, for all possible sequences $Z_1,\dots,Z_m$ of any arbitrary length $m$.  For multiplex structures however,  losing information of $\ell(t)$ in general breaks Markovianity and therefore $X(t)$ is typically non-Markovian. Accordingly, $X(t)$ and 
$Y(t)$ now share the same joint distributions only up to
blocks of size $m=2$. For blocks of size $m\geq 3$ their statistics may differ $P_X (Z_1,\dots,Z_m) \neq P_Y(Z_1,\dots,Z_m)$. To quantify such
difference, we make use of the $m-$th order Kullback-Leibler
divergence rate~\cite{roldan2010,roldan2012} between data blocks
of size $m\geq 1$:
\begin{equation}
  \mathcal{D}(m)\coloneqq \frac{1}{m}\sum_{{\cal B}(m)}
  P_X(Z_1,\dots,Z_m) \log \frac{P_X(Z_1,\dots,Z_m)}{P_Y(Z_1,\dots,Z_m)}\quad,
  \label{KLD}
\end{equation}
where ${\cal B}(m)$ enumerates all the blocks of size $m$. The statistic $\mathcal{D}(m)$ is semi-positive definite and vanishes only when the joint probabilities coincide \cite{cover}. Thus by
construction $\mathcal{D}(1)=\mathcal{D}(2)=0$. The Markovianity-breaking criterion
implies that if $\mathcal{D}(m)>0$ for $m\geq 3$ then the underlying dynamics is multiplex~\footnote{Even if $X(t)$ is Markovian, the matrix ${\bf Q}$ cannot be computed exactly, but estimated from the observations. Similarly, one can only estimate accurately the probabilities $P_X(Z_1,\dots,Z_m)$ for moderate values of $m$. Therefore, in practical scenarios the decision rule on whether the underlying network is multiplex or not requires to introduce an error threshold $\epsilon>0$ such that $\mathcal{D}(3) > \epsilon$ implies multiplexity. Such discrimination criterion may be formally described as a statistical test and error bounds may be obtained under mild assumptions.}.\\

 As a proof of concept, we first consider the simple
scenario where a random walker navigates over a two-layer multiplex
ring (each layer is a cycle graph of $K$ nodes), a model compatible with a discrete flashing ratchet as commented before. In the first layer, we define a
Markov chain with homogeneous transition probabilities $T^{(1)}_{\alpha+1,\alpha}=2/3;
\ T^{(1)}_{\alpha,\alpha+1}=1/3$ and $T^{(1)}_{\alpha\beta}=0$ if $\beta\neq \alpha+1\mod K$ or $\beta \neq (\alpha-1)\,\text{mod}\,K$. A random
walker diffusing in this layer will have an induced current in the
 direction of decreasing node indices. In the second layer, we define a different Markov
chain with homogeneous transition probabilities $T^{(2)}_{\alpha+1,\alpha}=1/2;
\ T^{(2)}_{\alpha,\alpha+1}=1/2$ and $T^{(2)}_{\alpha\beta}=0$ otherwise, i.e., an unbiased random walk. 
While we can always estimate ${{\bf Q}}$ numerically from the observed time series, in this simple case
it is easy to derive it analytically: $Q_{\alpha\beta}=W_1 T^{(1)}_{\alpha\beta} + W_2 T^{(2)}_{\alpha\beta}$, where $W_1$ ($W_2$) is the probability of finding the walker in
layer $\ell=1$ ($\ell=2$). Now, since in this case $R_{ij}=R_{ji}=r$, the
system is symmetric with respect to the switching process and the
walker spends on average the same amount of time in each of the
two layers, $W_1=W_2=1/2$, and then 
$Q_{\alpha\beta}= T^{(1)}_{\alpha\beta}/2  + T^{(2)}_{\alpha\beta}/2$. 
For this specific example, we thus find $Q_{\alpha+1,\alpha}=7/12,
\ Q_{\alpha,\alpha+1}=5/12$.\\
\noindent The left panel of Fig.~\ref{fig1} shows numerical results of $\mathcal{D}(m)$ as a function of the block size $m$  for different
switching rates $r$. In order to deal with finite-size effects (which
increase exponentially with $m$), we systematically increase the size
of the walker series under study as a function of $m$, taking series
of size $N(m)=N_0\cdot 2^m$ data extracted from the original system
and from the corresponding Markovian surrogates. We used $N_0=10^5$ although smaller values yield qualitatively equivalent results. 
As expected, $\mathcal{D}(1)=\mathcal{D}(2)=0$, meaning
that $Y(t)$ is a faithful Markovian surrogate of $X(t)$. Furthermore
$\mathcal{D}(m)>0$ for $m\geq 3$, meaning that $X(t)$ is non-Markovian~\cite{roldan2012} and hence the underlying network
is correctly identified as a multiplex. This result is robust for a quite large range of values
of $r$ (see Appendix B, and note that $r=0.5$ is a trivial exception), meaning that the method works even if the walker makes fast switches between layers. 
{A similar scenario is found if we tune the transition
probabilities such that no net induced current is found (see appendix B), pointing out that multiplexity can be unraveled
even in that case. In the right panel in Fig.~\ref{fig1} we plot $\mathcal{D}(3)$ for different series sizes to showcase how finite size effects vanish as the series gets larger. Notably with a few thousands data points we can already accurately detect multiplexity.}\\ 

\noindent Furthermore  we have demonstrated the flexibility and robustness of our method to detect multiplexity in a range of additional scenarios including (i) layers with increasingly different and disordered topologies controlled by both rewiring and edge addition, (ii) Erdos-Renyi graphs and (iii) similar scenarios on larger graphs. For all cases we find a correct multiplexity detection and good scalability (see Appendix B).

\subsection{Method to quantify multiplexity} 

The Markovianity-breaking phenomenon which we have exploited only provides a means to discriminate whether hidden layers do exist in the multiplex, but not to quantify the number underlying layers. In order to bridge this gap, we now make use of statistical inference tools to define a model selection scheme {\cite{PRLNewman}}. We assume that two models are different if they have a different number of layers. Accordingly, the number of layers $L$ is now modelled as a random variable with prior probability mass function $P_0(L)$. Given the value of $L$, the motion of the random walker is determined by the Markov-switching of layers, with transition matrix ${\bf R}_L$, and Markov walks within each layer characterised by ${\bf T}_L=\{ {\bf T}^{(\ell)} \}_{l=1}^L$. Assuming prior probability density functions  for these parameters $p_{0,R}({\bf R}_L)$ and  $p_{0,T}({\bf T}_L)$, the likelihood of a given model with $L$ layers conditional on the observed data $\{ X(t) \}_{t=1}^N$ reads
\begin{eqnarray}
&&P(\{X(t)\}_{t=1}^N | L) = \int P(\{X(t)\}_{t=1}^N | {\bf T}_L, {\bf R}_L ) \times \nonumber \\
&&\times p_{0,T}({\bf T}_L) p_{0,R}({\bf R}_L) \mu(\sd {\bf T}_L \times \sd {\bf R}_L), \label{eqLkhd}
\end{eqnarray}
where $\mu$ is a suitable reference measure for ${\bf T}_L$ and ${\bf R}_L$ (see appendices C and D for technical details). In general, this multidimensional integral cannot be computed exactly {and needs to be approximated numerically.} 
Then, the number of layers $L$ in the system that generated the data $\{ X(t) \}_{t=1}^N$ can be detected using a maximum a posteriori (MAP) criterion, i.e., 
\begin{equation}
\hat L^{\text{MAP}} = \arg \max_{L \in \{1, 2, \ldots, L_+\}} P(\{X(t)\}_{t=1}^N|L) P_0(L),
\label{eqMAP}
\end{equation}
where $\hat L^{\text{MAP}}$ is the estimate of the actual number of layers, $2\leq L_+<\infty$ is the (assumed) maximum possible value of $L$, and  
$P( L | \{X(t)\}_{t=1}^N) \propto P(\{X(t)\}_{t=1}^N | L) P_0(L)$ 
is the a posteriori probability mass function of $L$ given the observations $\{X(t)\}_{t=1}^N$.\\

\begin{figure}[h]
\centering
\includegraphics[width=0.95\columnwidth]{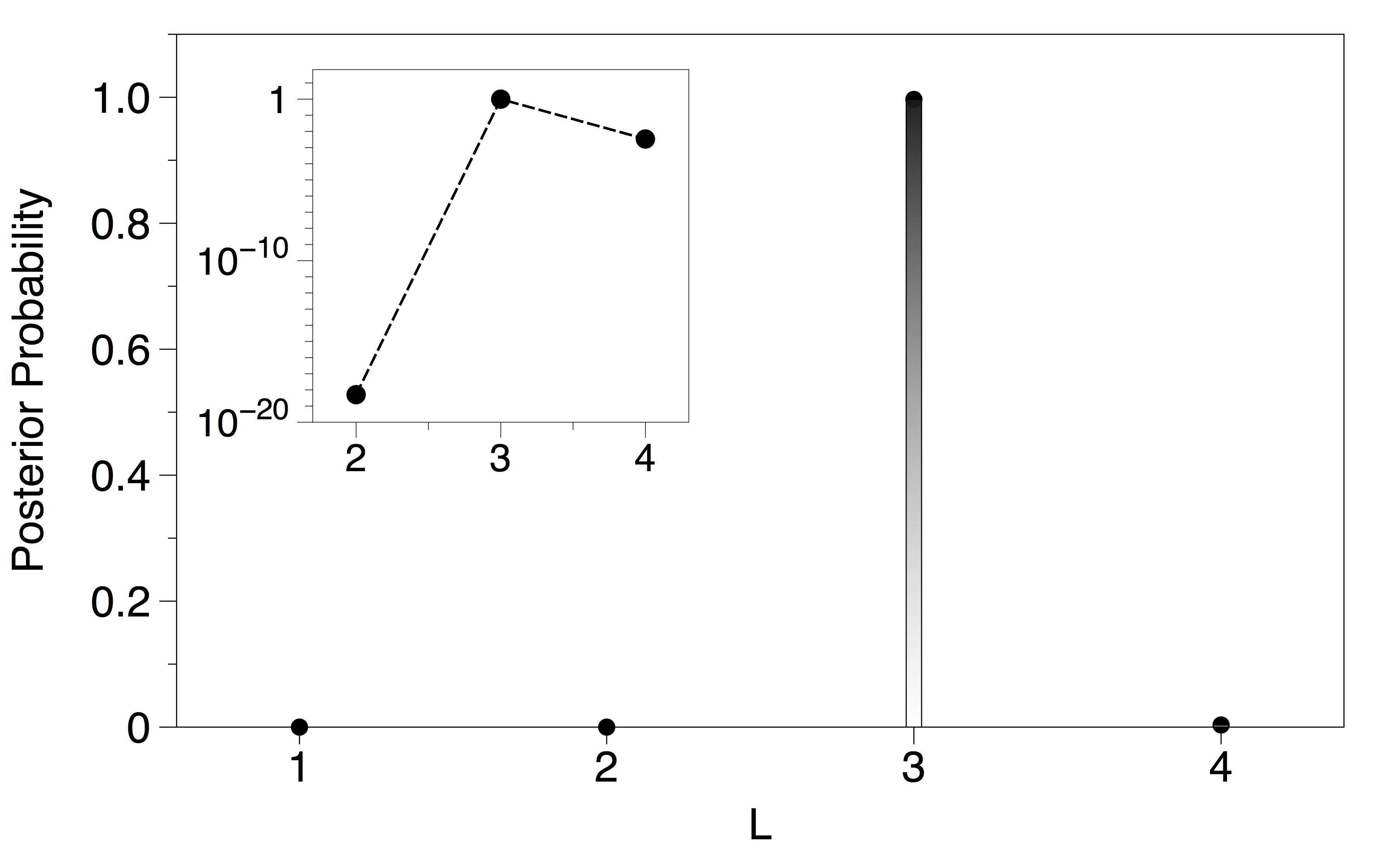}
\caption{Posterior probability $P(L | \{X(t)\}_{t=1}^N)$ as a function of the number of layers $L$, computed from a trajectory of $2 \times 10^4$ time steps generated using a model with $L=3$ cycle graphs ($K=3$) and parameters $R_{ii}=1-r=0.84$ for $i=1,2,3$,  $T^{(1)}_{\alpha,\alpha+1} = 0.16$, $T^{(2)}_{\alpha,\alpha+1} = 0.76$ and $T^{(3)}_{\alpha,\alpha+1} = 0.24$ (see the text and appendix B for details). The algorithm easily estimates the correct model $L=3$. (Inset) A linear-log plot of the same graph. Note that the probability for $L=1$ is zero up to the computer's accuracy, hence this point is not shown in logarithmic scales.}
\label{Post}
\end{figure}

\noindent {The practical computation of the MAP estimator in Eq.~\eqref{eqMAP} can be addressed in different ways. The classical literature on hidden Markov models (HMMs)~\cite{Rabiner89,Fessler94,Ghahramani01} suggests the use of the expectation-maximisation~(EM) algorithm (in various forms) to compute approximate maximum likelihood (ML) estimators, $\hat {\bf T}_L$ and $\hat {\bf R}_L$, of the parameters and then assume that $P(\{X(t)\}_{t=1}^N | L) \approx P(\{X(t)\}_{t=1}^N | \hat {\bf T}_L, \hat {\bf R}_L)$ in order to compare the models {(i.e., one tries to optimise the parameters instead of averaging over them)}. {This approach has also been applied to jump Markov affine systems \cite{Ozkan}} and relies on standard techniques but it has several drawbacks {and is therefore not adopted here}. Actually, the equation of the {\it model} likelihood with the {\it parameter} likelihood easily breaks down when the parameter estimates are poor (e.g., because of overfitting). {Another major disadvantage of EM is that it converges locally, and thus performs badly when the parameter likelihood is multimodal or when the parameter dimension varies significantly for different models}. More sophisticated parametric schemes have been proposed (see, e.g., \cite{Siddiqi07}) however they are still subject to these fundamental limitations. {Integration in \eqref{eqLkhd}, which we adopt here,} has been favoured theoretically but criticised practically because of the computational cost of approximating $P(\{X(t)\}_{t=1}^N | L)$ numerically \cite{Ghahramani01}. However, we have found that state-of-the-art variational Bayes \cite{McGrory09} or adaptive importance sampling \cite{Cappe08} methods can be applied effectively up to moderate values of $L$. See Appendices C3 and D for further discussion, including examples of using both deterministic integration and the adaptive Monte Carlo sampler from \cite{Koblents15}.} \\

\noindent To illustrate {the MAP model selection method given by Eq.~\eqref{eqMAP},} we consider again the discrete flashing ratchet model formed now by $L=3$ ring-shaped layers with $K=3$ nodes and homogeneous transition probabilities (see Fig.~\ref{fig0}b for an illustration and Appendix D for other { examples). In this  example, the probability to stay in the same layer is $R_{\ell\ell} = 1-r = 0.84$ and the probabilities for the walker to move from node $\alpha\to \alpha+1$ are $T_{\alpha,\alpha+1}^{(1)} = 0.16$, $T_{\alpha,\alpha+1}^{(2)} = 0.76$ and $T_{\alpha,\alpha+1}^{(3)} = 0.24$ respectively ($T^{(\ell)}_{\alpha+1,\alpha} = 1 - T_{\alpha,\alpha+1}^{(\ell)}$ for every $\ell$, and $T^{(\ell)}_{\alpha\beta}=0$ for any other $\alpha$ and $\beta$).\\

\noindent In order to evaluate the likelihood function $P( \{X(t)\}_{t=1}^N | L)$, we need to approximate the integral in \eqref{eqLkhd} for each value $L=1,\dots,4$ as discussed previously. Here we assume $L_+=4$. {A very simple strategy is to evaluate it via numerical integration over a deterministic grid of 19 points on the interval $(0,1)$ for each unknown parameter.} For the case $L=1$ this reduces to a single unknown, $T^{(1)}_{\alpha,\alpha+1}$, and for $L>1$ there are $L+1$ unknowns: $r$ and $T^{(\ell)}_{\alpha,\alpha+1}$ for $l=1, \ldots, L$.  Note that the particular choice of transition probabilities was taken to make the problem more challenging, as these values are not commensurate with the integration grid points. {We use an unbiased} prior probability density function $p_{0,R}({\bf R}_L)$ given by a uniform probability density function on $(0,1)$ for the unknown parameter $r$, while the prior $p_{0,T}({\bf T}_L)$ is used to penalise system configurations with two or more identical layers \footnote{Note that a multiplex with $L=2$ and ${\bf T}^{(1)}={\bf T}^{(2)}={\bf T}$ is equivalent to a monoplex with transition matrix ${\bf T}$.}. For this particular numerical experiment the penalising prior is $p_{0,T}({\bf T}_L) \propto \min_{(\ell,\ell')} | T_{\alpha,\alpha+1}^{(\ell)} - T_{\alpha,\alpha+1}^{(\ell')} |$, i.e., the prior probability density function of a given configuration ${\bf T}_L$ is proportional to the minimum distance between any pair of matrices ${\bf T}^{(\ell)}$ and ${\bf T}^{(\ell')}$. Since in this example each scalar ${\bf T}^{(\ell)}$ fully characterises layer $\ell$, this prior simply penalises configurations where two or more layers are very similar and thus avoids overfitting. {In general, the prior $p_{0,T}({\bf T}_L)$ is set to penalise models where pairs of layers have similar transition matrices, to avoid redundancy, so $p_{0,T}({\bf T}_L)\propto \min_{(\ell,\ell')} || {\bf T}^{(\ell)} - {\bf T}^{(\ell')} ||$, for some chosen norm $||\cdot||$.}\\
\noindent Figure~\ref{Post} shows that the true model is easily estimated with the proposed scheme  as the posterior probability emphatically peaks at $L=3$, which implies that $\hat L^{\text{MAP}}=3$. Without penalisation --i.e., with a uniform prior $p_{0,T}$-- we obtain multiple equivalent solutions involving layers with identical values of the estimated parameters.\\

\noindent {Now, it is well known that direct, grid-based deterministic integration of the posterior probability is intuitive but computationally inefficient. Accordingly, we have further considered alternative approximations of the integral in \eqref{eqLkhd} using a nonlinear population Monte Carlo (NPMC) algorithm \cite{Koblents15}, and effectively reduced the runtime by a factor close to $100$ on the same computer  for the example of Fig. \ref{Post} (see Appendices C and D for details). Actually, our NPMC also includes an importance sampling procedure by which an efficient grid of the parameter space is obtained. This procedure meshes in a tight way the regions where the posterior probability density of the parameters is high, and in a sparse way in the regions where the posterior probability density is low. Accordingly, Monte Carlo sampling this mesh is guaranteed to sample parameters with high likelihood (this is a global optimization process, at odds with EM), and a simple inspection of the likelihoods of each sample allows us to robustly decide which is the one with higher likelihood. Accordingly, we are able to infer that the maximum of the posterior probability density $P(\{X(t)\}_{t=1}^N | {\bf T}_L, r ) p_{0,T}({\bf T}_L)$ (i.e., the integrand of Eq. \eqref{eqLkhd} with uniform prior for $r$) is indeed attained {\it at the true value} of ${\bf T}_3$ (see appendix D5 for details). We therefore conclude that our model selection scheme correctly estimates not only the number of layers, but also the transition probabilities within each layer, i.e. the full architecture}.\\ 
Finally, we have also verified that the posterior probability density is smooth close to its maximum as perturbations $\tilde {\bf T}_3={\bf T}_3+\delta {\bf T}$ systematically yield a smaller posterior probability density function for sufficiently large $N$ (see figure 21).

\begin{figure*}[ht]	
\centering
\includegraphics[width=0.9\columnwidth]{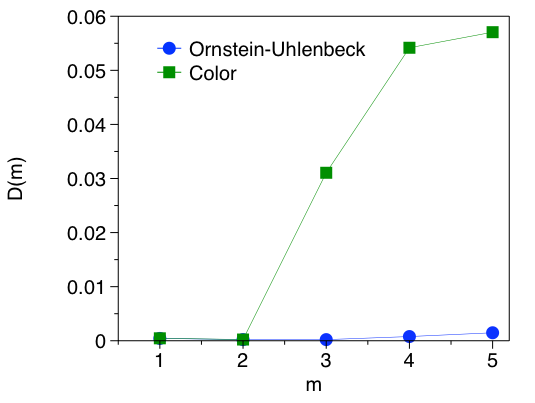}
\includegraphics[width=0.9\columnwidth]{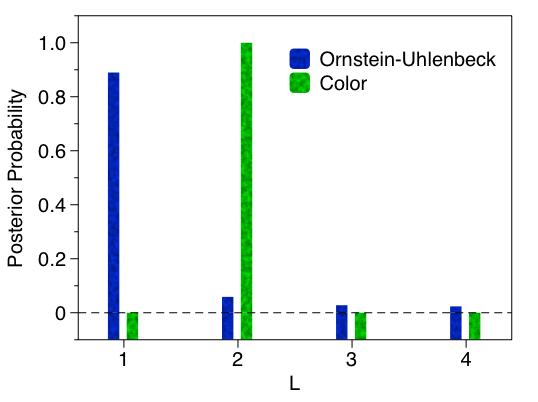}
\includegraphics[width=0.9\columnwidth]{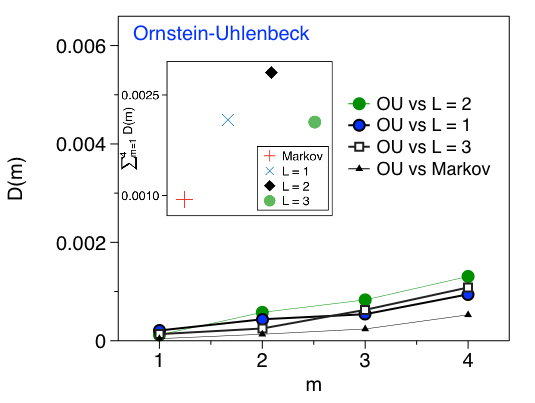}
\includegraphics[width=0.9\columnwidth]{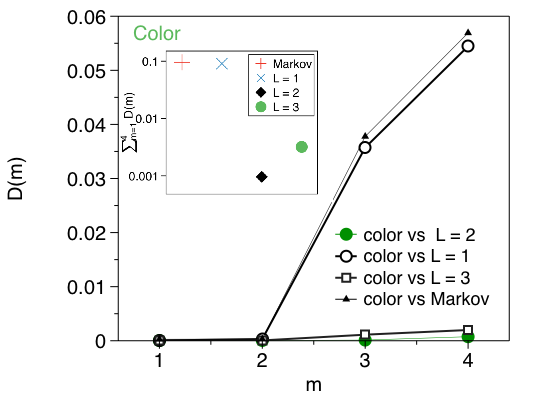}
\caption{(Top, Left panel) Multiplexity detection statistic $\mathcal{D}(m)$, applied to a trajectory generated by (blue) an Ornstein-Uhlenbeck process (Markovian, see the text) and (green) a Langevin equation with coloured noise (non-Markovian, see the text) after embedding the trajectories in a ring topology via eq. \ref{discre}. The multiplexity detection is negative in the Markovian case and positive for the Langevin equation with a non-white noise term, as expected. (Top, Right panel) Posterior probability of a multiplex model with $L$ layers, confirming that $L=1$ (i.e. monoplex) is the most probable model for the Markovian case (Ornstein-Uhlenbeck) and that a model with $L=2$ is the most likely for the case of a Langevin equation with correlated noise. (Bottom, Left panel) $\cal D$(m) (Order-m Kullback Leibler divergence) between the Ornstein-Uhlenbeck (Markovian) process and a synthethic series generated by the most likely model with $L$ layers (for $L=1,2,3$). Estimation of the model parameters (the architecture) is possible thanks to the importance sampling procedure which pre-selects parameter configurations with high likelihood, in such a way that one can find the global optimum by searching in this reduced set.
The synthetic series which shows highest similarity with the original process is found for the model $L=1$ (only improved by the comparison with a Markovianised time series) coinciding with the prediction found by the model selection scheme. (Bottom, Right panel) Similar measures than the left panel, performed on the series extracted from the Langevin equation with coloured noise (non-Markovian). In this case the synthetic series that shows more similarity with the actual non-Markovian series is the one generated by the most likely model with $L=2$ layers.}
\label{OU}
\end{figure*}

\section{A multiplex decomposition of non-Markovian dynamics}


As it happens for any Bayesian inference-based method \cite{CNM, GSP, PRLNewman}, our approach in principle would provide conclusive indication of the hidden multiplexity in the case the prior --intralayer dynamics are diffusive-- is a reasonable assumption. Notwithstanding, the methodology proposed here actually extends above and beyond the reconstruction of hidden multiplex architectures using walkers with partial information. As a matter of fact, a similar approach can be considered even when the architecture is {\it truly} single-layered. Suppose for instance that a given observed time series was  truly the outcome of a non-Markovian dynamics running on a physical single-layered network. In that case, our multiplex reconstruction method would still provide the most probable multiplex model, with $L>1$ due to lack of walker's Markovianity. The key difference is that now layers in the hidden multiplex would be  providing the most probable {\it effective} multiplex reconstruction with Markovian intralayer dynamics that would yield such complex dynamics. Incidentally, note that this brand of {\it effective} models is used in community detection in single-layered networks, which result from finding the optimal number of effective groups of nodes which maximise a certain likelihood function~\cite{PRLNewman}.\\ 
{In what follows, we first capitalise on this new interpretation to extend our previous analysis on random walks on graphs to continuous stochastic processes, and then we present two existence theorems which guarantee that this stochastic decomposition is universally applicable to random sequences with arbitrary memory}.

\subsection{{Extension to continuous processes}}


\noindent {When the original dynamics is continuous we can discretise motion and embed the original time series into a simple graph topology,} for example via the transformation $\{X(t)\}\to \{\tilde{X}(t)\}$ given by $\tilde{X}(0)=0$ and
\begin{equation} 
 \tilde{X}(t+1)=\begin{cases} 
      \tilde{X}(t)+1\mod K, \ \text{if}\  X(t+1)>X(t) \\
      \tilde{X}(t)-1\mod K, \ \text{if}\  X(t+1)<X(t), 
   \end{cases}
\label{discre}
\end{equation}
and apply our multiplexity detection methods to the discretised trace $\{\tilde{X}(t)\}$. 
To illustrate and validate this extension, we have considered two continuous  stochastic processes: (i) an Ornstein-Uhlenbeck process (Markovian) governed by the  stochastic differential equation 
\begin{equation}
\dot{x}=-x + \xi, \label{Eq:Lang}
\end{equation}
 where $\xi$ is a  Gaussian white noise  with zero mean $\langle \xi (t) \rangle=0$, amplitude $\sigma$  and autocorrelation $\langle \xi(t)\xi(t')\rangle=\sigma^2 \delta(t-t')$ and (ii) a generalisation of the preceding Langevin equation where the noise term is not white anymore but has some colour. Such process can be described by the following Langevin equation
 \begin{equation}
\dot{x}=-x + \eta, \label{Eq:LangCol}
\end{equation}
  with $\eta(t)$ being itself defined by an Ornstein-Uhlenbeck process such that $\langle \eta(t)\rangle=0$ and $\langle \eta(t)\eta(t')\rangle \propto \exp(-|t-t'|/\tau)/\tau$ with $\tau>0$ the correlation time of the noise. Note that in Eqs.~(5) and (6) the dot denotes time derivative.   When the noise term $\eta$ is not white anymore, the Langevin equation (6) generates non-Markovian trajectories for the variable $x$. Interestingly, Eq. (6) is attracting considerable attention in soft matter as a minimal model for the  non-Markovian dynamics of the position of a passive particle immersed in an active (e.g. bacterial) bath \cite{active1, active2}.\\
  
\noindent We perform numerical simulations of these processes using an Euler-Mayurama algorithm and subsequently embed the resulting trajectories of $10^4$ data in a cycle-graph topology (see Fig.~\ref{fig0}b) via Eq.~\eqref{discre} with $K=4$, and then we have applied our multiplex detection and estimation protocol. In the top panels of figure \ref{OU} we plot ${\cal D}(m)$ and the outcome of a layer estimation. For the Ornstein-Uhlenbeck process the method does not detect multiplexity (and a layer estimation confirms that a model with $L=1$ layer is the most probable one), in good agreement with the fact that the Ornstein-Uhlenbeck process is indeed a Markovian process. On the other hand, the Langevin equation with coloured noise has a nontrivial memory kernel and generates non-Markovian dynamics, correctly captured by the fact that ${\cal D}(3)>0$. In this case the dynamics optimally decomposes into a Markov switching combination of $L=2$ layers.\\

\noindent To further validate these results,  we now generate synthetic trajectories from the estimated multiplex models with $L$ layers. Importantly, we should recall that our layer estimation method makes use of an importance sampling algorithm to concentrate the Monte Carlo search in the regions of the parameter space with large likelihood. As a byproduct, we are are able to estimate --given the number of layers $L$-- the parameters of the most likely model, and therefore we can now generate synthetic series from this fitted model. 
We then compare the statistics of synthetic series from the most likely model with $L$ layers with those of the original Ornstein Uhlenbeck and non-Markovian processes, respectively. In the bottom panels of Fig.~\ref{OU} we plot the $m$-th order Kullback-Leibler divergence between the original (discretised) series and the synthetic series generated by the reconstructed multiplex model. We confirm that the series generated by the model with $L=1$ and $L=2$ respectively are the ones that show higher similarity  with the Ornstein-Uhlenbeck (Markovian) and non-Markovian series, respectively. 
Finding $L=2$ layers in the non-Markovian case is indeed {reasonable having in mind} that Eq.~\eqref{Eq:LangCol} allows a well-known decomposition as the following 2D (Markovian) stochastic differential equation
 \begin{eqnarray}
\dot{x}&=&-x +  y,\nonumber \\
\dot{y}&=&-\beta y + \xi \label{Eq:LangCol2}
\end{eqnarray}
with $\xi$ a Gaussian white noise process and $\beta>0$ a positive constant.\\

\noindent That is to say, in even more general terms, our methodology provides a mathematically sound solution for the stochastic projection of non-Markovian dynamics onto a base of simple diffusive dynamical modes. {For random sequences which were originally generated by a random walker navigating a hidden multiplex network, the method easily reconstructs such hidden architecture, whereas for general random sequences (such as the ones generated by Eqs. \ref{Eq:Lang} and \ref{Eq:LangCol}) the effective multiplex model provides a good reconstruction of the original dynamics. Hence the question: is this type of hidden jump-Markov model able to fully (i.e., exactly) reconstruct any random sequence (i.e. high order Markovian and fully-non Markovian)? These questions are responded affirmatively in the next subsection, where we state and prove two theorems which address these matters.}



\subsection{\cred{Exact decomposition of Markovian and non-Markovian dynamics via multiplex models}}

\cred{In this subsection we pose the question of whether the statistics of arbitrary random sequences $\{X(t)\}_{t\geq0}$, possibly with long memory, can be recovered {\it exactly} using a hidden jump-Markov model of the class described in Section \ref{smethods} (from now on we refer to it as a multiplex model). The answer is positive (in a probabilistic sense to be made precise), as summarised by two representation theorems.}\\

\noindent \cred{In particular, we first state and prove that every Markov model of finite (but {arbitrarily} large) order $h$ can be recast into a multiplex model. Then we address the representation of models with infinite memory, by letting $h\rightarrow\infty$, and show that, under mild regularity assumptions, they admit a compact representation in the form of a model with an uncountable (continuous) set of hidden layers. For readability, we have relegated as many technical details as possible (including all the mathematical proofs) to Appendices F and G, and only discuss here the key results and implications.}

\subsubsection{\cred{Representation of Markov models of order $h$}}

\cred{
Let us consider a discrete state space with $K$ elements, $\mbK=\{1,\ldots,K\}$ (this will be the node set of the multiplex). A Markov sequence of (integer) order $h\geq1$ is defined by the set of $K^{h+1}$ probability masses 
\begin{equation}
P_{\mbK}^h(i_t|i_{t-h:t-1}):= P( X(t)=i_t|X(t-h:t-1)=i_{t-h:t-1} )  \nonumber
\end{equation}
{where $X(t-h:t-1) = \{ X(t-h), \ldots, X(t-1) \}$ and $i_{t-h:t} = \{ i_{t-h}, \ldots, i_t \} \in\mbK^{h+1}$ is a sequence of $h+1$ state space observations.} If we fix $i_{t-h:t-1}$, then $P_\mbK^h(i_t|i_{t-h:t-1})$ is a probability mass function (pmf) and, as a consequence, $P_{\mbK}^h(i_t|i_{t-h:t-1})\in[0,1]$ and $\sum_{i_t\in\mbK}P_{\mbK}^h(i_t|{i_{t-h:t-1}})=1$.\\
}

\noindent \cred{
Now, any Markov model of finite order $h$ can be transformed into an {\it equivalent} multiplex model with a sufficiently large, but finite, number of layers $L$, as formally stated below.} 

\cred{
\begin{teorema} \label{th_A}
For every Markov model of order $h<\infty$ on the state space $\mbK=\{1,\ldots,K\}$ there exists an equivalent multiplex model, with observation space $\mbK$, and $L=K^h$ layers.
\end{teorema}
}

\noindent \cred{
The proof of Theorem \ref{th_A} is technical and is therefore put in Appendix \ref{ap_A}. Theorem \ref{th_A} guarantees that every random Markov sequence with finite memory $h$ can be represented by a multiplex model with $K^h$ layers (i.e., this is an existence theorem). The theorem does not state that this representation is unique, though, and it does not state that it is minimal either. According to the numerical evidence given in the previous sections, we conjecture that a suitable selection of ${\bf R}_L$ and ${\bf T}_L$ (e.g., using the estimation methods described in this paper) can yield an accurate representation of a sequence of order $h$ with considerably less than $K^h$ layers.}\\
{Let us also remark that there is no contradiction between Theorem \ref{th_A} and our earlier claim that multiplex models can represent systems with infinite memory. Indeed, depending on the choice of its parameters ($L$, ${\bf R}_L$ and ${\bf T}_L$), a multiplex system can yield random sequences with either finite or infinite memory. For example, in the proof of Theorem \ref{th_A} we explicitly construct a multiplex model that matches the transition probabilities of a Markov model of order $h$. On the other hand, multiplex models where the transitions between layers are independent of the walker's past trajectory yield sequences with infinite memory (except for pathological cases).}

\subsubsection{\cred{Representation of sequences with infinite memory}}

\cred{In a second part, we extend the previous existence theorem to a wide class of infinite memory (i.e. fully non-Markovian) models.
Let $\{X_h(t)\}_{t\geq0}$ denote a random sequence on $\mbK=\{1,\ldots,K\}$ that can be represented exactly by a Markov model of order $h$. We consider here the class of sequences with infinite memory that can be obtained from Markov models as $h\rightarrow\infty$ and refer to them as ``Markov-$\infty$" sequences. To be precise, we say that $\{X(t)\}_{t\geq0}$ is the limit of $\{X_h(t)\}_{t\geq0}$ as  $h\rightarrow\infty$, and write
\begin{equation}
X \stackrel{d}{=} \lim_{h\rightarrow\infty} X_h
\end{equation}
when we can approximate the transition probabilities of the sequence $\{X(t)\}_{t\geq0}$ with an arbitrarily small error using a Markov model of sufficiently large order. Specifically, we need to introduce the following technical definition:
\begin{definicion}
The random sequence $\{X(t)\}_{t\geq0}$ is Markov-$\infty$ if it satisfies the regularity conditions below:
\begin{itemize}
\item[(C1)] The joint probability of any sequence of states vanishes uniformly with the length of the sequence, i.e.,
$$
\lim_{t\rw\infty} \sup_{ i_{0:t} \in \mbK^{t+1} } P(X(0:t) = i_{0:t}) = 0.
$$
\item[(C2)] There exists a sequence of Markov models $\{X_h(t)\}_{t\geq0}$, $h=1,2,\ldots$, such that for any $\epsilon>0$, arbitrarily small, there exists $h' \in \mbN$, sufficiently large, that guarantees
\begin{eqnarray}
\sup_{i_{0:t} \in \mbK^{t+1}} &|& P( X(t)=i_t | X(0:t-1) = i_{0:t-1} )  - \ldots \nonumber \\
&& \ldots - P_\mbK^h(i_t|i_{t-h:t-1}) \quad | < \epsilon, \nonumber
\end{eqnarray}
for every $t > h$ and
\begin{eqnarray}
\sup_{i_{0:t} \in \mbK^{t+1}} &|& P( X(t)=i_t | X(0:t-1) = i_{0:t-1} )  - \ldots \nonumber \\
&& \ldots - P_\mbK^h(i_t|i_{t-h:t-1}) \quad | = 0, \nonumber
\end{eqnarray}
for every $t \le h$, whenever $h > h'$.
\end{itemize}
\end{definicion}
}

\noindent \cred{Let $\{ X_L(t) \}$ represent a random sequence generated by a multiplex model with $L$ layers. Since every Markov-$\infty$ model is the limit of a sequence of Markov systems with increasing order, $h \rw \infty$, they can also be obtained (via Theorem \ref{th_A}), as the limit of a sequence of multiplex systems as the number of layers grows, $L\rw\infty$. Moreover, it turns out that the limit $\lim_{L\rightarrow\infty} \{ X_L(t) \}_{t\ge 0}$ can be interpreted itself as a multiplex model with an \textit{uncountable set of layers} and a first-order Markov system on $\mbK$ associated to each layer. This is made precise by the following theorem:}

\cred{
\begin{teorema} \label{th_B}
Let $\{X(t)\}_{t\geq0}$ be a Markov-$\infty$ random sequence on $\mbK$. There exists a Markov kernel
\begin{equation}
\mM:\mB([0,1))\times[0,1)\rightarrow[0,1],
\end{equation}
where $\mB([0,1))$ denotes the Borel $\sigma$-algebra of subsets of $[0,1)$, a probability measure $\mM_0:\mB([0,1))\rightarrow[0,1]$, and an uncountable family of $K\times K$ transition matrices 
\begin{equation}
{\bf T}(y)= 
\begin{bmatrix}
T_{11}(y)    &\ldots       & T_{iK}(y) \\
\vdots	&\ddots	&\vdots\\
T_{K1}(y)    &\ldots       & T_{KK}(y)  \\
\end{bmatrix}, \quad y\in[0,1),
\end{equation}
such that
\begin{eqnarray}
&P(X(t)=i_t | X(0:t-1) = i_{0:t-1} )  =& \nonumber \\
&=P_{i_0}\int\dots\int\prod_{k=1}^{t-1}T_{i_{k+1}i_k}(y_k)\mM(dy_k|y_{k-1})\mM_0(dy_0), & \nonumber
\end{eqnarray}
for every $i_{0:t} \in \mbK^{t+1}$, where $P_{i_0}=P( X(0) = i_0)$.
\end{teorema}}

\noindent {See Appendix G for a proof. Theorem \ref{th_B}} \cred{indicates that multiplex network models can be generalised to obtain probabilistic systems with an infinite and uncountable number of layers, which are flexible enough to represent (i.e., exactly recover) a broad class of random sequences with infinite memory. 
}

\begin{figure}[htp]	
\centering
\includegraphics[width=0.9\columnwidth]{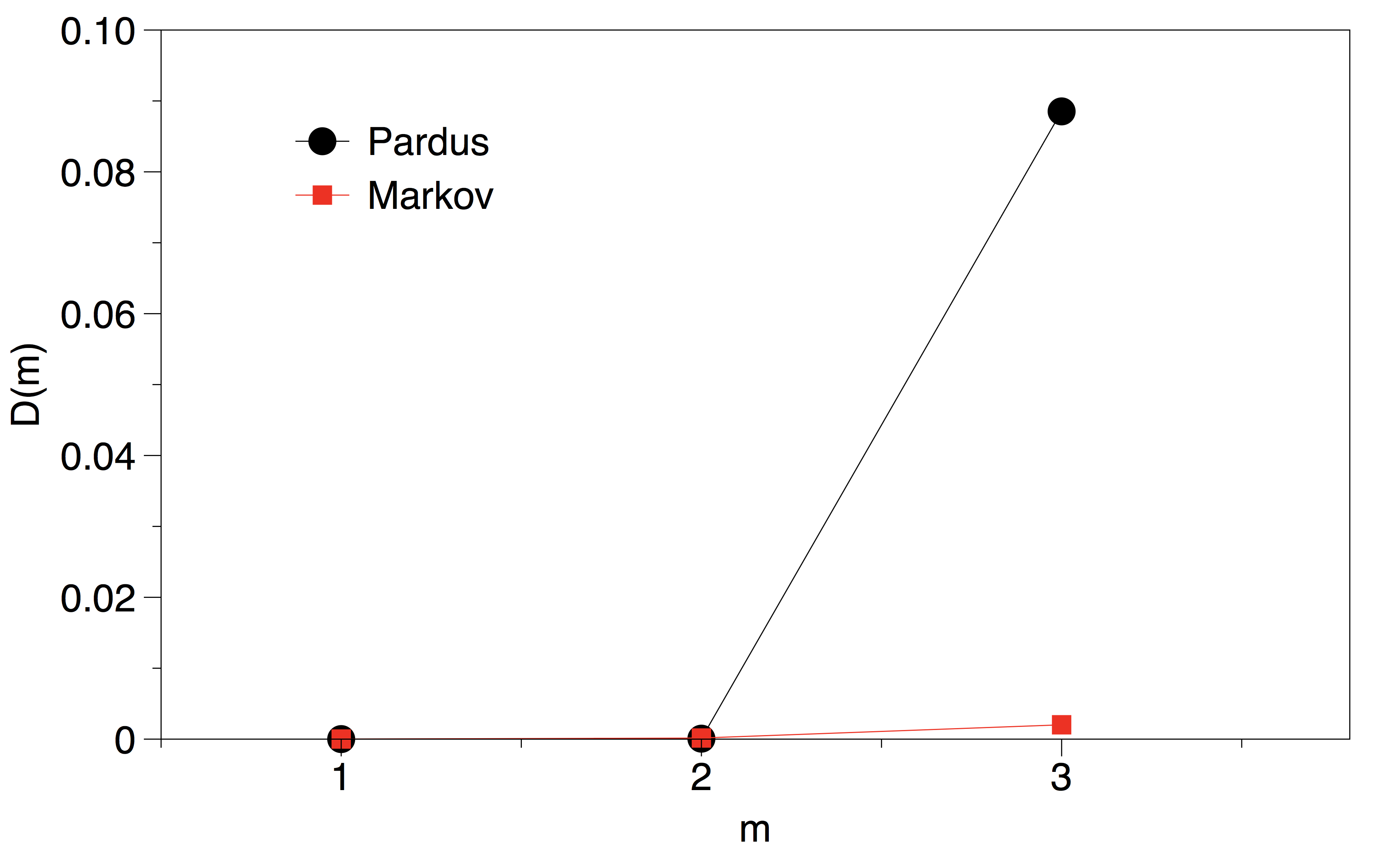}
\caption{Multiplexity detection statistic $\mathcal{D}(m)$, applied to a experimental mobility trajectories on the Pardus universe \cite{Szell}. We find $\mathcal{D}(3)>0$ which suggests that mobility series are non-Markovian, in agreement with independent evidence \cite{Szell}.}
\label{f1}
\end{figure}

\section{Applications to experimental data}
 We now apply our methodology to two experimental recordings of completely different nature: the mobility dynamics of human players of an online videogame and the dynamics of polymerases during RNA transcription. 

\subsection{Application to mobility: the Pardus universe }

Human and animal mobility~\cite{mobilityJesus1, mobilityJesus2, mobilityJesus3} are often described by dynamical processes  on single-layer networks, and interestingly, has been found to {signatures of memory~\cite{mobility, mobility2, Salnikov}, which can be interpreted as a deviation from Markovianity}. Can such lack of Markovianity  be interpreted as being the result of a Markovian dynamics taking place on a hidden
multiplex network? In the case where mobility takes place across (hidden) multimodal transportation systems (as when we collect GPS traces of urban mobility), layers could be physical (underground, bus, car, etc). On the other hand, animal foraging dynamics are clearly different and alternate during day and night \cite{day}. In a similar vein, human mobility patterns change when switching from work/leisure styles \cite{leisure}: these would be cases where layers would be effective, rather than physical. There are a large variety of problems involving the aforementioned scenarios which could be amenable to our approach. To guarantee computational efficiency, we would only require the network over which the agents move not to be too large, something that in the general case can be achieved by coarse-graining the network via community detection. Mathematically, within this context a non-Markovian process running on a monoplex vs a Markov switching process running on a multiplex are equally valid models,  much in the same way a function is equivalent to its Fourier series representation. Still, we consider this new interpretation not just suggestive but also parsimonious from a cognitive point of view, and therefore might be of relevance in the study of memory in search processes.

To illustrate this type of application, we consider experimental mobility trajectories performed by players (agents) in a virtual environment: the Pardus universe (see \cite{Szell} for details). 
The Pardus universe is a multiplayer online role-playing videogame which is used as a large scale socio-economic laboratory to study mobility in a  controlled way. It consists of an (online) physical network with $K=400$ nodes, a networked universe with social and economic activities, where the players the game move around. The mobility traces of these players can then be naturally symbolised by coarse-graining the original network via community detection into a network of 20 non-overlapping communities wired by a $20 \times 20$ weighted adjacency matrix with complex topology \cite{Szell}.
 
 Szell {\it et al}  analysed online player's trajectories  and reported evidences of long-term memory in the diffusing patterns of agents mobility in this universe~\cite{Szell}. Here we apply the multiplexity detection statistic to a long time series obtained by concatenating individual traces. In Fig.~\ref{f1} we show the numerical results (black dots), along with the null case of a Markovian diffusion (red squares) over the same network. We find $\mathcal{D}(3)>0$ as a footprint of multiplexity, which can be here interpreted as the hidden presence of several {\it effective} layers. Szell {\it et. al.} introduced a long-term memory model to account for the mobility dynamics, and the heterogeneity of players. An alternative interpretation  is that players are performing simple diffusion dynamics, but are switching stochastically between different effective dynamical regimes. The challenge, which we leave as an open question for future work, would be to assign a social or perhaps cognitive meaning to the different layers. In this sense, our method is here closer to an unsupervised clustering paradigm than to a supervised one. 
Since our experimental trace is given by the  concatenation of traces of different walkers,  
 the different effective layers could also be revealing here a taxonomy of different game strategies.\\
\begin{figure*}[ht]	
\includegraphics[width=0.64\columnwidth]{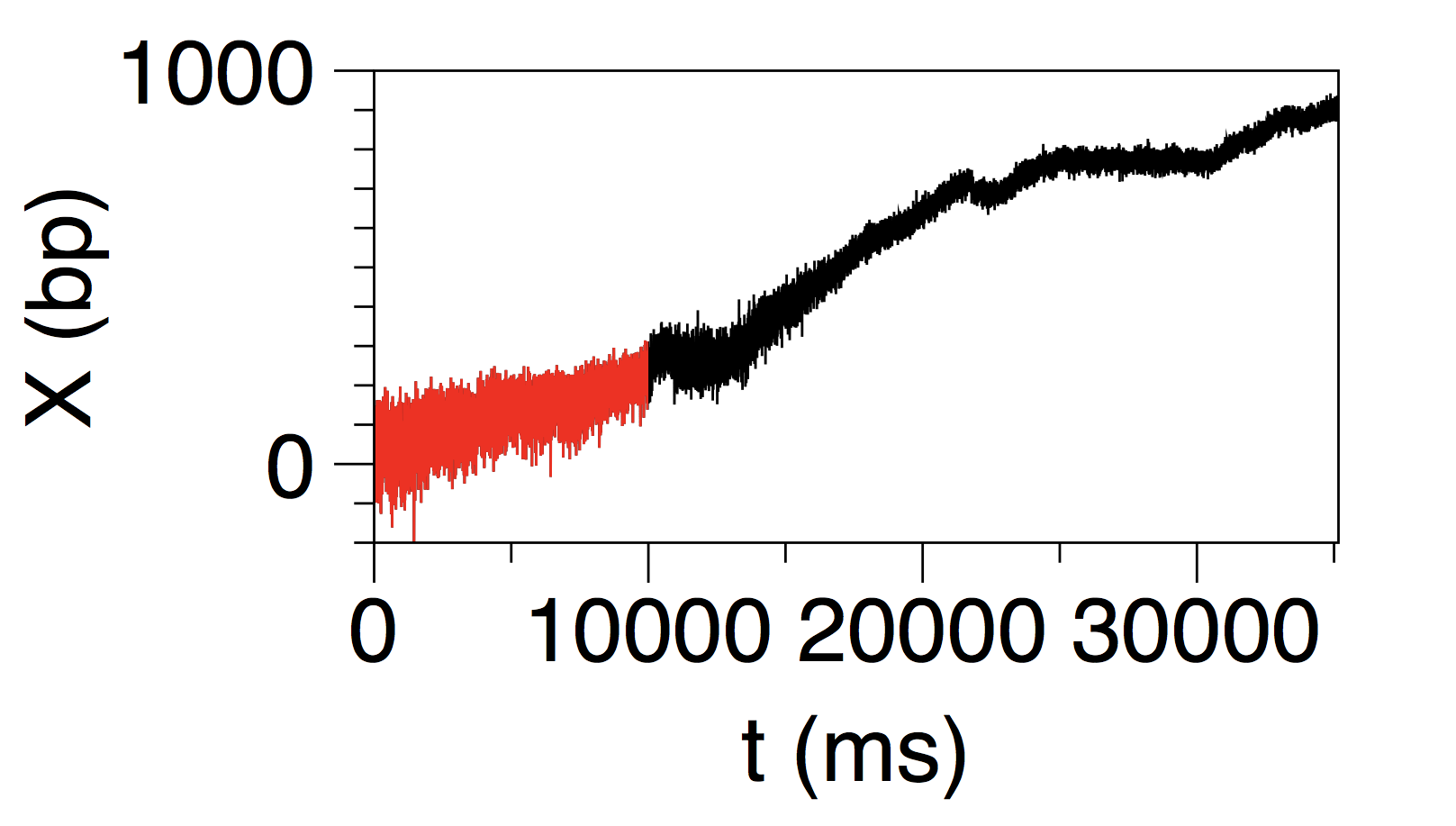}
\includegraphics[width=0.6\columnwidth]{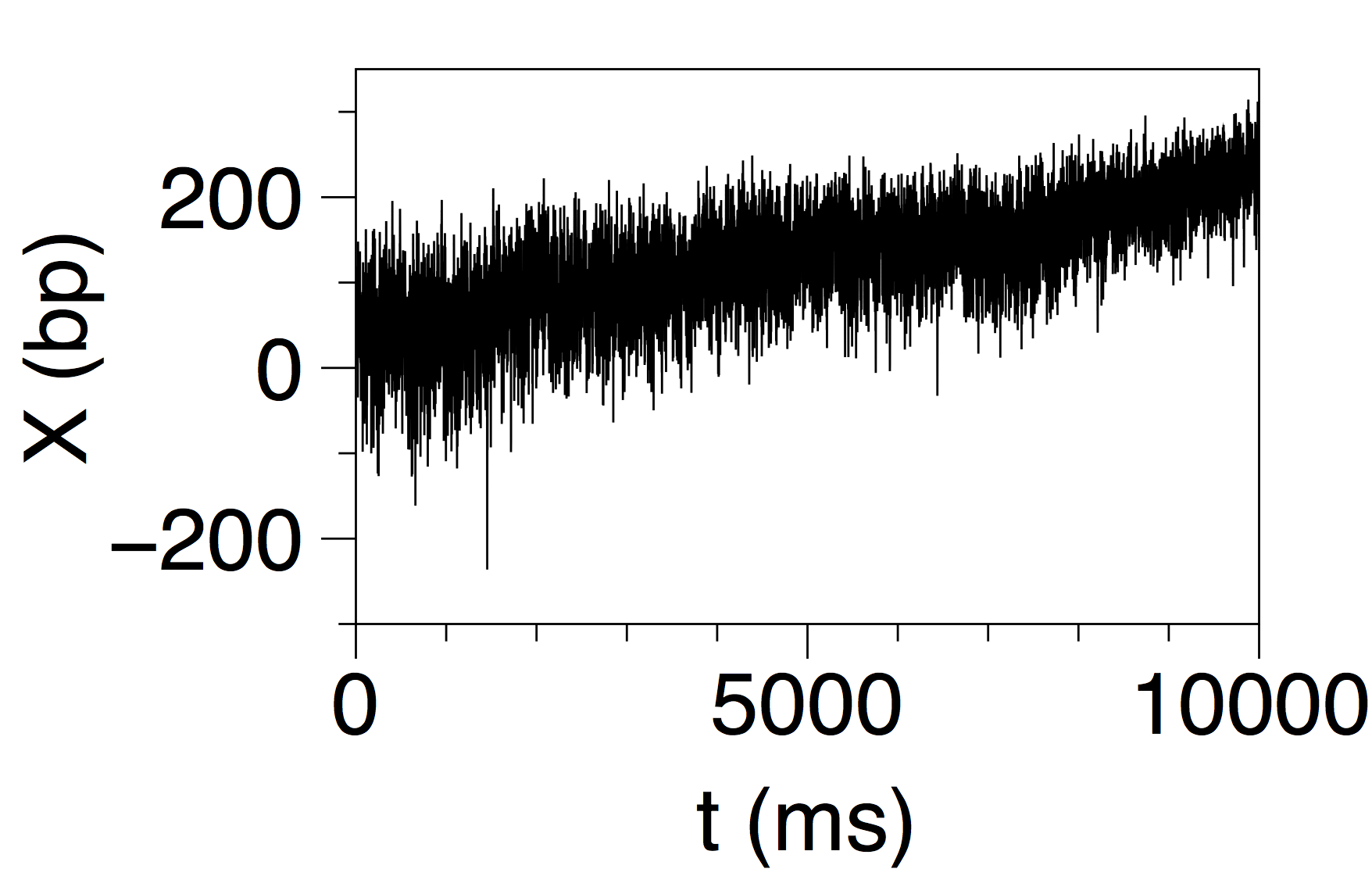}
\includegraphics[width=0.6\columnwidth]{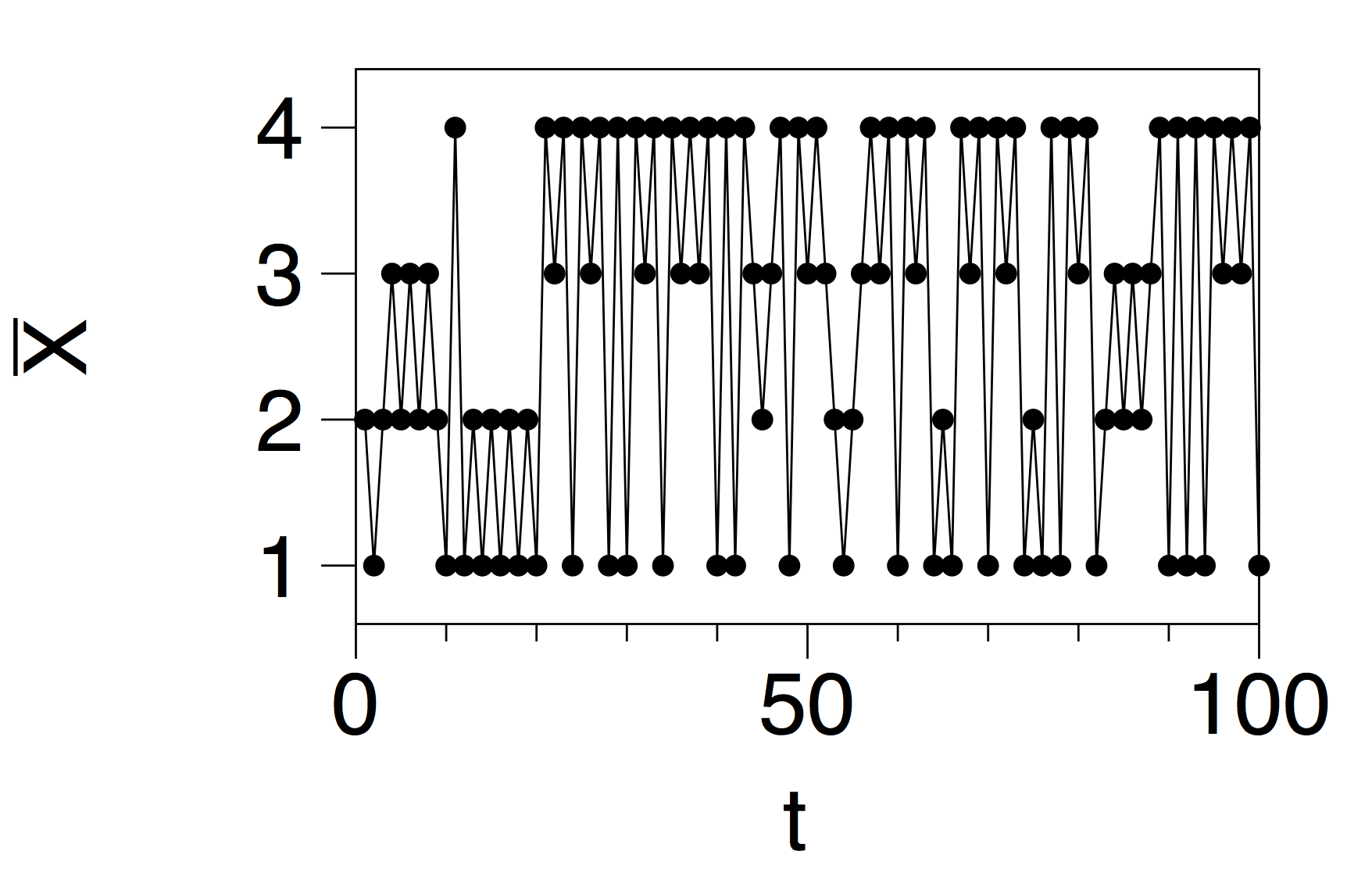}
\includegraphics[width=0.9\columnwidth]{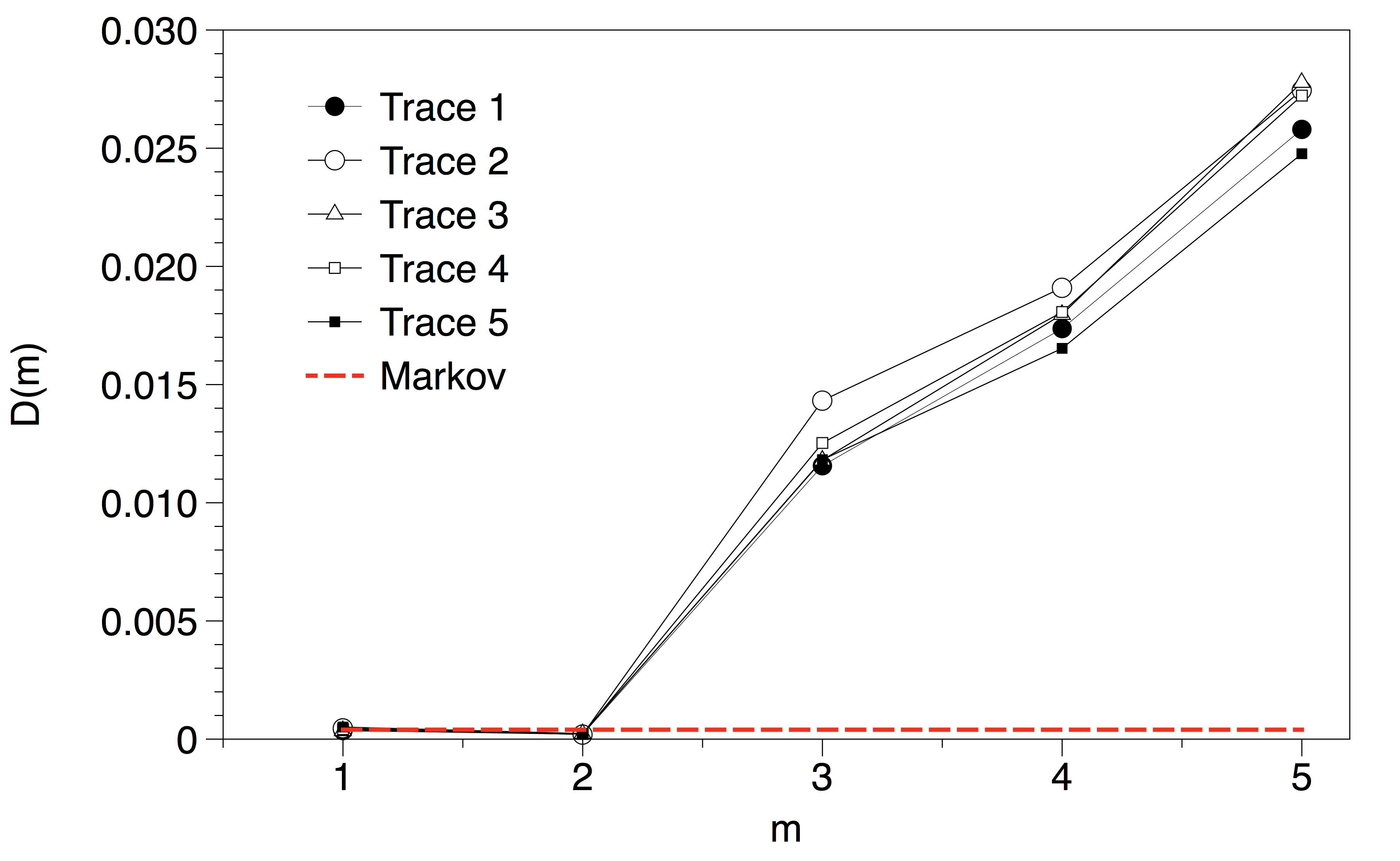}
\includegraphics[width=0.9\columnwidth]{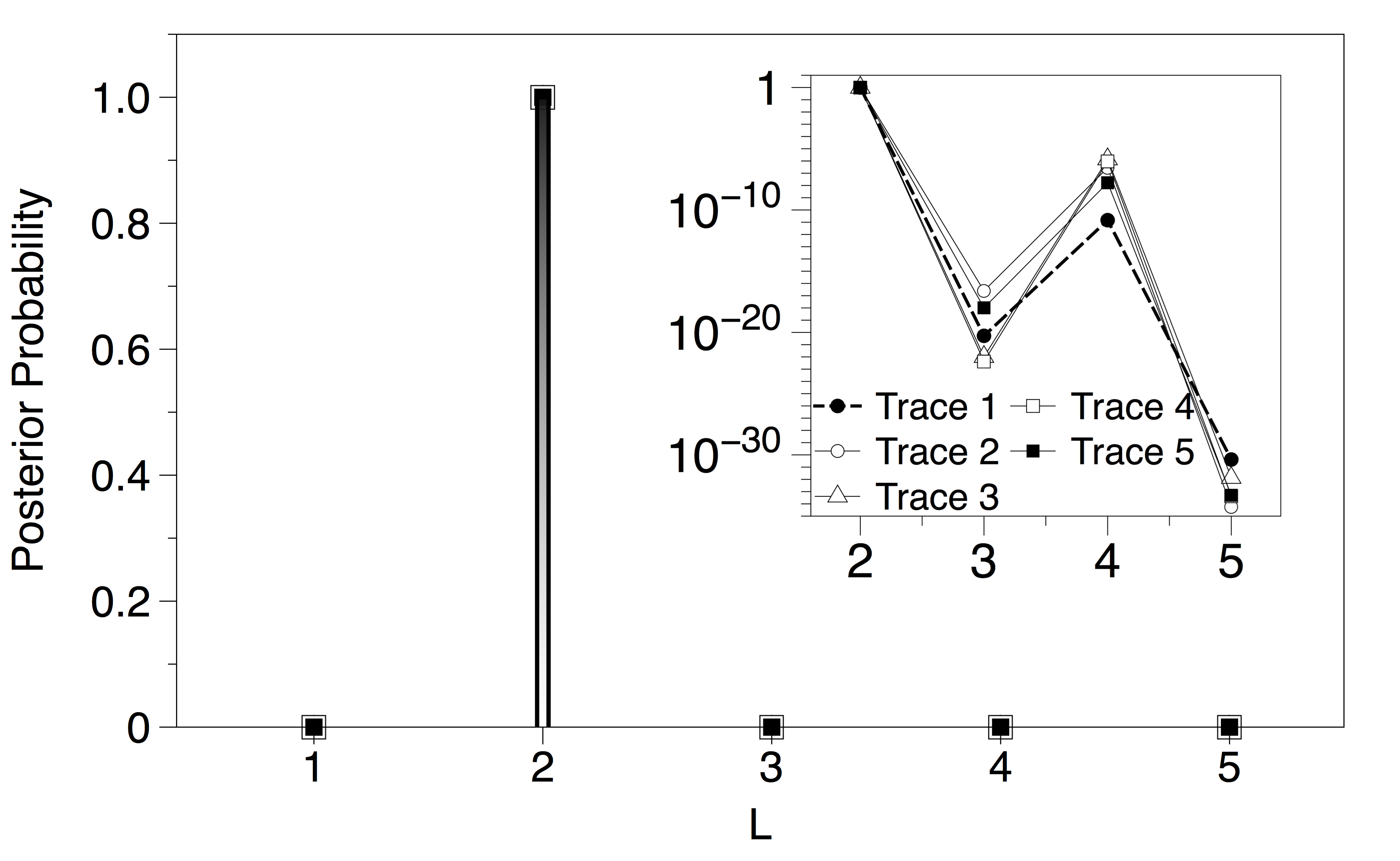}
\caption{(Upper panels) Experimental RNA Polymerase I trace $X(t)$ (in basepair units) from which we extract an excerpt of the first $10^4$ time steps (highlighted in red in the original trace, shown in the middle panel), and a sample of the first 100 data of its symbolised trace $\tilde{X}(t)$ (right). Experimental traces of Polymerase I are extremely noisy and one cannot always easily distinguish different dynamical regimes (see appendix E for additional traces studied in this work).
(Bottom, left) Multiplexity detection statistic $\mathcal{D}(m)$ applied to five (symbolised) experimental RNAP traces. We consistently find $\mathcal{D}(3)>0$, which suggests that to correctly decribe the dynamics of RNA Polymerase I at least we need two diffusive layers.
(Bottom, right panel) Posterior probability for the layer estimation applied to five experimental series. We confirm that, with overwhelming probability, the most likely number of layers is $L=2$. {One possibility is that the identification of $L>1$ is the consequence of the presence of coloured noise in the single-molecule optical tweezer transcription experiment, whereas another possibility is that $L=2$ correspond to the elongation (active state) and backtracking (passive state) dynamical modes (in the inner panel the same results are shown in log-linear scales, the probability associated to $L=1$ is zero so is not defined in a log scale).}}
\label{f2}
\end{figure*}

\subsection{Application to biology: transcription by eukaryotic RNA polymerases} 
It has been shown that the ratchet mechanism plays an important role in active transport by molecular motors in living cells~\cite{ratchet2}.
A paradigmatic example of a molecular motor that can be well described by a ratchet mechanism is RNA polymerase (RNAP). RNAPs are macromolecular enzymes responsible for the transcription of genetic information encoded in the DNA into RNA ~\cite{CramerReview}. During transcription the spatial location of RNAPs along the DNA template exhibits noise due to thermal fluctuations. In addition the dynamics of polymerases exhibits switching between two different dynamical regimes: an active polymerisation state called \textit{elongation} and a passive or diffusive state called \textit{backtracking}~\cite{galburt}. While elongating, RNAP moves along the DNA template with a net velocity of the order of $\sim 20$ nucleotides per second with $0.34$nm being the distance between two nucleotides. In the backtracking regime, the polymerisation reaction is stalled and RNAPs perform passive Brownian diffusion due to several types of noise (e.g. thermal and chemical). Optical tweezers enable measuring the motion of a single RNAP during transcription of a single DNA template at a basepair resolution~\cite{abbon,galburt}. Single-molecule experimental RNAP traces are however extremely noisy and the problem of identifying the possible mixture of different underlying dynamical processes, including transitions from elongation to backtracking as well as the hidden presence of other regimes, from a single experimental trace is a challenging task.

At the single-molecule level, the dynamics  RNAP corresponds to a molecular motor producing a non-Markovian dynamics which can thus be modelled as a Markov switching random walk dynamics in a biochemical multiplex with  $L=2$ layers (corresponding to elongation and backtracking respectively)~\cite{dang}. This suggests that our methodology (using cycle graphs with homogeneous transition rates as the prior topology) is applicable and under the aforementioned assumptions we should find both $\mathcal{D}(3)>0$ and the model with $L=2$ should have a clear maximum likelihood. To test such prediction, we have applied our complete methodology to five single-molecule experimental traces of the position $X$ of RNA polymerase I (Pol I) from yeast \textit{S. cerevisiae}, obtained with a dual-trap optical tweezer setup in the assisting force mode~\cite{lisica}. We systematically choose the first $10^4$ data points at $1$kHz sampling rate (see Figure \ref{f2} and Appendix E) to keep the time series short and make the inference problem harder (for instance, for the series shown in Figure \ref{f2} it is not easy to visually distinguish the elongation and backtracking regimes). Since the original traces have continuous state space, we
discretise these seried by embedding the experimental recordings into a cycle graph via the transformation proposed in equation \ref{discre} for $K=4$.
An example is shown in the top-right panel in Fig.~\ref{f2}. In the left-bottom panel of the same figure we 
plot $\mathcal{D}(m)$ applied to $\{\tilde{X}(t)\}$ for each of the five experimental series, consistently showing $\mathcal{D}(3)>0$. For comparison and control of finite size effects,  a similar measure is computed on a null model: a time series of $10^4$ data points generated by a (monoplex) Markov chain whose transition matrix has been estimated from $\tilde{X}(t)$ (red dashed line). We can conclude that the underlying dynamics requires at least two alternating dynamics --a stochastic alternation between two diffusive layers--, hence the projection onto an effectively multiplex model. 

Subsequently, in the bottom right panel of the same figure we provide the results on the layer estimation using our nonlinear population Monte Carlo algorithm (convergence after 12 iterations, $10^2$ samples per iteration). The algorithm directly provides $\log( P( {\cal O} | L ) )$, so assuming a uniform prior on the number of layers  $\log( P( {\cal O} | L ) )\propto \log( P( L | {\cal O} ) )$, i.e. the logarithm of the a posteriori probability of model $L$.
The true probability of the model (in natural units) is subsequently extracted and plotted accordingly. We find that the probability is essentially one for model with $L=2$ layers and negligible for the rest. Our algorithm thus reveals that the optimal hidden multiplex has $L=2$ effective layers. 
{One possibility is that the identification of $L>1$ is the consequence of the presence of coloured noise in the single-molecule optical tweezer transcription experiment. Quite intriguingly, additional evidence points to the presence of non-Markovianity also in the backtracking regime (see Appendix E), what might suggest the presence of coloured noise even in the backtracking mode, as in the example in Sec. III.
Another possibility is, as previously discussed, that the two effective layers  correspond to the biochemical mechanisms of elongation (active state) and backtracking (passive state) dynamical modes.}
Finally, we expect our approach to be also applicable to more complicated scenarios such as to identify the number of different nucleotides in copolymerisation processes  of templates with strong  disorder~\cite{gaspard}.  

\section{Discussion and Conclusion}

 In this work we have introduced a method that both detects and {quantifies the degree of multiplexity} in the hidden underlying structure of a networked system by only having access to local and partial statistics of a random walker. {Our working hypothesis (prior) is that there is a hidden multiplex where walkers diffuse, switching layers stochastically and diffusing over each layer. 
Under these circumstances, any random walker for which we only see a projection of such trajectory in the aggregated network is necessarily non-Markovian, if the number of layers is larger than one. Hence our algorithm for multiplexity detection exploits such breaking of Markovianity as a means to detect multiplexity. \\
 Incidentally, here we have focused in the specific case of multiplex networks, where every layer has the same number of (replica) nodes. Actually, in a multiplex one can even have the same topology in each layer, where only transition weights differ, as in the case of the multiplex cycle graphs considered above. On the other hand, in a generic multi-layer network each layer will have in general different number of nodes and different topology. This latter situation can be reinterpreted as having a multiplex where in each layer we can have isolated nodes which are never reached by a walker, or forbidden transitions. Accordingly, we envisage that layers in a multiplex would be in general harder to distinguish via our method than layers in a generic multi-layer network, and as such we expect that the detection method to be easily generalisable to the multi-layer case, something that should be studied in the future.}\\

\noindent {In a second step,} we have introduced a probabilistic scheme to estimate the most probable number of layers composing the hidden multiplex. Note that probabilistic model selection is not new in network inference, for instance in \cite{roger} a similar concept is used to estimate the most probable combination of basic block models that accounts for a certain network topology ({see also \cite{tiago1, tiago2, tiago4, tiago5}}), {whereas in \cite{PRLNewman} a probabilistic framework was developed to estimate the most probable number of communities in a single-layer network}. {Formally similar strategies to estimate model parameters based on $\epsilon$-machines or jump-Markov system identification have also been put forward in the Nonlinear Dynamics \cite{Crutchfield} and Control Theory \cite{Ozkan} communities, respectively.}
{In our case, the posterior probabilities quantify the likelihood of having a hidden multiplex with $L$ layers, {i.e. our approach for multiplex model estimation is purely Bayesian}.  We were able to demonstrate the validity and accuracy of this second part in simple synthetic networks (multiplex cycles) up to $L=5$ layers, as reported in the main text and appendices. Since the model selection protocol can be seen as {a multidimensional Bayesian inference problem, these schemes --similarly to Hidden Markov Models and other methods in statistical inference-- suffer from poor scalability:} essentially the computation of the model posterior probability explodes with the number of unknowns. This is a limitation of the method, and its optimisation is therefore an open problem for future work. As a matter of fact, a simple and intuitive (although inefficient) way to estimate these posteriors is to use a (deterministic) grid integration scheme. In an effort to improve scalability and optimise such calculations we {have proposed a nonlinear population Monte Carlo algorithm (described in full detail in appendices C and D) which reduces the computer runtime by a factor close to $10^2$, without performance loss, for all the examples where we had previously used deterministic integration.}} Interestingly enough, our model selection scheme not only selects the most probable number of hidden layers: by capitalising on an importance sampling routine that focuses in a small region of the parameter space where the likelihood is concentrated we can also provide a Bayesian estimation of the model parameters (i.e. the topology and transition rates), thereby estimating not only the most probable number of layers but, given that number of layers, the full architecture.\\

{We have thoroughly illustrated the validity and scalability of the whole method by addressing several synthetic systems of varying complexity, as depicted in Sec. II  and the Appendices, where we show that we can reconstruct the full architecture of the hidden multiplex network by analysing the statistics of random walks over the projected network.\\}

\noindent {Interestingly, our method can be applied to signals of arbitrary origin, extending the formalism to deal also with continuous processes --i.e. time series which are not necessarily random walkers navigating a network--, {after a simple graph embedding (series discretisation).} Under this extension, our hidden jump-Markov model provides a decomposition of a given non-Markovian dynamics into a Markov switching combination of diffusive modes, and thus enjoys larger generality: the reconstructed multiplex model embeds the originally non-Markovian signal into a random walk navigating an effective multiplex network, where each layer accounts for a different type of diffusive dynamics. We validated this extension by analysing canonical continuous processes (Ornstein-Uhlenbeck process and a Langevin equation with coloured noise). Furthermore, we have proved two existence theorems which guarantee that an exact reconstruction is always possible, for any type of (arbitrary) finite order Markovian and fully non-Markovian (i.e. infinite memory) process.} \\

\noindent Finally, we applied this methodology in two experimental scenarios (a case of human mobility in an online universe and the analysis of the dynamics of RNA Polymerase) and hence showcased its applicability in real, experimental data.\\

\noindent To conclude, starting from the question on whether it is possible to disentangle the hidden multiplex architecture of a complex system if one only has experimental access to a projection of this architecture, in this work we have elaborated a mathematical and computational framework that actually deals more generally with the decomposition of non-Markovian dynamics. When these series are indeed traces of walkers navigating a network, {under the premise of having intralayer diffusion} our approach provides a workable solution for the unfolding of a multiplex network from its aggregated projection. {We should make clear that, generally speaking, it is not possible to assert that the effective multiplex representation is {\it the true architecture}, much like one cannot typically claim that there exist true communities in an observed network --but rather, one says that observations are better reproduced by a multiplex model than by a monoplex one, in a similar spirit as models of networks with structure sometimes reproduce better the observations that models that lack such community structure.}\\
In the general case our method provides a potentially useful approach to disentangle combination of dynamical regimes that appear intertwined in noisy dynamics, this being for instance the case of RNA polymerase moving in a noisy environment and stochastically switching between an active and a passive state. {This suggests that applications of this work not only include Network Science but extend to other fields in Biophysics, Condensed Matter Physics, gambling or Mathematical Finance, where non-Markovian signals pervade.}\\

\noindent {\bf Acknowledgments. }We sincerely thank Michael Szell, Roberta Sinatra and Vito Latora for sharing data on Pardus universe and for fruitful discussions, and anonymous referees for useful comments. L.L. acknowledges funding from EPSRC Early Career Fellowship EP/P01660X/1. I.P.M. acknowledges the financial support of the Spanish Ministry of Economy and Competitiveness (projects TEC2015-69868-C2-1-R ADVENTURE and TEC2017-86921-C2-1-R CAIMAN).
J.M. acknowledges the financial support the Spanish Ministry of Economy and Competitiveness (project TEC2015-69868-C2-1-R ADVENTURE) and the Office of Naval Research (ONR) Global (Grant Award no. N62909-15-1-2011).\\

\noindent LL, IPM and JM contributed equally to this work.

\onecolumngrid

\appendix

\section{A FEW NUMERICAL CONSIDERATIONS}

\noindent {\bf Bounds on switching rate. }Numerically, the problem of detecting multiplexity via statistical differences between the Markovian surrogate $Y(t)$ and $X(t)$ should, in principle, be easier when the switching rate $r$ is small enough, such that we allow trends associated to each layer to build up in the series, but large enough such that $D(1)=D(2)=0$ (that is, large enough such that the one and two step joint distributions are still equivalent). A simple theoretical lower bound is $r<1/m$ (as the average size of a trend is $1/r$, and this should be at least as large as the block size). Note that this bound is not tight in what refers to real-world scenarios (as for human mobility), where switching rate
is normally low in order to avoid unnecessary delays, or has characteristic timescales
which are much lower than the diffusion timescales (as for changing foraging mode
between day and night, in animal mobility).\\

\noindent {\bf Finite size bias in KLD. }As a technical remark, note that $\text{KLD}(p||q)$ diverges if the distributions $p$ and $q$ have different supports (i.e., if $q(m)=0, p(m)\neq 0$ or $p(m)=0,\  q(m)\neq 0$ for some value $m$). In order to take appropriately weight this possibility while maintaining the measure finite in pathological cases, a common procedure \cite{roldan2012} is to introduce a small bias that allows for the possibility of having a small uncertainty for every contribution. Here we introduce a bias of order ${\cal O}(1/n^2)$ where $n$ is the series size (i.e., we replace all vanishing frequencies with $1/n$, and we normalise the frequency histogram appropriately). 


\section{INFERRING MULTIPLEXITY IN SOME ADDITIONAL SCENARIOS}
\noindent We start by providing in figure \ref{fig1_sss} an additional analysis similar to Figure 2 in the main manuscript, but where there is no induced current. The method correctly detects multiplexity in this arguably more complicated scenario. From now on we use ${\cal D}(m)$ and $\text{KLD}_m/m$ indistinctively to express the Kullback-Leibler divergence of order $m$ between a series $X$ and its Markovianised surrogate.\\

\begin{figure}[ht!]
\centering
\includegraphics[width=0.5\columnwidth]{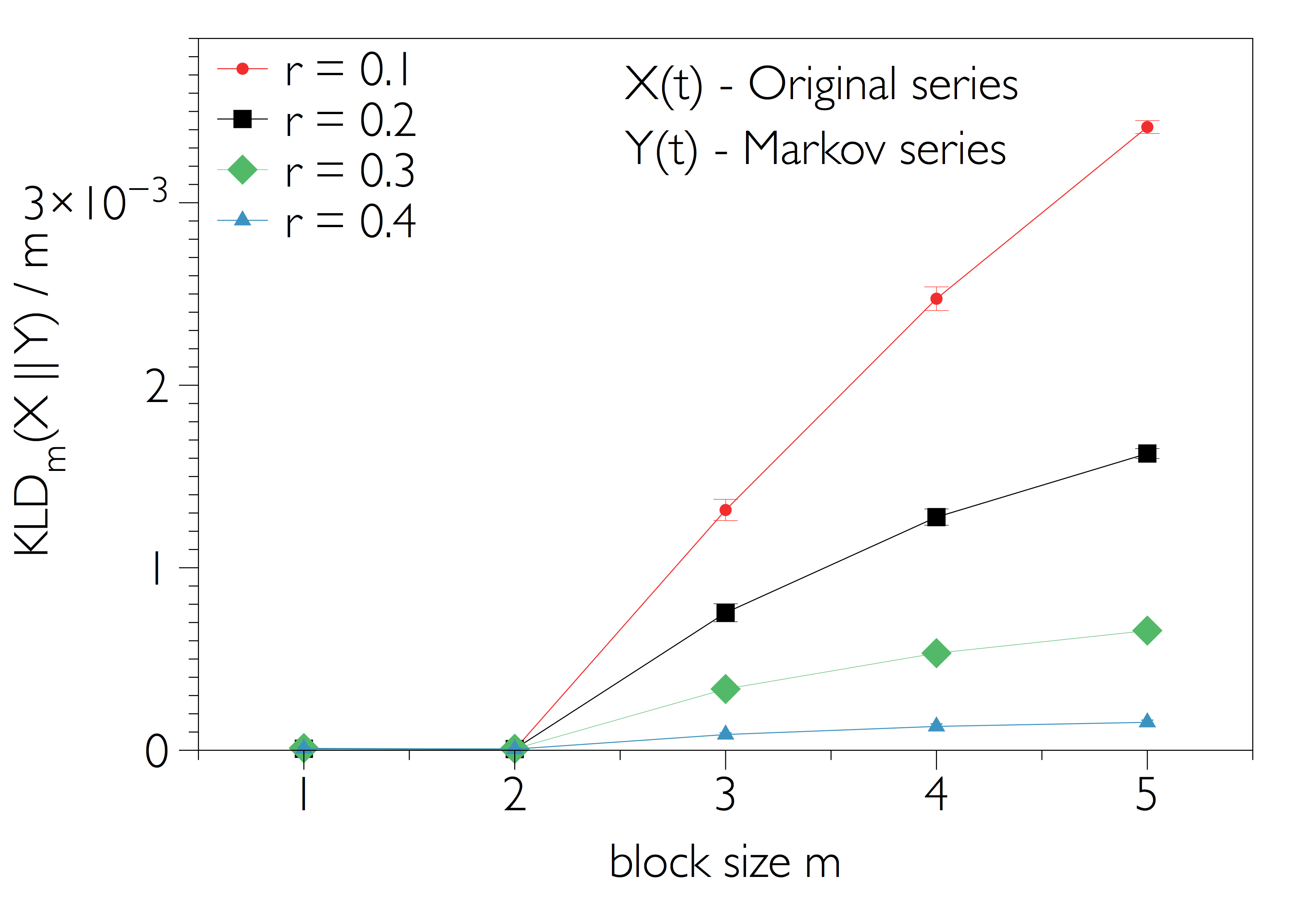}
\caption{The normalised Kullback-Leibler divergence
  $D(m)$ between $X(t)$ and its Markovian surrogate $Y(t)$, in the
  case where $X(t)$ does not show any induced current, for different values
  of the switching rate $r$. The series $X(t)$ records the position of
  a walker diffusing over two layers, with transition probabilities
  in the layers are $T^{(1)}_{i,i+1}=1/3$ and $T^{(2)}_{i,i+1}=2/3$. We
  correctly find that $X(t)$ is non-Markovian even for large values of
  $r$ as $D(m>2)>0$, what suggests an underlying multiplex structure.}
\label{fig1_sss}
\end{figure}

\noindent In the next sections we extend the initial study on inferring multiplexity to the case where layers have different complex topologies, departing from the situation where each layer is a cycle graph (ring). Note that when each layer has a different topology we expect the algorithm to detect more easily the underlying multiplex character (in this sense, extension of this formalism to the more general case of a multi-layer network is very promising).
In particular,  we explore the following additional scenarios:
\begin{itemize}
\item Scenario 2: two complete graphs with different transition matrices.
\item Scenario 3: two layers with cycle graphs where in one of the layers we introduce a shortcut which is crossed with a probability $\epsilon$. This scenario allows us to explore small perturbations in the transition matrices with respect to the original scenario studied in the main text.
\item Scenario 4: each layer has a different topology and dynamics, the first being a cycle graph (ring) with positive net current and the second layer is a complete graph with null net current.
\item Scenario 5: initially having two identical layers (two cycle graphs), we introduce a number of additional edges (shortcuts) in the second one, to explore topological perturbations on our initial scenario. We also explore the scalability of this general scenario by considering the effect of increasing the number of nodes.
\item Scenario 6: initially we consider two identical layers formed by Erdos-Renyi graphs, and then rewire in one of the layers a percentage of the edges. Transition matrices are obtained here by unbiasing the walker $T_{ij}=A_{ij}/k_i$. We also explore the scalability of this scenario by considering the effect of increasing the size of the graphs.
\end{itemize}

\subsection{Scenario 2: Complete graphs}
In this scenario we consider two layers with identical topology but different transition matrices. Here we build two replicas of a complete graph. In the first layer we define an unbiased random walker with transition matrix
\[
\textbf{T}^{(1)}= 
\begin{bmatrix}
0    & 1/3        & 1/3                 &    1/3 \\
1/3    & 0        & 1/3                 & 1/3  \\
1/3   & 1/3      & 0     & 1/3     \\
  1/3   & 1/3 & 1/3     & 0  \\
\end{bmatrix}
\]
whereas in the second layer we introduce a parametric deviation 
\[
\textbf{T}^{(2)}= 
\begin{bmatrix}
0    & 1/3  +2\epsilon      & 1/3 -\epsilon                &    1/3 -\epsilon \\
1/3 +2\epsilon   & 0        & 1/3 -\epsilon                 & 1/3 -\epsilon \\
1/3 +2\epsilon  & 1/3 -\epsilon      & 0     & 1/3 -\epsilon     \\
  1/3 +2\epsilon  & 1/3 -\epsilon & 1/3 -\epsilon    & 0  \\
\end{bmatrix}
\] 
The larger $\epsilon$, the more different the statistics of a Markov Chain over each layer separately. For $\epsilon=0$ both layers are identical and a walker diffusing over the multiplex (switching layers at a constant rate $r$) reduces to a Markov Chain over one layer, whereas for $\epsilon>0$ the process is non-Markovian if we only have access to the state $X(t)$. In figure \ref{fig2_bis} we plot the results for $D(m)$ which show that multiplexity can always be detected, and such detection is easier as $\epsilon$ increases.

\begin{figure}
\centering
\includegraphics[width=0.4\columnwidth]{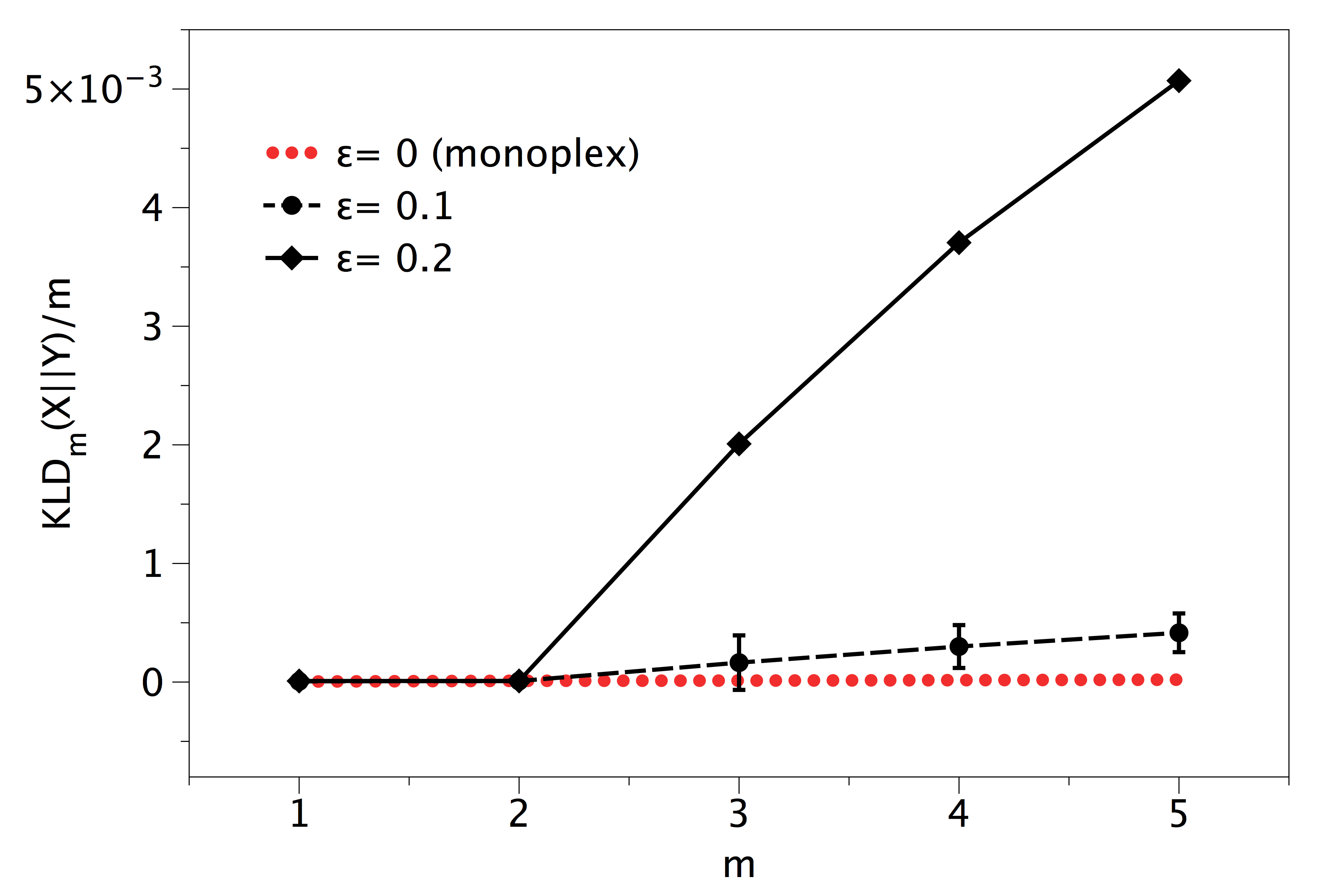}
\caption{{\bf Scenario 2. }Normalised Kullback-Leibler divergence $D(m)$ between $X(t)$ and its Markovianised surrogate $Y(t)$, in the case where both layers are complete graphs with different transition matrices and the switching rate is $r=0.1$. The difference is parametrised by $\epsilon$ (when $\epsilon=0$ both matrices are identical, and they increasingly differ with increasing values of $\epsilon$). The method detects multiplexity ($D(m>2)>0$) with larger values as $\epsilon$ increases.}
\label{fig2_bis}
\end{figure}

\subsection{Scenario 3: Controlled perturbation on one layer}
In this scenario we explore the effect of a controlled perturbation in the topology of one of the layers. We start by defining $L=2$ identical layers (two rings with the same transition matrix, with an homogeneous probability to flow from $i\to i+1$ $T_{i,i+1}=1/3$). 
In the second replica we introduce a shortcut between two nodes, weighting the probability of traversing this node as $2\epsilon$, and biasing accordingly the rest of the edges. Accordingly the transition matrices of both layers read

\[
\textbf{T}^{(1)}= 
\begin{bmatrix}
0    & 1/3        & 0                 &    2/3 \\
2/3    & 0        & 1/3                 & 0  \\
0   & 2/3      & 0     & 1/3     \\
  1/3   & 0 & 2/3     & 0  \\
\end{bmatrix}
\] 
\[
\textbf{T}^{(2)}= 
\begin{bmatrix}
0    & 1/3-\epsilon        & 2\epsilon                  &    2/3-\epsilon \\
2/3+\epsilon    & 0        & 1/3-\epsilon                 & 0  \\
2\epsilon   & 2/3-3\epsilon          & 0     & 1/3+\epsilon      \\
  1/3+\epsilon   & 0 & 2/3-\epsilon     & 0  \\
\end{bmatrix}
\]
 When $\epsilon=0$ this extra edge has no effect and we should expect that the multiplex is effectively monoplex, whereas for $\epsilon>0$ both layers show gradually different structure and as such $X(t)$ is non-Markovian. Similarly as in the previous case, our methodology predicts that the network is multiplex when $D(m\leq 2)=0$ and $D(m>2)>0$. Intuitively, as $\epsilon$ increases the effect of the shortcut should be higher and thus detecting the multiplex nature of the network should be easier. In figure \ref{fig2} we plot $D(m)$ for a Markov Chain walking over this topology with a constant switching rate $r=0.1$ and different values of $\epsilon$. Results support the accuracy of the method, even for small values of $\epsilon$ the multiplex nature of the network is detected. 
 
\begin{figure}
\centering
\includegraphics[width=0.4\columnwidth]{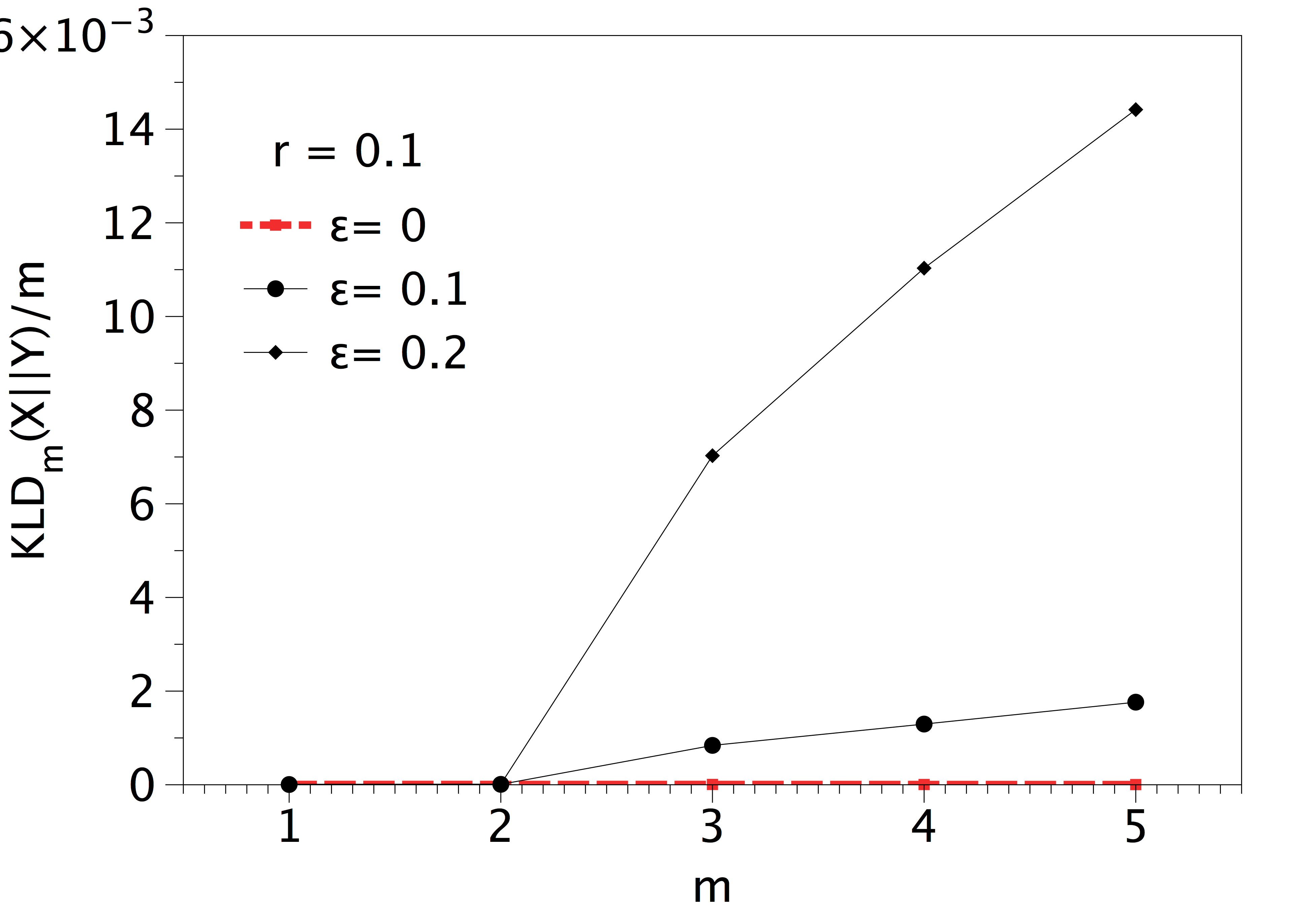}
\caption{{\bf Scenario 3. }Normalised Kullback-Leibler divergence $D(m)$ between $X(t)$ and its Markovianised surrogate $Y(t)$, in the case where the second layer has a shortcut which is traversed with probability $\epsilon$ (scenario 3) and the switching rate is $r=0.1$. For $\epsilon=0$ both layers are identical and the walker is essentially navigating over a monoplex and therefore $X(t)$ is Markovian and $D(m)=0 \ \forall m$. For $\epsilon>0$ the graph is multiplex and therefore $D(m>2)>0$, and such feature is detected with quantitative larger fingerprints as $\epsilon$ increases.}
\label{fig2}
\end{figure}

\subsection{Scenario 4: Ring versus complete graph}
\begin{figure}
\centering
\includegraphics[width=0.4\columnwidth]{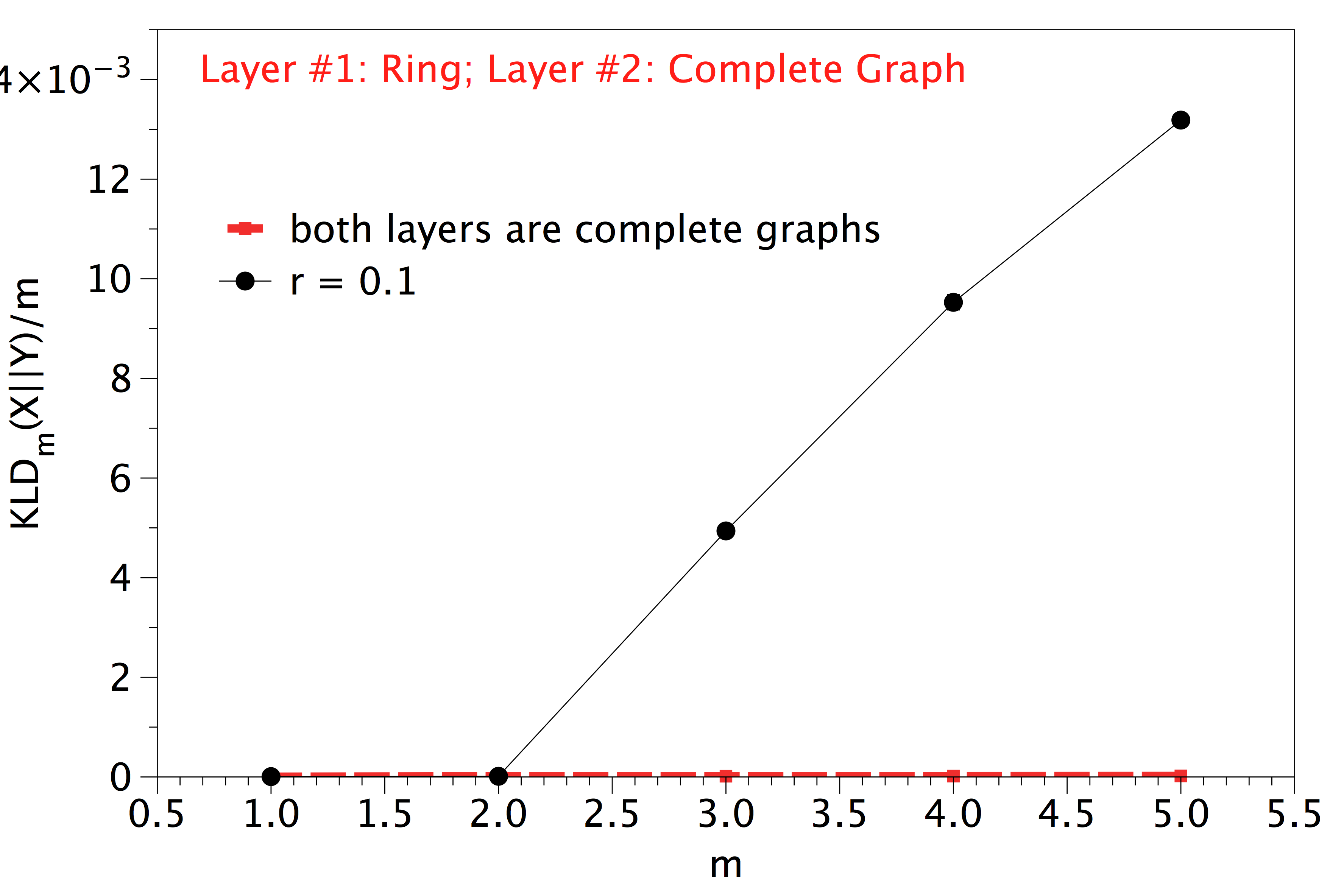}
\caption{{\bf Scenario 4. }Normalised Kullback-Leibler divergence $D(m)$ between $X(t)$ and its Markovianised surrogate $Y(t)$, in the case where the second layer is a complete graph. }
\label{fig3}
\end{figure}

In this scenario we focus on a multiplex with $L=2$ where each layer is totally different. The first layer is a ring with a net current described by the transition matrix
\[
\textbf{T}^{(1)}= 
\begin{bmatrix}
0    & 1/3        & 0                 &    2/3 \\
2/3    & 0        & 1/3                 & 0  \\
0   & 2/3      & 0     & 1/3     \\
  1/3   & 0 & 2/3     & 0  \\
\end{bmatrix}
\]
whereas the second layer is a complete graph with no net current and detailed balance everywhere:
\[
\textbf{T}^{(2)}= 
\begin{bmatrix}
0    & 1/3        & 1/3                 &    1/3 \\
1/3    & 0        & 1/3                 & 1/3  \\
1/3   & 1/3      & 0     & 1/3     \\
  1/3   & 1/3 & 1/3     & 0  \\
\end{bmatrix}
\] 
Again, we find that the method can clearly detect the multiplex nature of the network as $D(m>2)>0$.

\subsection{Scenario 5: Sequentially introducing shortcuts}

In this scenario we consider two identical replicas (a multiplex with $L=2$ layers) where in the second layer we add a certain number of additional edges. Originally, both replicas are rings with detailed balance (unbiased walker)
\[
\textbf{T}^{(1)}=\textbf{T}^{(2)}= 
\begin{bmatrix}
0    & 1/2        & 0                 &    1/2 \\
1/2    & 0        & 1/2                 & 0  \\
0   & 1/2      & 0     & 1/2     \\
  1/2   & 0 & 1/2     & 0  \\
\end{bmatrix}
\]
In this initial case we expect $D(m)=0, \forall m$ (the multiplex is effectively monoplex).
We then add to this benchmark a different number of edges (and accordingly we expect $D(m>2)>0$, and larger values when the number of added edges increases), effectively interpolating between scenario 2 and scenario 3. 
The adjacency matrices for three concrete cases (with no added edges, one added edge and two added edges - this latter case being equivalent to a complete graph) are depicted in figure \ref{fig6}, and in figure \ref{fig7} we show the values of $D(m)$.

\begin{figure}[b]
\centering
\includegraphics[width=0.3\columnwidth]{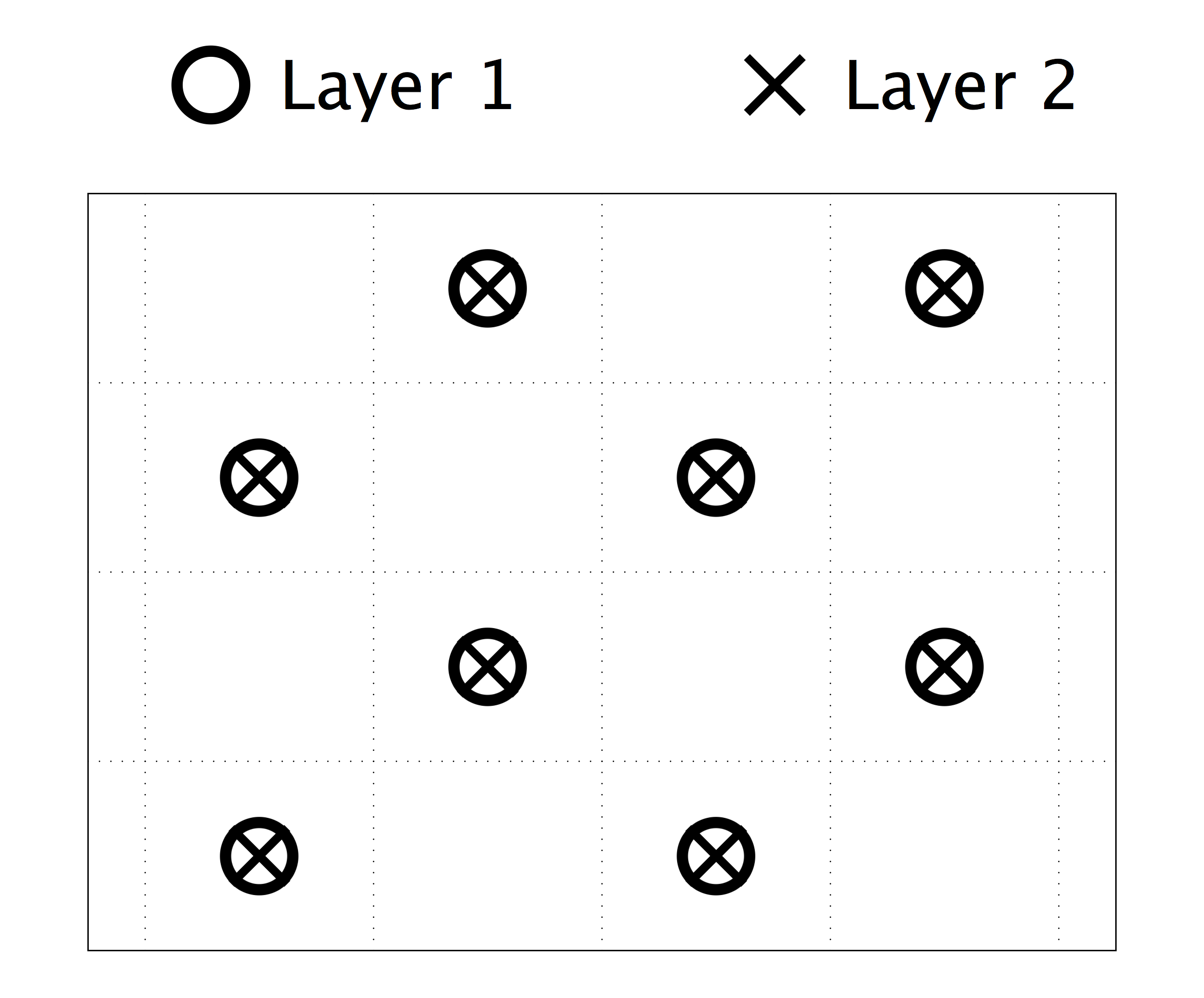}
\includegraphics[width=0.3\columnwidth]{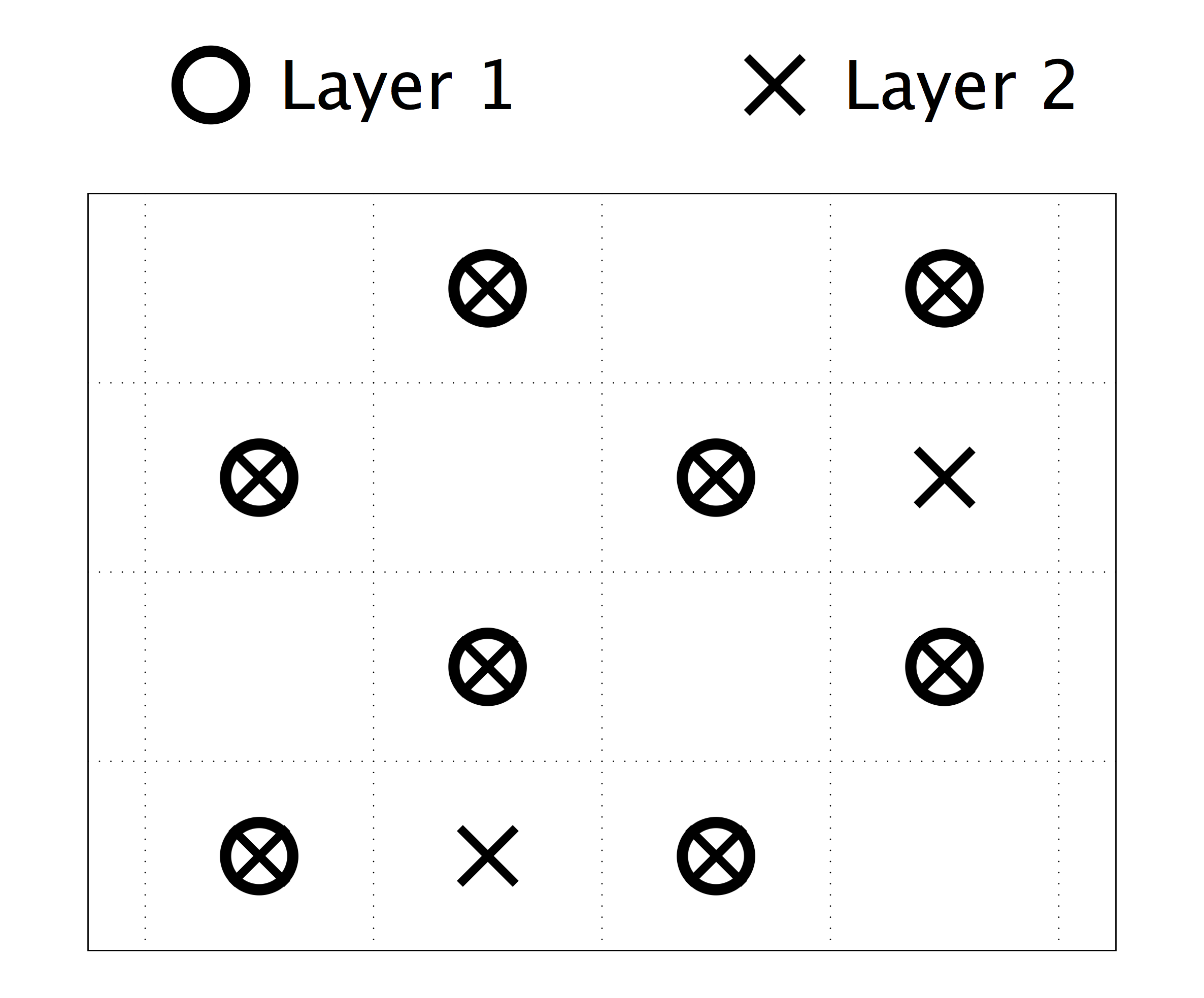}
\includegraphics[width=0.3\columnwidth]{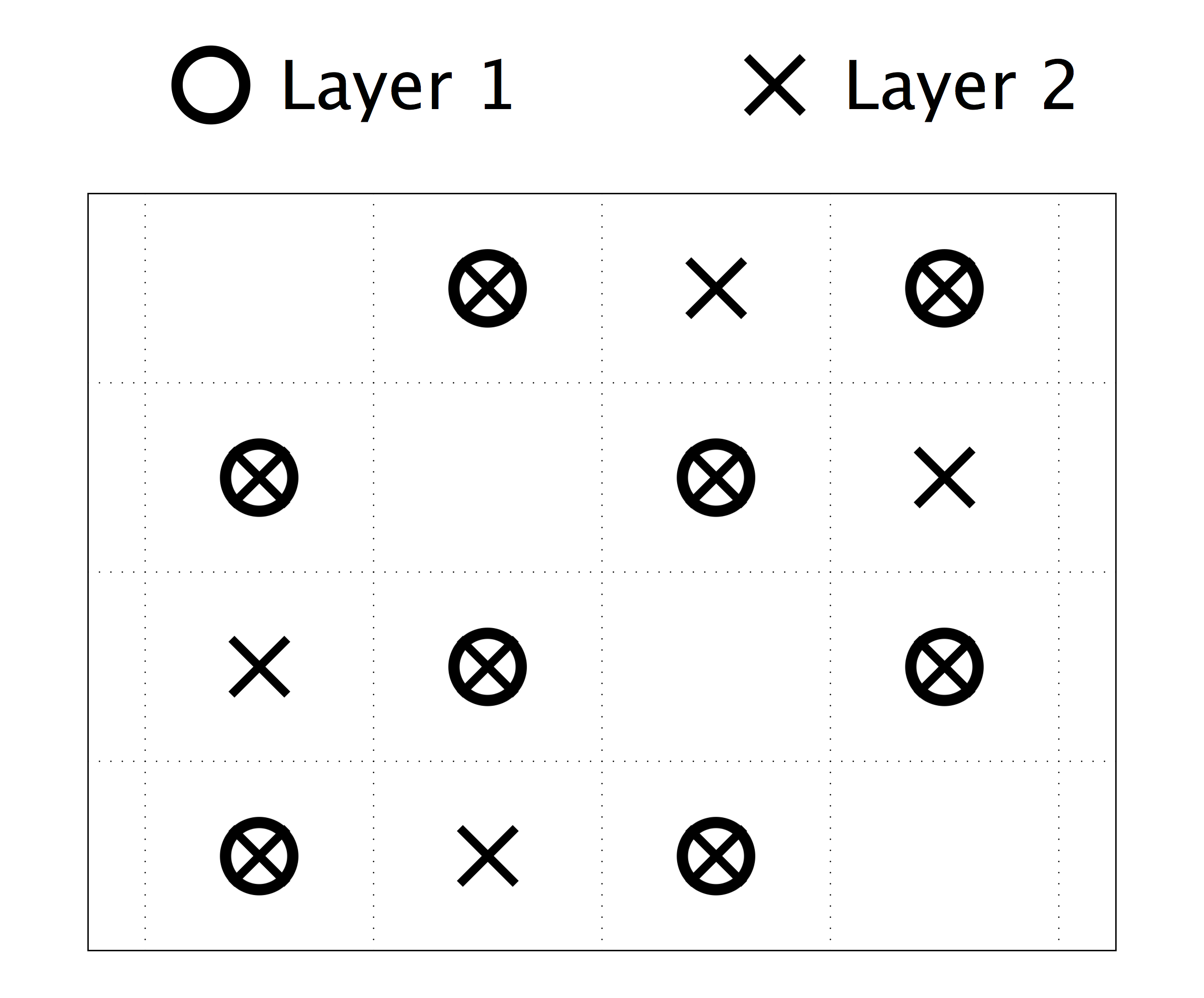}
\caption{{\bf Scenario 5. }Adjacency matrices of each layer for three cases: (i, left) no shortcuts are introduced and both layers are identical; (ii, middle) one shortcut has been introduced; (iii, right) two shortcuts have been rewired. The transition matrices for each case are $T_{ij}=A_{ij}/k_i$, where $k_i$ is the degree of node $i$.}
\label{fig6}
\end{figure}
\begin{figure}
\centering
\includegraphics[width=0.4\columnwidth]{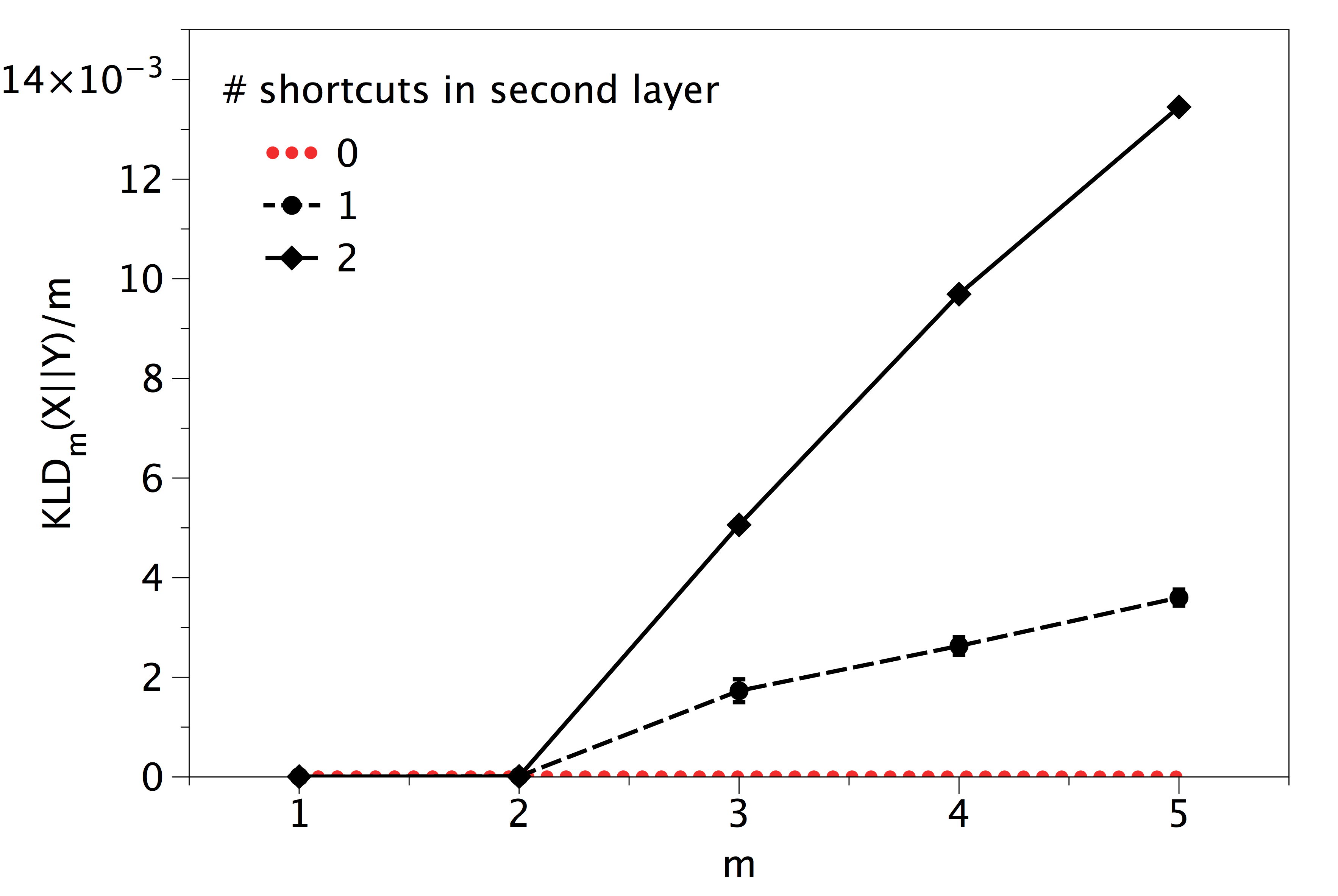}
\caption{{\bf Scenario 5. }Normalised Kullback-Leibler divergence $D(m)$ between $X(t)$ and its Markovianised surrogate $Y(t)$, for a two layer multiplex where the second layer is a replica of the first layer where a number of shortcuts have been introduced. When no edges have been introduced both replicas are identical, $X(t)$ is Markovian and $D(m)=0 \ \forall m$, otherwise the process is non-Markovian and the algorithm detects multiplexity by finding $D(m>2)>0$. Such detection improves as the topology of both layers is increasingly different.}
\label{fig7}
\end{figure}

\subsection{Scenario 5b: Introducing edges on larger graphs}
Here we investigate the scalability of the scenario 5 and in particular we explore (i) the effect of increasing the number of nodes $K$ in each layer and (ii) the effect of increasing the number of rewired edges on the detectability. We start by defining two identical replicas of a ring with $K$ nodes with unbiased transition matrix $T_{ij}=1/2$ for $i\ne j$.
In the second layer we introduce a number of shortcuts and we analyse two particular behaviors as it follows.\\

\noindent {\bf Effect of node increase. }
We fix the number of shortcuts $p=2$ and vary the number of nodes $K$, and explore the dependence of $D(m)$ on $K$. As we keep the series size ${\cal N}(m)=10^5\cdot2^m$ being independent from $K$, we expect that as $K$ increases the statistics are poorer as we need larger series to capture an equivalent number of transitions.\\
\begin{figure}
\centering
\includegraphics[width=0.4\columnwidth]{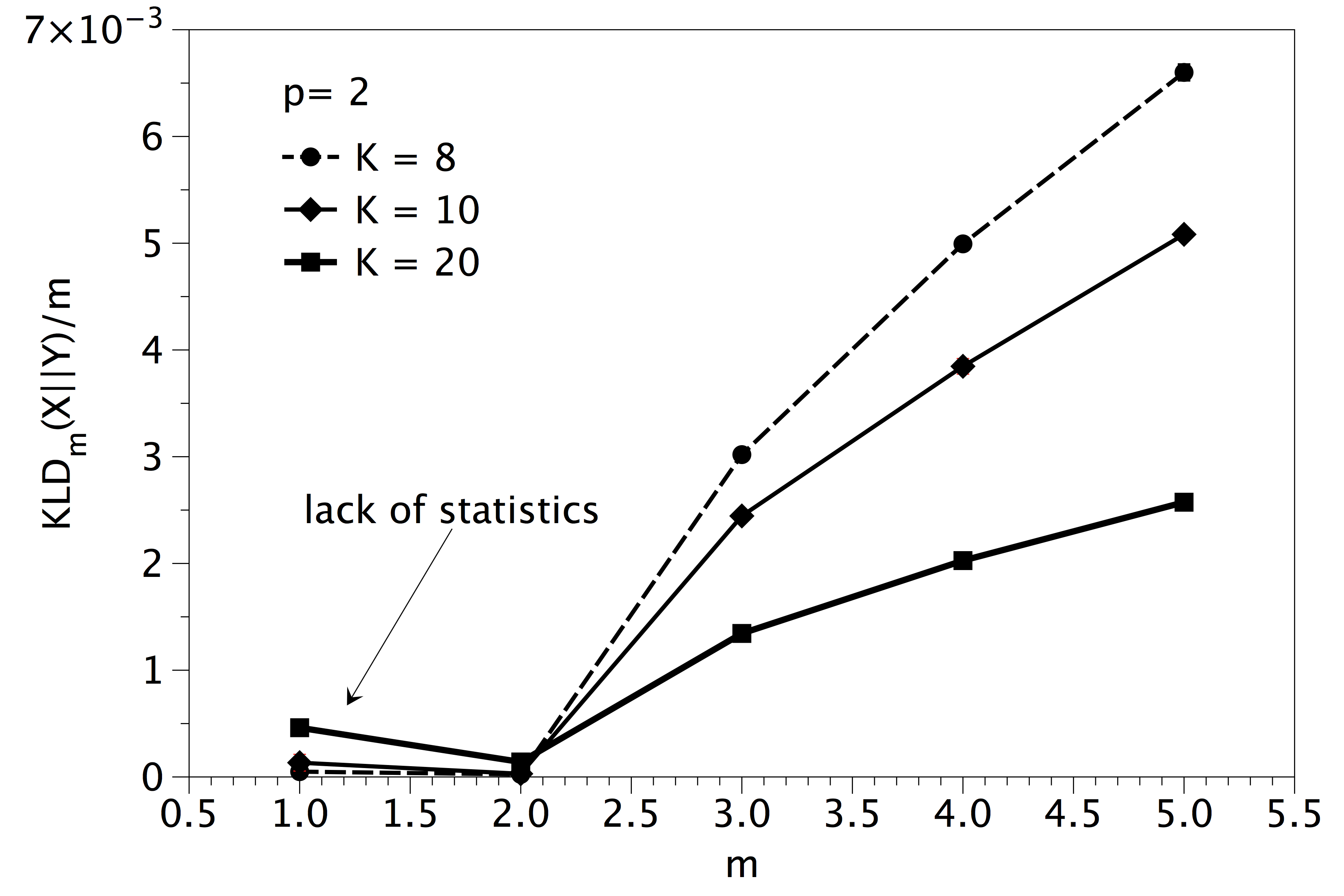}
\caption{{\bf Scenario 5b. }Normalised Kullback-Leibler divergence $D(m)$ between $X(t)$ and its Markovianised surrogate $Y(t)$, for a two layer multiplex where each layer has $K$ nodes, where the second layer is a replica of the first layer where $p=2$ shortcuts have been introduced.}
\label{fig7b}
\end{figure}

\noindent {\bf Detectability as a function of the number of shortcuts introduced. } Here we fix $K=10$ and explore the multiplex detectability as the number of shortcuts $p$ is increased in the second layer. Multiplexity is detected when $D(3)>0$, and the larger $D(3)$ the easier is such detection. In figure \ref{fig7bb} we plot $D(3)$ as a function of the number of shortcut edges $p$.
\begin{figure}
\centering
\includegraphics[width=0.4\columnwidth]{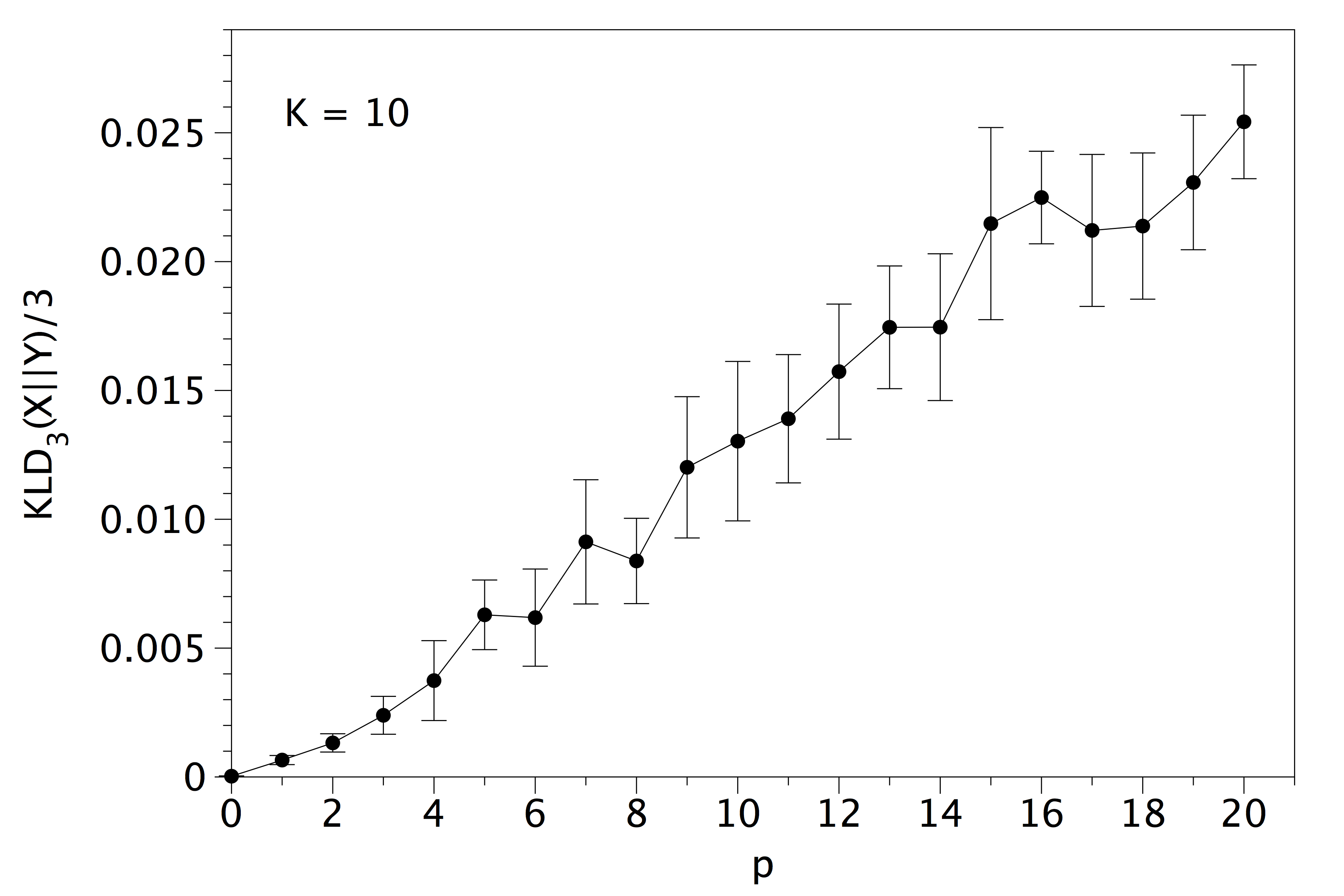}
\caption{{\bf Scenario 5b. }Detectability measure $D(3)$ for a two layer multiplex with $K=10$ nodes per layer where the second layer is a replica of the first layer where a number of shortcuts $p$ have been introduced. Results have been averaged over 10 network realisations for each case.}
\label{fig7bb}
\end{figure}

\subsection{Scenario 6: Rewiring edges}
\begin{figure}
\centering
\includegraphics[width=0.3\columnwidth]{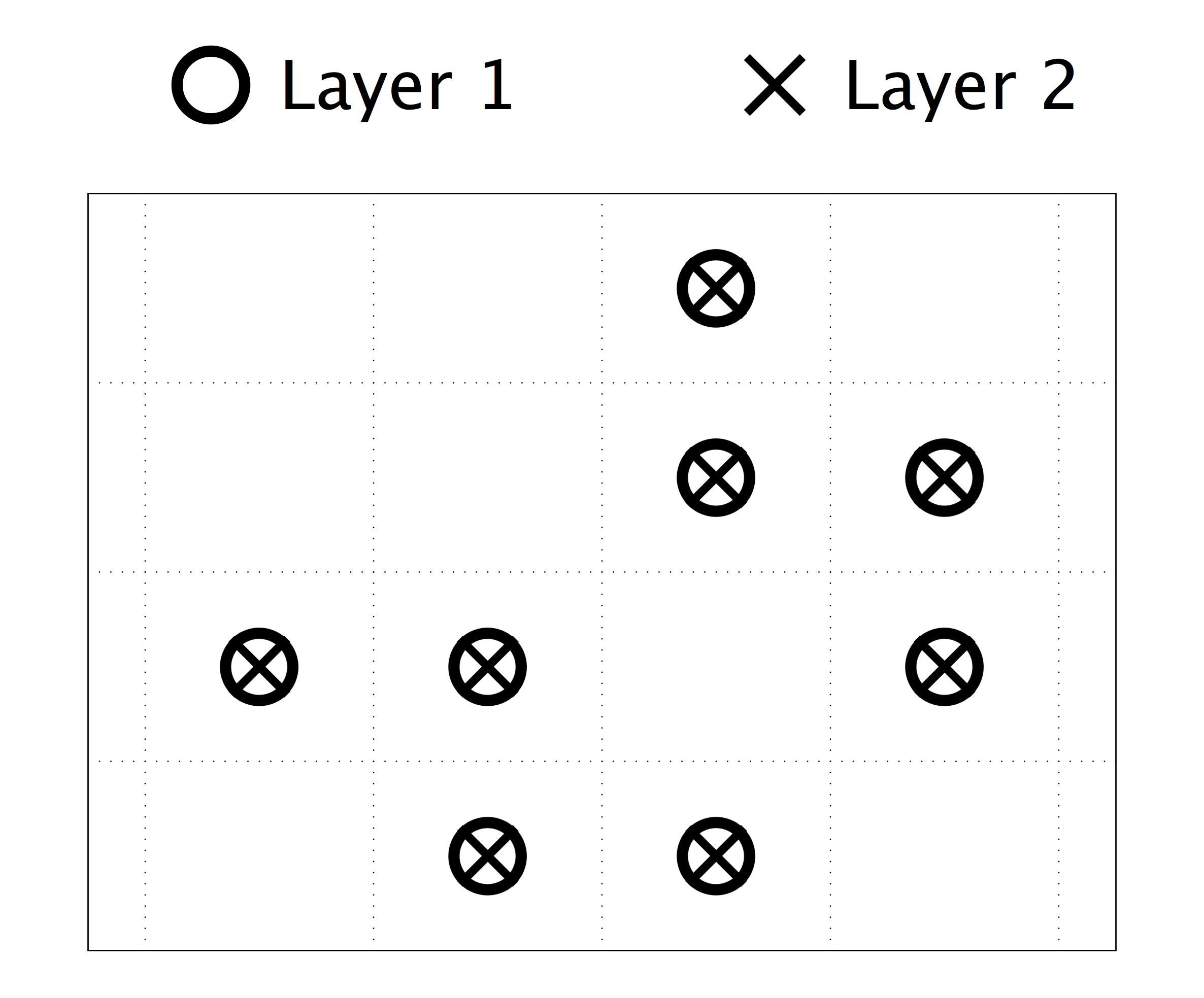}
\includegraphics[width=0.3\columnwidth]{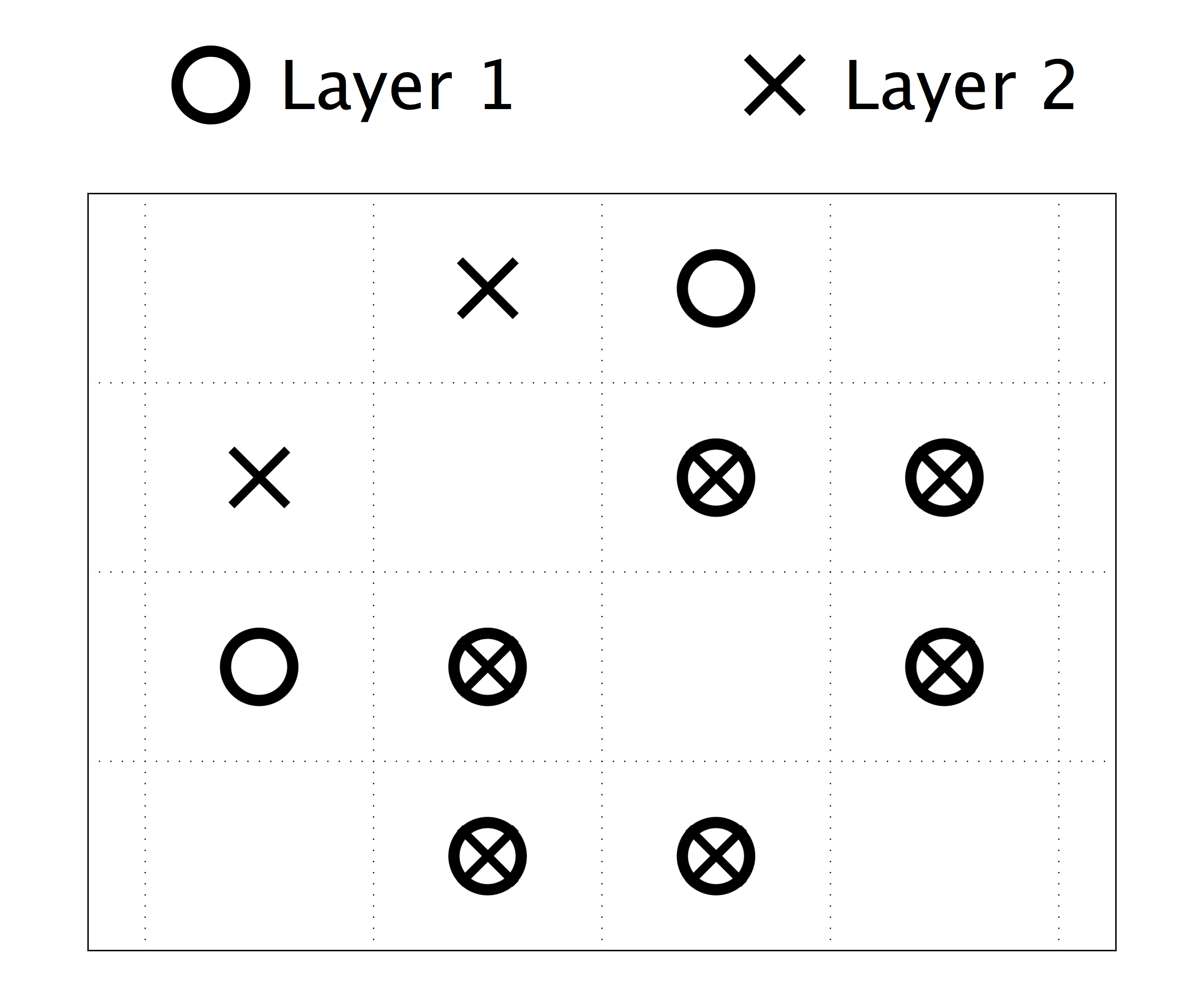}
\includegraphics[width=0.3\columnwidth]{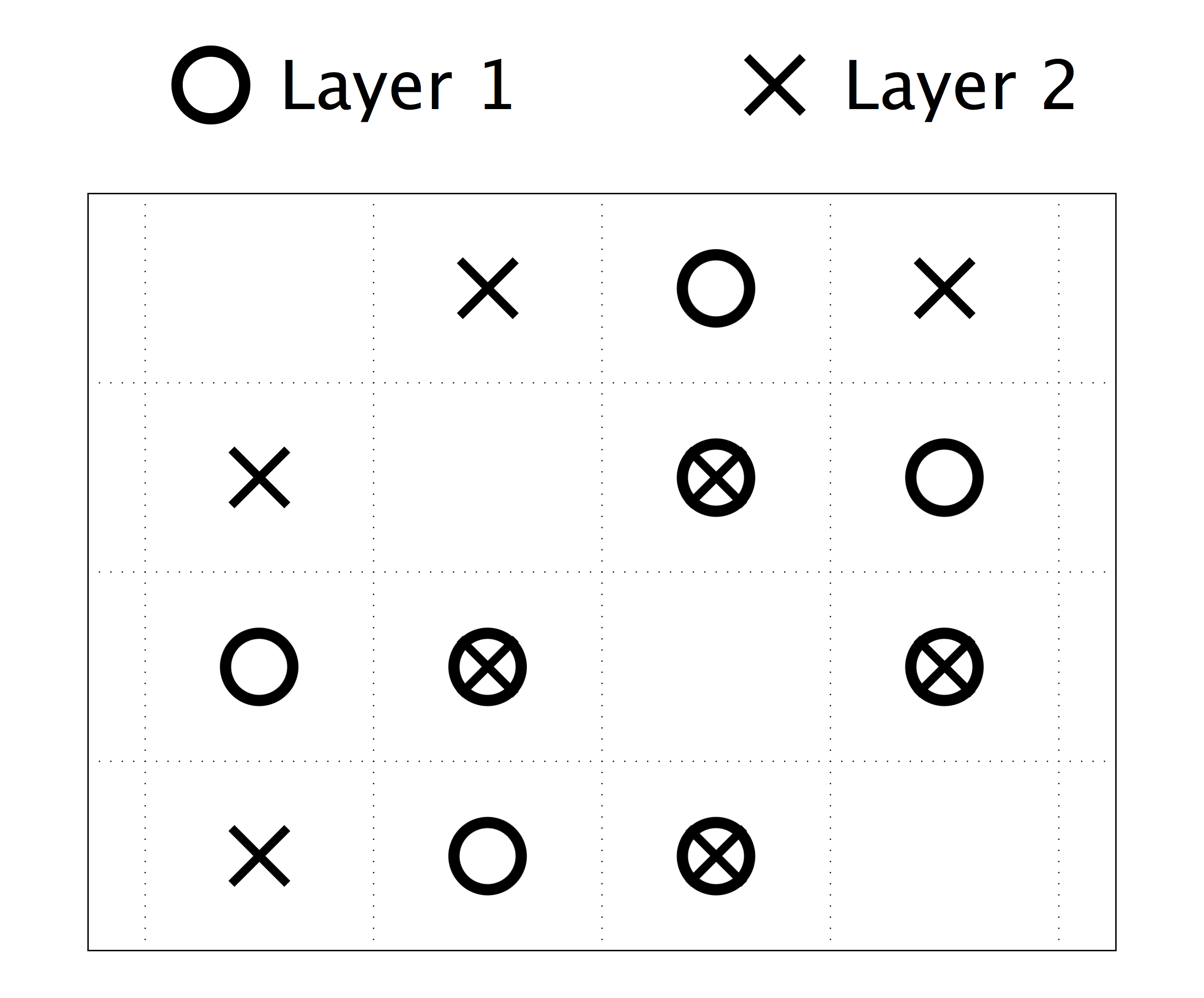}
\caption{{\bf Scenario 6. }Adjacency matrices of each layer for three cases: (i, left) no rewiring takes place and both layers are identical; (ii, middle) one edge has been rewired; (iii, right) two edges have been rewired. The transition matrices for each case are $T_{ij}=A_{ij}/k_i$, where $k_i$ is the degree of node $i$.}
\label{fig4}
\end{figure}

In this scenario we initially consider two replicas of the same Erdos-Renyi graph (where nodes $i$ and $j$ are connected with probability $p=0.8$, above the percolation threshold to have a connected graph). We consider three different situations, namely: (i) both layers are maintained identical, (ii) we rewire at random one edge, (iii) we rewire at random two edges. The adjacency matrices of each layer for these three cases are represented in figure \ref{fig4}, 
and we choose unbiased random walkers with layer transition matrices $P_{ij}=A_{ij}/k_i$.
In figure \ref{fig5} we plot the values of $D(m)$ for these three cases. As expected, when the layers are identical we find $D(m)=0 \ \forall m$, whereas when we rewire edges from a layer the network converts into a multiplex one and $D(m>2)>0$. Also, $D(m>2)$ take larger values -and thus multiplex detection is easier- when the layers are increasingly different.

\begin{figure}
\centering
\includegraphics[width=0.4\columnwidth]{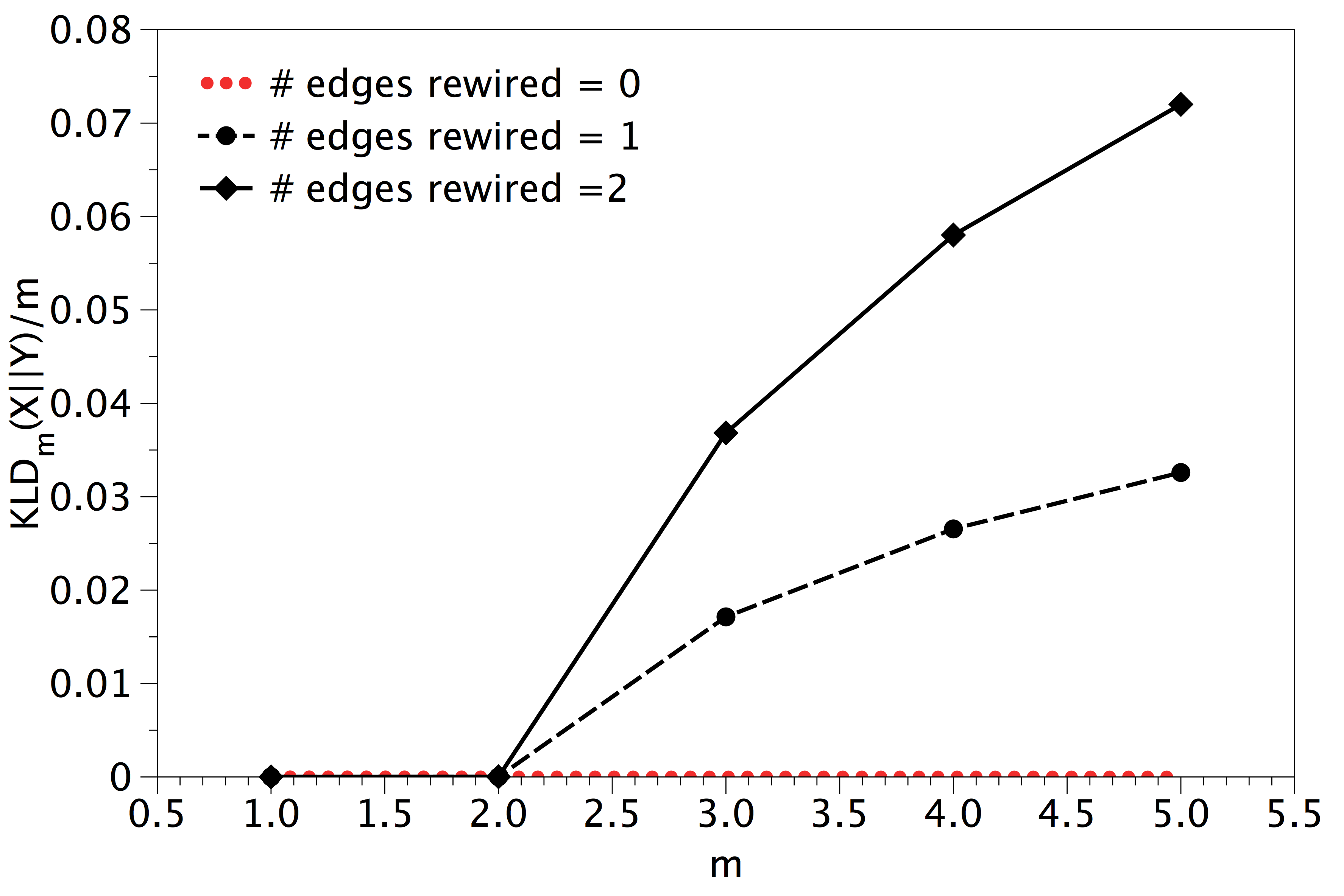}
\caption{{\bf Scenario 6. }Normalised Kullback-Leibler divergence $D(m)$ between $X(t)$ and its Markovianised surrogate $Y(t)$, for a two layer multiplex where the second layer is a replica of the first layer where a number of edges have been rewired. When no edges have been rewired both replicas are identical, $X(t)$ is Markovian and $D(m)=0 \ \forall m$, otherwise the process is non-Markovian and the algorithm detects multiplexity by finding $D(m>2)>0$. Such detection improves as the topology of both layers is increasingly different.}
\label{fig5}
\end{figure}

\subsection{Scenario 6b: Rewiring edges on larger graphs}
Finally, we consider Erdos-Renyi graphs (linking probability 0.65) with $K=10$ nodes per layer and explore the multiplex detectability as we rewire a percentage of nodes $p$ in the second layer.
Multiplexity is detected when $D(3)>0$, and the larger $D(3)$ the easier is such detection. In figure \ref{fig5b} we plot $D(3)$ as a function of $p$. Dots are the result of an ensemble average over ER graphs realisations. For an ensemble we keep fixed the number of rewired edges and compute the effective average percentage of rewired edges (which fluctuates as each realisation of an ER graph will have a different total number of edges).

\begin{figure}
\centering
\includegraphics[width=0.4\columnwidth]{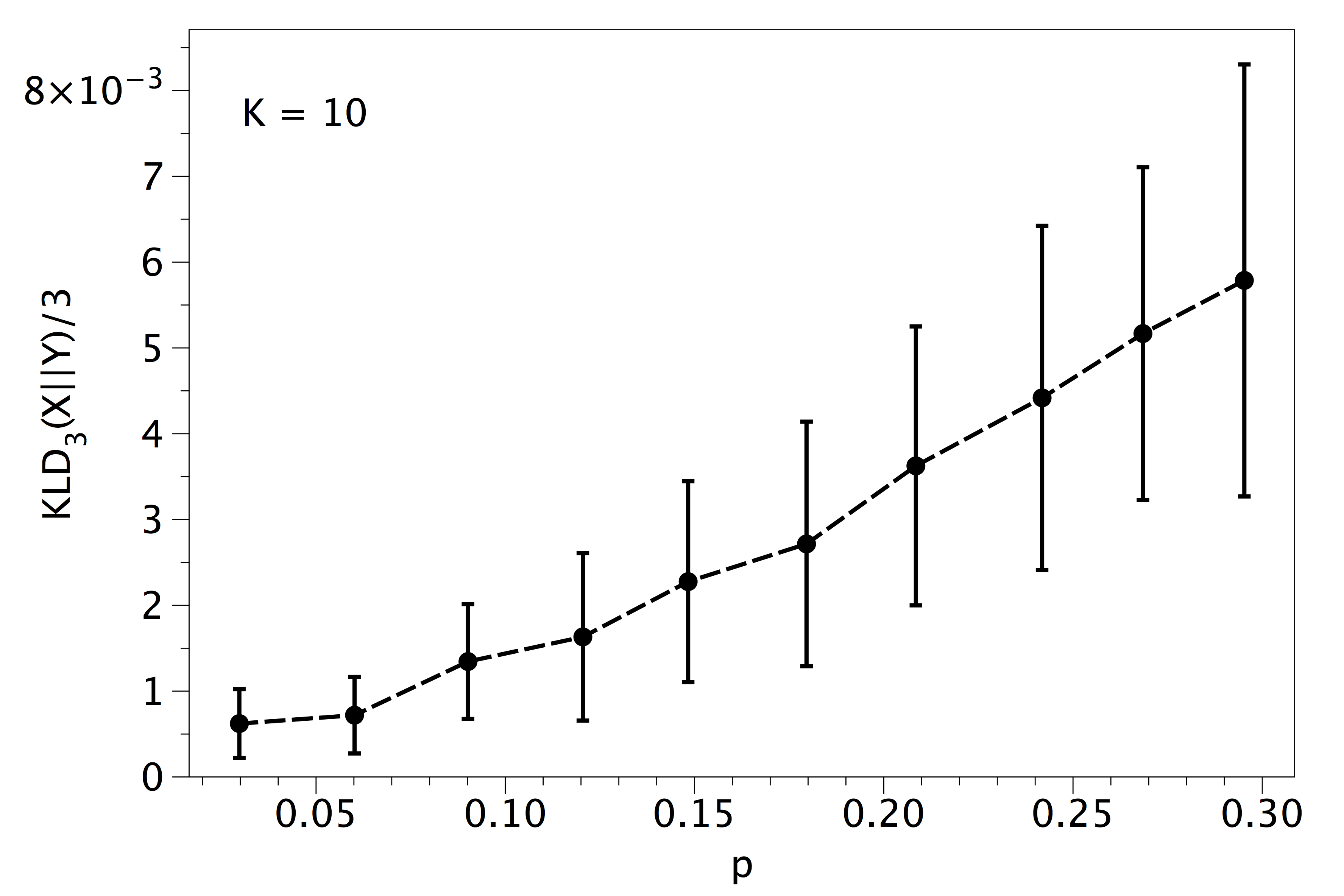}
\caption{{\bf Scenario 6b. }Detectability measure $D(3)$ for a two layer multiplex with $K=10$ nodes per layer, where each layer is a connected Erdos-Renyi graph (link probability 0.65). The second layer is a replica of the first layer where a percentage $p$ of nodes have been rewired. Results have been averaged over $10^2$ network realisations for each case.}
\label{fig5b}
\end{figure}

 \section{MATHEMATICAL AND ALGORITHMIC FRAMEWORK FOR LAYER ESTIMATION}

Here we formalise the problem and provide a detailed derivation of the model, the probabilistic model selection scheme and some possible algorithmic implementations of this scheme. Let us remark that the probabilistic framework described herein includes the case in which the path followed by the walker across the multiplex network cannot be observed exactly (i.e., there are observation errors) even if this is not addressed in the main text.  

We use an argument-wise notation to denote probability mass functions (pmf's) and probability density functions (pdf's). If $X$ and $Y$ are discrete random variables (r.v.'s) then $P(X)$ and $P(Y)$ are the pmf of $X$ and the pmf of $Y$, respectively, possibly different. Similarly, $P(X,Y)$ and $P(X|Y)$ denote the joint pmf of the two r.v.'s and the conditional pmf of $X$ given $Y$, respectively. We use lower-case $p$ for pdf's. If $X$ and $Y$ are continuous r.v.'s, then $p(X)$ and $p(Y)$ are the corresponding densities, possibly different, and $p(X,Y)$ and $p(X|Y)$ denote the joint and conditional pdf's. We may have a pdf of a continuous r.v. $X$ conditional on a discrete r.v. $Y$, $p(X|Y)$, as well as a pmf of $Y$ given $X$, $P(Y|X)$. Most r.v.'s are indicated with upper-case letters, e.g., $X$. If we need to denote a specific realisation of the r.v., then we use the same letter but lower-case, e.g., $X=x$ or $Y=y$. Matrices and vectors are indicated with a bold-face font, e.g., ${\bf T}$.

\subsection{The model}
We assume that a walker travels through a multiplex network taking random moves between neighbouring nodes and, occasionally, between layers. Let $L$ denote the number of layers in the multiplex network and let $K$ be the number of nodes per layer. At discrete time $t$, the random variable (r.v.) $X(t)$ denotes the in-layer walker position. Therefore, $X(t) \in \{0,\ldots,K-1 \}$ and $X(t) = k$ means that the particle is located at node number $k$ at time $t$, irrespective of the layer. The r.v. $\ell(t)$ indicates the layer at time $t$, i.e., $\ell(t) \in \{ 1, 2, ..., L \}$ and $\ell(t) = l$ means that the walker is found in layer $l$ at time $t$. The state of the walker, therefore, is given by the $2 \times 1$ vector ${\bf Z}(t) = [X(t), \ell(t)]^\top$.

\noindent At each time step, the walker may jump across layers. This motion is assumed to be Markov and hence it can be characterised by an $L\times L$ stochastic transition matrix ${\bf R}_L$, where the entry $R_{ij}$, $(i,j)\in\{0, \ldots, L-1\}^2$, represents the probability of moving from layer $i$ to layer $j$. Subsequently, the particle diffuses within the new layer to one of its neighbours. Within each single layer, the motion of the walker is also assumed to be Markov. Hence, in the $l$-th layer it is governed by a $K \times K$ transition matrix ${\bf T}^{(l)}$, such that $T_{ij}^{(l)}$ is the probability of a particle lying in layer $l$ to diffuse from node $i$ to node $j$. These probabilities are constant over time $t$. The complete Markov model is characterised by the set of matrices $\{ {\bf R}_L, {\bf T}^{(0)}, \ldots, {\bf T}^{(L-1)} \}$ and we denote $\mT_L=\{ {\bf T}^{(l)} \}_{l=0}^{L-1}$ in the sequel for convenience.

 
We can think of this model as a discrete-time state-space dynamical system, where the state variables at time $t$ are
\begin{equation}
{\bf X}(t) = \{ {\bf R}_L,\mT_L,X(t),\ell(t) \}, \quad t=0, 1, 2, ...
 \end{equation}
and we assume there is a sequence of observations $\{ Y(t) \}_{t \ge 1}$ taking values in the set of node labels $\{ 0, 1, \ldots, K-1 \}$. In the main text we assume that the observations are exact and, therefore, $Y(t)=X(t)$. However, we can handle a more general class of models in which $Y(t)$ is a r.v. with conditional pmf $P(Y(t)|X(t))$ (independently of the current or past layers). If the observation is exact, then 
$$
P(Y(t)=j | X(t)=i ) = \left\{
	\begin{array}{ll}
	1, &\mbox{if $j=i$,}\\
	0, &\mbox{otherwise,}\\
	\end{array}
\right.
$$ 
however the proposed model (and related numerical methods), admit the cases in which observation errors may occur and, hence, $P(Y(t)|X(t))$ is a non-degenerate pmf. We assume there are known and independent {\em a priori} pmf's for the node and layer at time $t=0$, $P_0(X(0),\ell(0))=P_0(X(0))P_0(\ell(0))$. In practical problems, the parameters $\mT_L$ and ${\bf R}_L$ are unknown and we also endow them with prior pdf's $p_0(\mT_L,{\bf R}_L)=p_0(\mT_L) p_0({\bf R}_L)$ with respect to (wrt) a suitable reference measure $\mu(\sd\mT_L \times \sd{\bf R}_L)$. Most often, and for a general scenario, $\mu$ can be the Lebesgue measure on $\Real^{L\times K \times K} \times \Real^{L \times L}$, but other choices may be posible if we wish to impose constraints on $\mT_L$ and ${\bf R}_L$. For the case of the network with ring-shaped layers in the main text, $\mu$ can be reduced to the Lebesgue measure on $\Real^{L+1}$.

%
\subsection{Bayesian model selection} \label{sPost}

Assume that we have collected a sequence of $N$ observations which we now label 
$$Y(1:N) = \{ Y(1), Y(2), \ldots, Y(N) \} \in \{0,1, \ldots, K-1\}^N.$$ 
We wish to make a decision as to {\em what model} is the best fit for that sequence. We adopt the view that two models are different when they have a different number of layers, hence if model $A$ has $L$ layers and model $B$ has $L'$ layers, $A = B \Leftrightarrow L = L'$. A convenient way to tackle this problem is to model the total number of layers $L$ as a r.v., in such a way that each possible value of $L$ corresponds to a different model. If we define 
\begin{itemize}
\item a prior probability mass function for $L$, say $P_0(L)$, for $L \in \{1, 2, \ldots, L_+ \}$, where $L_+$ is the maximum admissible number of layers, and
\item a likelihood function 
\begin{equation}
P\lb Y(1:N)|L \rb = \int P\lb
	Y(1:N) | \mT_L, {\bf R}_L
\rb 
p_0(\mT_L) p_0({\bf R}_L) \mu(\sd \mT_L \times \sd {\bf R}_L),
\label{eqLikelihoodOriginal}
\end{equation}
\end{itemize}
then we can aim at computing the posterior pmf of the number of layers
$$
P\lb 
	L|Y(1:N) 
\rb \propto P\lb 
	Y(1:N)|L 
\rb P_0(L)
$$
and choose the model according to the {\em maximum a posteriori} (MAP) criterion
\begin{eqnarray}
\hat L_{MAP} &=& \arg \max_{L \in \{1, 2, \ldots, L_+ \}} P\lb L|Y(1:N)\rb \nonumber\\
&=& \arg \max_{L \in \{1, 2, \ldots, L_+ \}} P\lb Y(1:N)|L\rb P_0(L), \label{eqMAPProblem}
\end{eqnarray}
i.e., we choose the value of $L$ that turns out more probable given the available observations. 

Expression \eqref{eqMAPProblem} yields the optimal solution to the problem of selection the number of layers in the multiplex from a probabilistic Bayesian point of view. As discussed below, one can find alternatives to this approach in the literature on hidden Markov models (HMMs) \cite{Rabiner89,Ghahramani01}, however the latter suffer from a number of theoretical and practical limitations and we strongly advocate the Bayesian solution \eqref{eqMAPProblem}.

\subsection{Connections with hidden Markov model estimation theory}

The problem of selecting the number of layers $L$ in the multiplex model can be cast as one of selecting a hidden Markov model (HMM) where the complete state is ${\bf Z}(t) = [ X(t), \ell(t) ]^\top$ and the transition from time $t$ to time $t+1$ is governed by the (unknown) stochastic matrices ${\bf R}_L, {\bf T}^{(0)}, \ldots, {\bf T}^{(L-1)}$. Let us adapt the notation a bit in order to make it closer to the classical HMM theory. We only have partial observations of the Markov chain, namely the sequence $\mathcal{O} = Y(1:N)$, with ``emission probabilities'' $P\lb Y(t)|X(t) \rb$, while the sequence of layer labels $\{\ell(t)\}_{t=1}^N$ remains unobserved. The goal is to estimate the total number of layers $L$ from the observations $\mathcal{O}$.

The theory of HMMs has received considerable attention in the literature since the 70s, due to their application in a variety of fields, including speech processing, molecular biology, data compression  or artificial intelligence. The problem of fitting a HMM, i.e., estimating its unknown parameters has been thoroughly researched \cite{Rabiner89}. The classical technique is the Baum-Welch algorithm, which is actually an instance of the expectation-maximisation (EM) method \cite{Dempster77,Rabiner89}. Indeed, the general EM methodology, in several forms, is the standard approach to the problem of fitting HMMs, often combined with other techniques for its implementation, such as the Viterbi algorithm, the forward-backward algorithm or the Kalman smoother (see \cite{Ghahramani01} for an excellent survey). Many of these techniques adopt the general form of an space-alternating EM algorithm where the unobserved states and the unknown parameters are iteratively estimated, one at a time. The space-alternating generalised EM (SAGE) methodology was introduced in \cite{Fessler94} and provides a common framework for many current algorithms for fitting HMMs.

However, the estimation of the number of layers in the proposed multiplex scheme does not amount to HMM fitting. Modelling $L$ as a random variable, in order to solve problem \eqref{eqMAPProblem} we aim at computing the model posterior probabilities given the available data $\mathcal{O}$, i.e.,
\begin{equation}
P(L|\mathcal{O}) \propto P(\mathcal{O}|L) P_0(L) 
\label{eqBMS}
\end{equation}
where $P_0(L)$, $L=1, 2, ...$, are the a priori probabilities we attribute to models with different number of layers (e.g., we may deem models with many layers less probable than simpler models with a few layers) and $P(\mathcal{O}|L)$ is the model likelihood. The latter is an integral with respect to the probability distribution of the matrix-parameters $\mT_L=\{{\bf T}^{(0)}, \ldots, {\bf T}^{(L-1)}\}$ and ${\bf R}_L$ for an $L$-layer system, namely
\begin{equation}
P(\mathcal{O}|L) = \int P( \mathcal{O} | \mT_L, {\bf R}_L ) p_{0,T}(\mT_L) p_{0,R}({\bf R}_L) \mu({\sf d}\mT_L \times {\sf d}{\bf R}_L),
\label{eqMLkhd}
\end{equation}
which is equivalent to Eq. \eqref{eqLikelihoodOriginal}. Recall that $p_{0,T}(\mT_L)$ and $p_{0,R}({\bf R}_L)$ are a priori pdf's w.r.t. a reference measure $\mu$. These pdf's can be chosen differently for different values of $L$. EM methods for HMM fitting are tools to address the problem of estimating $\mT_L$ and ${\bf R}_L$ via the maximisation of the parameter likelihood $P( \mathcal{O} | \mT_L, {\bf R}_L )$ that appears in the integrand of \eqref{eqMLkhd}. 

We see from \eqref{eqBMS} and \eqref{eqMLkhd}, however, that what we need is to be able to {\em integrate} the likelihood $P( \mathcal{O} | \mT_L, {\bf R}_L )$, rather than {\em maximising} it. Nevertheless, most methods in the HMM literature tackle the model selection problem (in our case, selection of the number of layers $L$) by computing estimates of the parameters via the EM method and then comparing the likelihoods {\em of the optimised parameters} \cite{Ghahramani01,Siddiqi07}. In our setup, this means that, given two choices $L_1$ and $L_2$, we would estimate $\hat \mT_{L_1}, \hat {\bf R}_{L_1}$ and $\hat \mT_{L_2}, \hat {\bf R}_{L_2}$ (using an EM scheme to maximise $P( \mathcal{O} | \mT_{L_i}, {\bf R}_{L_i} )$ for $i=1,2$) and then compare the likelihoods $P( \mathcal{O} | \hat \mT_{L_1}, \hat {\bf R}_{L_1} )$ and $P( \mathcal{O} | \hat \mT_{L_2}, \hat {\bf R}_{L_2} )$. This approach has several problems:
\begin{itemize}
\item There is no guarantee $\hat \mT_{L_1}, \hat {\bf R}_{L_1}$ and $\hat \mT_{L_2}, \hat {\bf R}_{L_2}$ are accurate estimates (e.g., they may be overfitted). It may well happen that, e.g., $\hat \mT_{L_1}, \hat {\bf R}_{L_1}$ are poor estimates and, hence, $P( \mathcal{O} | \hat \mT_{L_1}, \hat {\bf R}_{L_1} ) < P( \mathcal{O} | \hat \mT_{L_2}, \hat {\bf R}_{L_2} )$, while $P(\mathcal{O}|L_1) > P(\mathcal{O}|L_2)$.

\item The EM framework yields {\em local} optimisation algorithms. Even if the EM scheme converges, it may yield a local maximiser of the likelihood for $L_1$ and, perhaps, a global maximiser for $L_2$. In this case, we may have, again, that  $P( \mathcal{O} | \hat \mT_{L_1}, \hat {\bf R}_{L_1} ) < P( \mathcal{O} | \hat \mT_{L_2}, \hat {\bf R}_{L_2} )$, while $P(\mathcal{O}|L_1) > P(\mathcal{O}|L_2)$.

\item Even if we manage to obtain accurate maximum likelihood estimates of $\mT_{L_1},  {\bf R}_{L_1}$ and $\mT_{L_2}, {\bf R}_{L_2}$, there is no guarantee that $P( \mathcal{O} | \hat \mT_{L_1}, \hat {\bf R}_{L_1} ) < P( \mathcal{O} | \hat \mT_{L_2}, \hat {\bf R}_{L_2} )$  must imply $P(\mathcal{O}|L_1) < P(\mathcal{O}|L_2)$.
\end{itemize}
Many authors have aimed at mitigating these flaws by introducing different heuristics in the way the models to be fitted are chosen (typically, heuristics for merging and splitting candidate states, in our case candidate layers) and producing sophisticated EM parameter estimation algorithms. See \cite{Siddiqi07} for examples. This approach does not attack the core of the problem, though. 

Instead, \cite{Ghahramani01} advocates Bayesian model selection as a framework to address problem \eqref{eqBMS} that automatically handles overfitting (by imposing prior probability distributions on the parameters) and the comparison of models of different complexity (by integrating over the parameters as in \eqref{eqMLkhd}). In \cite{Ghahramani01}, the term used for the MAP model selection method of \eqref{eqMAPProblem} is, actually, {\em Bayesian integration}, which makes reference to the need to solve, or numerically approximate, the integral in \eqref{eqMLkhd}. Some candidate methods to tackle this computation include:
\begin{itemize}
\item The Laplace approximation \cite{Bishop06}, which consists in searching the maximum of $P( \mathcal{O} | \hat \mT_L, \hat {\bf R}_L )$ and then approximating the integrand $P( \mathcal{O} | \mT_L, {\bf R}_L ) p_{0,T}(\mT_L) p_{0,R}({\bf R}_L)$ by a Gaussian with the adequate covariance structure. This approach ignores the fact that $P( \mathcal{O} | \hat \mT_L, \hat {\bf R}_L )$ is, in our case, multimodal.

\item The variational Bayes method \cite{Watanabe04,McGrory09} is an approximation scheme that relies on the use of surrogate probability distributions for the parameters (which need to be analytically tractable) in order to design an EM method that tackles the maximisation of the model likelihood $P(\mathcal{O}|L)$, i.e., the integral in \eqref{eqMLkhd}. It is a relatively ``inexpensive'' method in terms of computational cost, comparable to classical EM-based model fitting techniques. However, as any EM scheme, it performs a local optimisation and does not guarantee an optimal solution.

\item Deterministic integration of \eqref{eqMLkhd} using either deterministic regular grids on the space of the parameters $\{ \mT_L, {\bf R}_L \}$ or specific cubature methods for some convenient  family of functions \cite{Sobolev13}. While accurate, the complexity of these methods typically grows exponentially with the dimension of the parameters, hence they can be prohibitive for larger scale models. Examples for multiplex models with up to $L \ge 4$ layers are shown.

\item Conventional Monte Carlo integration suffers from a similar complexity limitation. Classical Markov chain Monte Carlo (MCMC) samplers \cite{Gilks96,Robert04} could be well-suited to solve integrals with respect to the posterior pdf $p(\mT_L,{\bf R}_L| \mathcal{O})$; however, the integral in \eqref{eqMLkhd} is actually the normalising constant of this posterior, which turns out to be hard to estimate via MCMC, which limits its application to model selection in general \cite{Robert04}.

The classical alternative to MCMC in Monte Carlo integration is importance sampling (IS) \cite{Robert04}. While conventional IS suffers from a problem called weight degeneracy, that translates into poor scaling with the dimension of the parameters in the integral, recently, families of much more efficient {\em adaptive} IS schemes have been introduced \cite{DelMoral06,Cappe08,Beskos14,Koblents15}. These techniques yield estimates of the integral in \eqref{eqMLkhd} in a simple way (unlike MCMC) and can potentially work in high dimensions \cite{Beskos14}.  
\end{itemize}

Below, we present a detailed description of the nonlinear population Monte Carlo (PMC) scheme of \cite{Koblents15} and show and example of model selection with up to 10 layers ($L \le 10$). While conventional (and even state-of-the-art) importance samplers are based on the computation of weights of the form $w(z) \propto \frac{p(z)}{q(z)}$, where $p(z)$ is the target pdf and $q(z)$ is a proposal density, the key feature of the nonlinear PMC scheme is to compute transformed weights $\bar w(z) \propto \phi\lb \frac{p(z)}{q(z)} \rb$, where $\phi(\cdot)$ is a nonlinear function, in order to reduce the variance of the weights (if $z$ is a random variable, then $U=w(Z)$ is random as well). This very simple transformation, if properly chosen, improves significantly the numerical stability of the algorithm when the dimension of $Z$ grows, while preserving the convergence properties of conventional IS. The examples presented below, for the nonlinear PMC and a deterministic scheme based on regular grids, show that this Monte Carlo integration scheme can be as effective as a deterministic integrator with just a fraction of the running time. 

%
\subsection{Computation of the posterior probabilities via Monte Carlo integration}

Let us return to the original notation where $Y(1:N)$ denotes de sequence of observations. In order to select the number of layers $L$ in the multiplex according to the Bayesian criterion in Eq. \eqref{eqMAPProblem}, we need the ability to evaluate the posterior probability
$$
P\lb L | Y(1:N) \rb \propto P\lb Y(1:N) | L \rb P_0(L),
$$
where the prior $P_0(L)$ is known (chosen by design) but the model likelihood $P\lb Y(1:N) | L \rb$ is an integral given by Eq. \eqref{eqLikelihoodOriginal}, namely
\begin{equation}
P\lb Y(1:N) | L \rb = \int P\lb Y(1:N) | \mT_L, {\bf R}_L \rb p_0(\mT_L) p_0({\bf R}_L) \mu({\sf d}\mT_L \times {\sf d}{\bf R}_L).
\label{eqLkAgain}
\end{equation}
Using the Bayes theorem, we realise that the integrand in \eqref{eqLkAgain} is proportional to the posterior density of the parameters given the observations $Y(1:N)$, i.e.,
\begin{equation}
p\lb \mT_L, {\bf R}_L | Y(1:N) \rb \propto P\lb Y(1:N) | \mT_L, {\bf R}_L \rb p_0(\mT_L) p_0({\bf R}_L).
\label{eqQQ0}
\end{equation}
Taken together, Eqs. \eqref{eqLkAgain} and \eqref{eqQQ0} indicate that the model likelihood $P\lb Y(1:N) | L \rb$ is the normalisation constant of the posterior pdf of the parameters, $p\lb \mT_L, {\bf R}_L | Y(1:N) \rb$. This normalisation constant is often termed the {\em model evidence} is the Bayesian terminology.

An efficient way of computing the normalisation constant of a target pdf via Monte Carlo integration is by using the importance sampling (IS) method.

\subsubsection{Importance sampling in a nutshell}

Let $p(z)$ be a target pdf that we can evaluate up to a normalisation constant $c$, i.e., we have the ability to compute
$$
\tilde p(z) = c p(z)
$$
point-wise, but $c$ is unknown. The IS method \cite{Robert04} enables the estimation of $c$ (actually, it enables the estimation of integrals of the form $\int f(z)p(z)dz$ in general, for any integrable test function $f$) by sampling from an alternative pdf, $q(z)$, often called {\em proposal} density or {\em importance function}. We assume that $q(z)$ is chosen to satisfy that
\begin{equation}
w(z) = \frac{
	\tilde p(z)
}{
	q(z)
} < \infty,
\label{eqAsWe}
\end{equation}
where $w(z)$ is the {\em weight function}. The inequality in \eqref{eqAsWe} typically implies, at least, that $q(z)>0$ whenever $p(z)>0$.

The basic IS algorithm proceeds as follows:
\begin{enumerate}
\item Draw $M$ independent samples $z^1, \ldots, z^M$ from $q(z)$.
\item Compute weights 
$$
\tilde w^i = w(z^i) = \frac{ \tilde p(z^i) }{ q(z^i) }, \quad \mbox{for $i=1, \ldots, M$.}
$$
\item Normalise the weights,
$$
w^i = \frac{
	\tilde w^i
}{
	\sum_{m=1}^M \tilde w^m
}.
$$
\end{enumerate}
It is a straightforward application of the strong law of large numbers \cite{Robert04} to prove that
$$
\lim_{M\rw\infty} \sum_{i=1}^M w^i f(z^i) = \int f(z)p(z)dz \quad \mbox{almost surely (a.s.)}
$$
for any square-integrable test function $f$. 

However, the most relevant result for the purpose of this paper is that
\begin{equation}
\hat c^M = \frac{1}{M} \sum_{i=1}^M \tilde w^i
\label{eqNormaEst}
\end{equation}
is an unbiased, consistent estimator of the normalisation constant $c$ since. By the strong law of large numbers again,
\begin{equation}
\lim_{M\rw\infty} \hat c^M = \int w(z)q(z)dz = \int \tilde p(z)dz = c \quad \mbox{a.s.}
\label{eqNormaEst2}
\end{equation}
since $\tilde p(z) = c p(z)$ and $p(z)$ is a pdf (hence, it integrates to 1).

In the Bayesian model selection problem at hand, the target non-normalised function is given by $\tilde p(z) \equiv P\lb Y(1:N) | \mT_L, {\bf R}_L \rb p_0(\mT_L) p_0({\bf R}_L)$, which we can evaluate (as will be shown below), and $c = \int \tilde p(z)dz \equiv P\lb Y(1:N) | L \rb$ is the model likelihood.

\subsubsection{Nonlinear population Monte Carlo}

The main drawback of standard IS is that, whenever there is a significant mismatch between $\tilde p(z)$ and $q(z)$, the variance of the weights becomes very large. As a consequence, estimators converge very slowly (with $M$) and they become of little use. This issue is usually referred to as {\em weight degeneracy} \cite{Kong94}. It typically happens when the dimension of the random variable $Z$ is large or when, simply, the target pdf is very narrow.

To mitigate degeneracy, a number of adaptive IS have been proposed, especially since the publication of \cite{Cappe04}. Here we resort to one such method, that introduces a nonlinear transformation of the weights to control their variability and, hence, degeneracy. The technique is called nonlinear population Monte Carlo (NPMC) and it was originally proposed in \cite{Koblents15}. It consists of $J$ iterations, each involving the computation of both conventional importance weights (IWs) and transformed importance weights (TIWs). The transformation is a clipping or truncation operation, denoted $\phi(\cdot,\cdot)$. For a set of $M$ ordered IWs, $\tilde w^{i_1} > \tilde w^{i_2} > \cdots > \tilde w^{i_M}$, we obtain a set of $M$ TIWs, with clipping of order $M_c < \sqrt{M}$, as 
\begin{equation}
\bar w^i = \phi\lb i,\{ \tilde w^m \}_{m=1}^M \rb = \left\{
	\begin{array}{ll}
	\tilde w^{i_{M_c}}, &\mbox{if $\tilde w^i \ge \tilde w^{i_{M_c}}$,}\\
	\tilde w^i, &\mbox{otherwise.}
	\end{array}
\right.
\nonumber
\end{equation}
This operation truncates the $M_c$ bigger weights. A general algorithm is outlined below, with $M$ samples per iteration and an instrumental Markov sampling kernel $K\lb \cdot,\cdot | \mT_L^i, {\bf R}_L^i \rb$ centred at $\mT_L^i, {\bf R}_L^i$.
\begin{framed}
\begin{enumerate}
\item \textbf{Initialisation.} 
	\begin{enumerate}
	\item Draw $M$ independent samples $(\tilde \mT_{L,0}^i, \tilde {\bf R}_{L,0}^i)$, $i=1, ..., M$,  from the prior pdf's $p_0(\mT_L)$ and $p_0({\bf R}_L)$.
	\item Compute non-normalised IWs, $\tilde w_0^i = P\lb Y(1:N) | \tilde \mT_{L,0}^i, \tilde {\bf R}_{L,0}^i \rb$, $i=1, ..., M$.
	\item Compute non-normalised TIWs, $\bar w_0^i = \phi\lb i, \{\tilde w_0^m\}_{m=1}^M \rb$, $i=1, ..., M$.
	\item Normalise the TIWs, 
	$$
	w^i_0 = \frac{
		\bar w^i_0
	}{
		\sum_{m=1}^M \bar w_0^m
	}, \quad i=1, ..., M.
	$$ 
	\item Resample $M$ times the set $\{ \tilde \mT_{L,0}^i, \tilde {\bf R}_{L,0}^i \}_{i=1}^M$, with replacement and using the normalised TIWs as probability masses, to yield an unweighted sample set $\{ \mT_{L,0}^i, {\bf R}_{L,0}^i \}_{i=1}^M$.
	\end{enumerate}
	
\item \textbf{Iteration.} For $j = 1:J$:
	\begin{enumerate}
	\item Draw $M$ independent samples 
	$$
	(\tilde \mT_{L,j}^i, \tilde {\bf R}_{L,j}^i) \sim K\lb \mT_L,{\bf R}_L | \mT_{L,j-1}^i, {\bf R}_{L,j-1}^i \rb, \quad i=1, ..., M.
	$$
	\item Compute non-normalised IWs, 
	$$
	\tilde w_j^i = \frac{
		P\lb Y(1:N) | \tilde \mT_{L,j}^i, \tilde {\bf R}_{L,j}^i \rb p_0(\tilde \mT_{L,j}^i) p_0(\tilde {\bf R}_{L,j}^i)
	}{
		K\lb \tilde \mT_{L,j}^i, \tilde {\bf R}_{L,j}^i | \mT_{L,j-1}^i, {\bf R}_{L,j-1}^i \rb
	}, \quad i=1, ..., M.
	$$
	\item Compute non-normalised TIWs, $\bar w_j^i = \phi\lb i, \{\tilde w_j^m\}_{m=1}^M \rb$, $i=1, ..., M$.
	\item Normalise the TIWs, 
	$$
	w^i_j = \frac{
		\bar w^i_j
	}{
		\sum_{m=1}^M \bar w_j^m
	}, \quad i=1, ..., M.
	$$ 
	\item Resample $M$ times the set $\{ \tilde \mT_{L,j}^i, \tilde {\bf R}_{L,j}^i \}_{i=1}^M$, with replacement and using the normalised TIWs as probability masses, to yield an unweighted sample set $\{ \mT_{L,j}^i, {\bf R}_{L,j}^i \}_{i=1}^M$.
	\end{enumerate}
\end{enumerate}
\end{framed}

After the $J$-th iteration, we have the IS estimator of the model likelihood
$$
P^M\lb Y(1:N) | L \rb = \frac{1}{M} \sum_{i=1}^M \tilde w_J^i.
$$ 
This is a consistent estimator that converges with optimal Monte Carlo error rates. In particular, assuming that the IWs are bounded and bounded away from zero, \cite[Theorem 1]{Koblents16} states that for any, arbitrarily small $\epsilon<\frac{1}{2}$ there exists an a.s. finite random variable $U^\epsilon$ such that
$$
\left| P^M\lb Y(1:N) | L \rb - P\lb Y(1:N) | L \rb \right| < \frac{U^\epsilon}{M^{\frac{1}{2}-\epsilon}}
$$
and, therefore, 
$$
\lim_{M\rw\infty} P^M\lb Y(1:N) | L \rb = P\lb Y(1:N) | L \rb \quad \mbox{a.s.}
$$

All that remains is to show that the parameter likelihood function $P\lb Y(1:N) |  \mT_{L,j},  {\bf R}_{L,j} \rb$, and hence the IWs and the TIWs, can be evaluated exactly.

\subsubsection{Exact calculation of the parameter likelihood $P\lb Y(1:N) | \mT_L, {\bf R}_L \rb$}

\noindent We start with the factorisation
\begin{equation}
P\lb Y(1:N)|\mT_L,{\bf R}_L \rb=\prod_{t=1}^N P\lb Y(t)|Y(1:t-1),\mT_L,{\bf R}_L \rb
\label{eqStartWith}
\end{equation}  
where each factor 
\begin{equation}
P\lb 
	Y(t)|Y(1:t-1), \mT_L, {\bf R}_L
\rb \label{uno}
\end{equation}
is a pmf that can be computed exactly using a recursive algorithm. We can write the function $P\lb Y(t)|Y(1:t-1),\mT_L,{\bf R}_L \rb$ as a marginal of the mass $P\lb Y(t),X(t),\ell(t)|Y(1:t-1),\mT_L,{\bf R}_L \rb$, namely
 \begin{equation}
P\lb 
	Y(t)|Y(1:t-1),\mT_L,{\bf R}_L 
\rb = \sum_{X(t)} \sum_{\ell(t)} P\lb 
	Y(t)|X(t) 
\rb P\lb 
	X(t),\ell(t)|Y(1:t-1),\mT_L,{\bf R}_L 
\rb
\label{eqPMFy}
\end{equation}
where we have used the fact that $P(Y(t)|{\bf X}(t))=P(Y(t)|X(t))$. The conditional observation pmf $P(Y(t)|X(t))$ is one of the building blocks of the state space model, so it can be readily evaluated. The predictive pmf $P(X(t),\ell(t)|Y(1:t-1),\mT_L,{\bf R}_L)$, on the other hand, can be decomposed as
 \begin{eqnarray}
 P\lb 
 	X(t),\ell(t)|Y(1:t-1),\mT_L,{\bf R}_L
\rb &=& \sum_{X(t-1)} \sum_{\ell(t-1)} P\lb
	X(t)|X(t-1),\ell(t-1), \mT_L
\rb \times \nonumber\\
&& \times  P\lb
	\ell(t)|{\ell(t-1)},{\bf R}_L
\rb \times \nonumber\\
&& \times P\lb
	X(t-1),\ell(t-1)|Y(1:t-1),\mT_L,{\bf R}_L
\rb
 \label{eqPredUV}
 \end{eqnarray}
 which depends on the filtering pmf at time $t-1$:
\begin{eqnarray}
 P\lb
 	X(t-1),\ell(t-1)|Y(1:t-1),\mT_L,{\bf R}_L
\rb &\propto& P\lb 
	Y(t-1)|X(t-1) 
\rb \times \nonumber \\
&& \times P\lb 
	X(t-1),\ell(t-1)|Y(1:t-2),\mT_L,{\bf R}_L 
\rb.
 \label{eqFiltUV}
\end{eqnarray} 

\noindent Note that Eqs. \eqref{eqPMFy}, \eqref{eqPredUV} and \eqref{eqFiltUV} are recursively related. If we start from the prior pmf $P_0(X(0),\ell(0))=P_0(X(0))P_0(\ell(0))$, then we can recursively compute the required sequence of pmf's for $t=1, 2, \ldots, T$ as
\begin{eqnarray}
P\lb
	X(t),\ell(t)|Y(1:t-1),\mT_L,{\bf R}_L
\rb &=& \sum_{X(t-1)} \sum_{\ell(t-1)} P\lb
	X(t)|X(t-1),\ell(t-1), \mT_L
\rb \times \nonumber \\
&& \times P\lb
	\ell(t)|\ell(t)-1,{\bf R}_L
\rb \times \nonumber\\
&& \times P\lb
	X(t-1),\ell(t-1)|Y(1:t-1),\mT_L,{\bf R}_L
\rb,\label{eqKK1}\\
P\lb
	Y(t)|Y(1:t-1),\mT_L,{\bf R}_L
\rb &=& \sum_{X(t)} \sum_{\ell(t)} P\lb
	Y(t)|X(t)
\rb \times \nonumber\\
&& \times P\lb
	X(t),\ell(t)|Y(1:t-1),\mT_L,{\bf R}_L
\rb,\label{eqKK2}\\
P\lb
	X(t),\ell(t)|Y(1:N),\mT_L,{\bf R}_L
\rb &\propto& P\lb
	Y(t)|X(t)
\rb P\lb
	X(t),\ell(t)|Y(1:t-1),\mT_L,{\bf R}_L
\rb. \nonumber \\
\label{eqKK3}
\end{eqnarray}
Substituting \eqref{eqKK2} into \eqref{eqStartWith} for each $t=1, ..., N$ yields the parameter likelihood.

%
%
\section{QUANTIFYING COMPLEXITY: ADDITIONAL RESULTS} \label{sIntegration}

In this section we present some additional computer simulation results that complement those shown in the main text. For the simulations, we assume (the same as in the main text) a simple ring topology for the layers of the multiplex, as well as a particular structure for the matrix ${\bf R}_L$, in such a way that it can be parametrised by a single scalar $r\in(0,1)$. This is described in Section \ref{sMul}. In Section \ref{sApprox} we show how, given the ring topology of the layer and assuming exact observations $Y(t)=X(t)$, it is possible to construct a relatively simple deterministic approximation to the posterior probabilities $P(L|Y(1:N))$. An NPMC algorithm for this specific model is also detailed. Finally, our numerical results are displayed and discussed in Section \ref{sNum}.

\subsection{Multiplex of ring-shaped layers} \label{sMul}
From now on we assume a ring topology (cycle graph) for every layer. In this particular case the particle may jump from one node to its neighbours (left or right on the ring) and from one layer to another layer. This motion is random, and we model it by way of the following probabilities,
\begin{eqnarray}
R_{ij} &=& P\lb \ell(t)=j | \ell(t-1)=i \rb, \\
\sT^{(l)} &=& P\lb
	X(t)=X(t-1)+1 \Mod{K} | \ell(t-1) = l
\rb,
\end{eqnarray}
where $R_{ij}$ represents the probability of moving from layer $i$ to layer $j$ and $\sT^{(l)}$ is the probability of moving rightwards in layer $l$, i.e., of jumping from node $i$ to node $i + 1 \Mod{K}$. The probability of moving leftwards within layer $l$ is, therefore, 
$$
1-\sT^{(l)} = 1 - P\lb 
	X(t)=X(t-1)-1 \Mod{K} |\ell(t-1)=l 
\rb.
$$ 
These rightwards-jump probabilities for layers $l=0, 1, \ldots, L-1$, are put together in the vector \footnote{In the main text, we use ${\bf T}_L$ to refer to the set of transition matrices $\mT_L = \{ {\bf T}^{(l)} \}_{l=0}^{L-1}$. For the multiplex network being described, where each layer is a ring, the probabilities $\sT^{(l)}$, $l=0, \ldots, L-1$, contain the same information as $\mT_L$, hence $\mT_L$ and ${\bf T}_L$ are equivalent.}
 \begin{equation}
  {\bf T}_L=\left[
  	\begin{array}{c}
	\sT^{(1)}\\
	\vdots\\
	\sT^{(l)}\\
	\end{array}
\right]_{L\times 1}\\
 \end{equation}

In order to simplify the structure of the stochastic matrix ${\bf R}$ we assume that, at each time $t$, the walker may stay at the same layer (as in $t-1$) with probability $1-r$ ($r\in(0,1)$) or jump to a different layer with probability $\frac{r}{L-1}$. Hence, the transition matrix ${\bf R}_L$ becomes
$${\bf R}_L=\left[\begin{array}{cccc}
1-r & \frac{r}{L-1} & \cdots & \frac{r}{L-1}\\
\frac{r}{L-1} & 1-r & \cdots & \frac{r}{L-1}\\
\vdots & \vdots & \ddots & \vdots\\
\frac{r}{L-1} & \frac{r}{L-1} & \cdots & 1-r \end{array} \right]$$
and there is a single unknown parameter, the probability to jump, $r$.

\subsection{Approximation of the posterior model probabilities}\label{sApprox}
In the main text we have focused on scenarios where the node visited at time $t$ can be observed exactly, i.e., $Y(t)=X(t)$ and 
$$
P(Y(t)=j | X(t)=i ) = \left\{
	\begin{array}{ll}
	1, &\mbox{if $j=i$,}\\
	0, &\mbox{otherwise.}\\
	\end{array}
\right.
$$
This constraint simplifies some of the general equations in Section \ref{sPost}. Specifically, the recursions given by eqs. \eqref{eqKK1}--\eqref{eqKK3} for the evaluation of the pmf $P(Y(t)|Y(1:t-1),{\bf T}_L,r)$ (note that we have reduced ${\bf R}_L$ and $\mT_L$ to the scalar $r$ and vector, ${\bf T}_L$)  become
\begin{eqnarray}
P(X(t)|X(1:t-1),{\bf T}_L,r) &=& \sum_{\ell(t-1)} P\lb
	X(t)|\ell(t-1),X(t-1),{\bf T}_L
\rb \times \nonumber\\
&& \times P\lb
	\ell(t-1)|X(1:t-1),{\bf T}_L,r 
\rb, \label{eqRec1}\\
P(\ell(t)|X(1:t),{\bf T}_L,r) &=& P\lb
	X(t) | \ell(t), X(1:t-1), {\bf T}_L, r
\rb P\lb
	\ell(t) | X(1:t-1), {\bf T}_L, r
\rb \nonumber\\
&=& P\lb
	X(t) | X(1:t-1), {\bf T}_L, r
\rb \times \nonumber \\
&& \times \sum_{\ell(t-1)} P\lb
	\ell(t) | \ell(t-1), r
\rb P\lb
	\ell(t-1) | X(1:t-1),{\bf T}_L,r
\rb. \nonumber \\ \label{eqRec2}
\end{eqnarray}
Eqs. \eqref{eqRec1} and \eqref{eqRec2} can be applied recursively with initial conditions given by the prior pmf's $P_0(X(0))$ and $P_0(\ell(0))$. In particular, note that 
\begin{equation}
P(X(1)|{\bf T}_L,r) = \sum_{X(0)}\sum_{\ell(0)} P(X(1)|X(0),\ell(0),{\bf T}_l)P(\ell(0))P(X(0))
\label{eqAux0}
\end{equation}
while 
\begin{equation}
P(\ell(1)|{\bf T}_l,r)= \sum_{\ell(0)} P(\ell(1)|\ell(0),r) P(\ell(0)).
\end{equation}
These priors are assumed known as part of the model specification (typically, they can be uniform pmf's, as they have been selected in the computer experiments).

Given Eq. \eqref{eqRec1} we can evaluate the likelihood of the parameters ${\bf T}_L$ and $r$. In particular,
\begin{equation}
P\lb
	X(1:N) | {\bf T}_L, r
\rb = P(X(1)|{\bf T}_L,r) \prod_{t=1}^N P(X(t)|X(1:t-1),{\bf T}_L,r)
\label{eqLaIntegral}
\end{equation}
with $P(X(1)|{\bf T}_L,r)$ computed as in \eqref{eqAux0}. Numerically, it is convenient to work with the log-likelihood 
\begin{equation}
\log P\lb
	X(1:N) | {\bf T}_L, r
\rb = \log P(X(1)|{\bf T}_L,r) + \sum_{t=1}^N \log P(X(t)|X(1:t-1),{\bf T}_L,r)
\end{equation}
and transform it into natural units only when strictly needed. 

The likelihood of the r.v. $L$ given by the observation record $X(1:N)$ is, therefore, given by the integral
\begin{equation}
P( X(1:N) | L ) = \int P( X(1:N) | {\bf T}_L, r ) p_0({\bf T}_L) p_0(r) \sd{\bf T}_L \sd r,
\end{equation}
where we have assumed prior densities wrt the Lebesgue measure. Indeed, in the absence of any prior knowledge, it is natural to choose $p_0(r)$ to be uniform over the interval $(0,1)$, i.e., $p_0(r) = 1$. The prior $p_0({\bf T}_L)$ is important because it is used to penalise system configurations with two or more identical layers. Note that a multiplex with $L=2$ and identical transition matrices ${\bf T}^{(1)} = {\bf T}^{(2)} = \tilde {\bf T}$ is equivalent to a monoplex with transition matrix $\tilde {\bf T}$. Similarly, any multiplex with $L$ layers out of which $L'\le L$ have identical transition matrices is fully equivalent to a reduced system with $L-L'+1$ layers \footnote{Along these lines, also note that a multiplex with $L$ ring-shaped layers and $r=\frac{1}{2}$ can be reduced to a monoplex with probability to move rightwards $T_{i,i+1} = \frac{1}{L}\sum_{l=0}^{L-1} \sT^{(l)}$.}. For our computer simulations, we have chosen
\begin{equation}
p_0({\bf T}_L) \propto \min_{l \ne l'} \left| \sT^{(l)} - \sT^{(l')} \right|,
\end{equation} 
i.e., we penalise systems with at least two layers that have similar transition probabilities.

\subsection{Deterministic integration}

One conceptually simple way to approximate the integral in Eq. \eqref{eqLaIntegral} is 
\begin{equation}
P( X(1:N) | L ) \approx \sum_{\sT^{(0)} \in \mG} \cdots \sum_{\sT^{(L-1)} \in \mG} \sum_{r \in \mG} P\lb X(1:N) | {\bf T}_L, r \rb p_0({\bf T}_L),
\end{equation}
where $\mG=\{ g_1, \ldots, g_H\}$ is a grid of $H$ points over the interval $(0,1)$, i.e., $0 < g_1 < g_2 < \cdots < g_H < 1$. The proportionality constant of the prior $p_0({\bf T}_L)$ can be approximated numerically as well (over the grid $\mG^L$). We have selected a uniform prior on $L$, i.e., assuming that $L \in \{ 1, 2, \ldots, L_+ \}$, we choose $P_0(L)=\frac{1}{L_+}$. In this way, the MAP selection criterion reduces to choosing the value of $L \in \{1, 2, \ldots, L_+\}$ that maximises the model likelihood $P( X(1:N) | L )$.

\subsection{NPMC algorithm for Monte Carlo integration}

Let us specify a practical NPMC algorithm for the multiplex composed of ring-shaped layers. The prior densities are $p_0(r) = U(0,1)$ and $p_0({\bf T}_L) \propto \min_{l\neq l'} | \sT^{(l)} - \sT^{(l')} |$. We approximate the normalisation constant $\hat C$ of the latter pdf by standard Monte Carlo integration (simulating the $\sT^{(l)}$'s uniformly over $(0,1)$).

The sampling kernels $K(\cdot,\cdot | {\bf T}_L,r)$ are truncated Gaussian pdf's. Specifically, if $TN(x|\mu,\sigma^2,a,b)$ denotes the truncated Gaussian density
$$
TN(x|\mu,\sigma^2,a,b) = \frac{
	\exp\left\{
		-\frac{1}{2\sigma^2} ( x - \mu )^2
	 \right\}
}{
	\int_a^b \exp\left\{
		-\frac{1}{2\sigma^2} ( u - \mu )^2
	 \right\} {\sf d}u
}, \quad \mbox{for $x \in (a,b)$,}
$$ 
then 
$$
K\lb {\bf T}_L, r | \breve {\bf T}_L, \breve r\rb = TN(r | \breve r, \sigma^2, 0, 1) \prod_{l=1}^L TN( \sT^{(l)} | \breve \sT^{(l)}, \sigma^2, 0, 1) 
$$
where $\sigma^2 = 0.005$.

The number of clipped weights is $M_c = \lfloor \sqrt{M} \rfloor$. Recall that function $\phi(i,\{ \tilde w_j^m \}_{m=1}^M)$ truncates the $M_c$ biggest weights to be equal (and identical to the $M_c$-th largest weight).

\begin{framed}
\begin{enumerate}
\item \textbf{Initialisation.} 
	\begin{enumerate}
	\item Draw $M$ independent samples $(\tilde {\bf T}_{L,0}^i, \tilde r_{0}^i)$, $i=1, ..., M$,  from the prior pdf's $p_0({\bf T}_L)$ and $p_0(r) = 1$ for $r\in (0,1)$.
	\item Compute non-normalised IWs, $\tilde w_0^i = P\lb X(1:N) | \tilde {\bf T}_{L,0}^i, \tilde r_{0}^i \rb$, $i=1, ..., M$.
	\item Compute non-normalised TIWs, $\bar w_0^i = \phi\lb i, \{\tilde w_0^m\}_{m=1}^M \rb$, $i=1, ..., M$.
	\item Normalise the TIWs, 
	$$
	w^i_0 = \frac{
		\bar w^i_0
	}{
		\sum_{m=1}^M \bar w_0^m
	}, \quad i=1, ..., M.
	$$ 
	\item Resample $M$ times the set $\{ \tilde {\bf T}_{L,0}^i, \tilde r_{0}^i \}_{i=1}^M$, with replacement and using the normalised TIWs as probability masses, to yield an unweighted sample set $\{ {\bf T}_{L,0}^i, r_{0}^i \}_{i=1}^M$.
	\end{enumerate}
	
\item \textbf{Iteration.} For $j = 1:J$:
	\begin{enumerate}
	\item Draw $M$ independent samples 
	\begin{eqnarray}
	\tilde r_j^i &\sim& TN(r | r_{j-1}^i, \sigma^2, 0, 1), \quad i=1, ..., M,\nonumber\\
	\tilde \sT_j^{(l),i} &\sim& TN( \sT | \sT_{j-1}^{(l),i}, \sigma^2, 0, 1 ), \quad l = 1, ..., L, \quad i = 1, ..., M. \nonumber 
	\end{eqnarray}
	\item Compute non-normalised IWs, 
	$$
	\tilde w_j^i = \frac{
		P\lb X(1:N) | \tilde {\bf T}_{L,j}^i, \tilde r_{j}^i \rb p_0(\tilde {\bf T}_{L,j}^i) p_0(\tilde r_{j}^i)
	}{
		TN(\tilde r_j^i | r_{j-1}^i, \sigma^2, 0, 1) \prod_{l=1}^L TN( \tilde \sT_j^{(l),i} | \sT_{j-1}^{(l),i}, \sigma^2, 0, 1 )
	}, \quad i=1, ..., M.
	$$
	\item Compute non-normalised TIWs, $\bar w_j^i = \phi\lb i, \{\tilde w_j^m\}_{m=1}^M \rb$, $i=1, ..., M$.
	\item Normalise the TIWs, 
	$$
	w^i_j = \frac{
		\bar w^i_j
	}{
		\sum_{m=1}^M \bar w_j^m
	}, \quad i=1, ..., M.
	$$ 
	\item Resample $M$ times the set $\{ \tilde {\bf T}_{L,j}^i, \tilde r_{j}^i \}_{i=1}^M$, with replacement and using the normalised TIWs as probability masses, to yield an unweighted sample set $\{ {\bf T}_{L,j}^i, r_{j}^i \}_{i=1}^M$.
	\end{enumerate}
\end{enumerate}
\end{framed}



\subsection{Numerical results}\label{sNum}
\noindent {
{\bf From deterministic integration to NPMC: efficiently validating the full architecture. }We first consider a different example than the one shown in the main text, where now the true model from which the observations are generated has only $L=2$ layers, and we aim at estimating this hidden model via posterior probability maximisation. As for the case considered in the main text, the posterior probabilities are computed via the approximation in \ref{sApprox}, in particular using a 19-point grid $\mG = \{0.05, 0.10, \ldots, 0.95\}$ for each unknown variable, which includes the probability to move rightwards, $\sT^{(l)}$, over each layer and the additional parameter, $r$, which is the probability to jump to a different layer at each iteration. For a maximum of $L=4$ layers, this makes a grid of $19^5$ nodes, plus a $19^4$ grid for $L=3$, plus a $19^3$ grid for $L=2$, plus a 19-point grid for $L=1$.\\}

\begin{figure}[h]
\centering
\includegraphics[width=0.4\columnwidth]{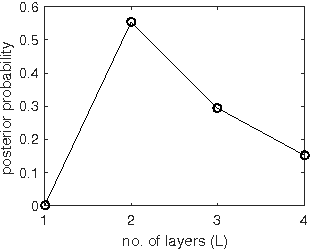}
\includegraphics[width=0.4\columnwidth]{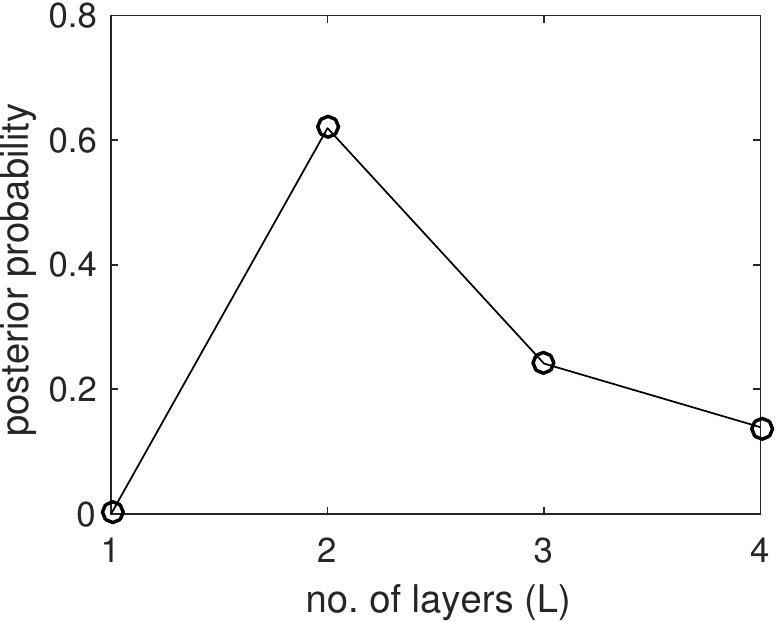}
\caption{(Left panel) Posterior probability $P(L | X(1:N) )$ as a function of the number of layers $L$, computed from a trajectory of $10^4$ time steps generated using a model with $L=2$ layers and parameters $1-r=0.89$, $\sT^{(1)} = 0.14$ and $\sT^{(2)} = 0.56$. Computation is performed using deterministic grid integration. The algorithm clearly picks out the correct model $L=2$. (Right panel) Similar results as for the left panel, but using the NPMC algorithm instead of the deterministic integration, which is much faster and allows us to reconstruct the full architecture.}
\label{additional_estimate}
\end{figure}

{In figure \ref{additional_estimate} we plot the posterior probability of four possible models with $L=1,\dots,4$ layers. It peaks out at the correct value, although less emphatically than for the case shown in the main text. The reason is that in this example, these probabilities are obtained using a record of only $N=10^4$ observations, while we used a larger time series of $N=2\times 10^4$ observations in the case reported in the main text. Nevertheless, the method correctly classifies the underlying architecture, what suggests that the estimation protocol is robust against short series size, something extremely relevant for real-world applications. Note also that the posterior probability of $L=1$ is zero up to the computer degree of accuracy. The observations $X(1:N)$, $N=10^4$, were generated from a model with $1-r=0.89$, $\sT^{(1)} = 0.14$ and $\sT^{(2)} = 0.56$. This is not a node of the grid $\mG^3$ and yet the model is identified unequivocally.\\} 

{In a second step, we explore this particular case with our NPMC algorithm,  with $I=10$ iterations, $M=900$ Monte Carlo samples per iteration and clipping parameter $M_c=\sqrt{M}=30$, to approximate the posterior probabilities $P(\tilde L|X(1:N))$ for $\tilde L=1,2,3$ and $4$.  We observe in the right panel of Fig. \ref{additional_estimate} that the posterior probability peaks at $\tilde L=L=2$, with similar values of the approximations. The runtime was $\approx68$ minutes, while the computations for deterministic grid integration took $\approx6,480$ minutes (a reduction by a factor of $\approx95$) on the same computer and with the same software.\\}

{\noindent Moreover, the NPMC also allows us to efficiently estimate, given $L=2$, the most probable architecture of the multiplex, i.e. $r$, $\sT^{(1)}$ and $\sT^{(2)}$. As NPMC performs an importance sampling and therefore grids parameter space tighter in the regions with high probability measure, then Monte Carlo sampling is guaranteed to look at parameters with high likelihood. As a matter of fact, since the importance sampling provides a complete representation of the probability measure, no configurations with high likelihood will be left out (multimodal densities and likelihoods are approximated accurately, including all extrema, if the number of Monte Carlo samples --the points in the grid-- is large enough), and in this sense this is a global optimization method, at odds with expectation-maximization, which converges locally.\\ 
In figure \ref{importance_q_r} we depict the parameters sampled by the NPMC (100 of Monte Carlo samples), showing that they are indeed very close to the ground truth. The probability densities of the parameters are shown in figure \ref{parameter_densities}. Actually, since the grid itself yields a random discrete probability measure, integrals with respect to this measure converge to integrals with respect to the posterior probability distribution of the parameters. Hence, we can compute various kinds of estimates of the parameters (and their expected errors). A simple inspection of the posterior probability of each of these configurations therefore allows to estimate the full architecture, which will coincide with our ground truth.\\}

\begin{figure}
\centering
\includegraphics[width=0.5\columnwidth]{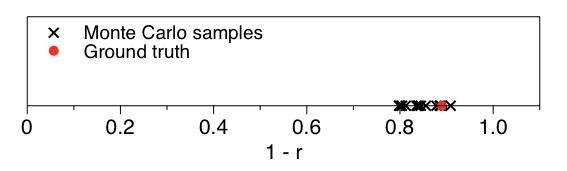}
\includegraphics[width=0.5\columnwidth]{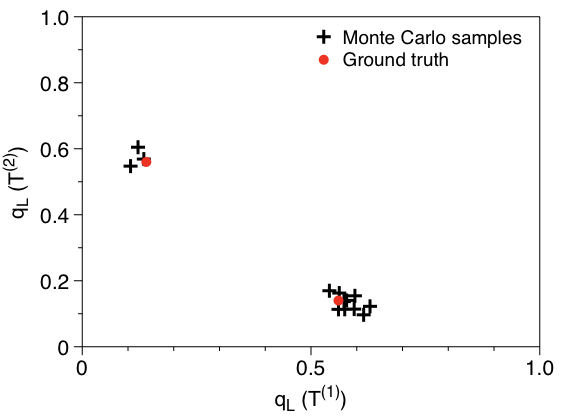}
\caption{{Scatter plots of 100 Monte Carlo samples of $1-r$ (top panel) and the interlayer transition probabilities $\sT^{(1)}$ and $\sT^{(2)}$ (bottom panel). All these are concentrated in a region close to the ground truth (note that there are two clouds in the bottom panel due to the symmetry $sT^{(1)} \leftrightarrow \sT^{(2)}$).}}
\label{importance_q_r}
\end{figure}

\begin{figure}
\centering
\includegraphics[width=0.3\columnwidth]{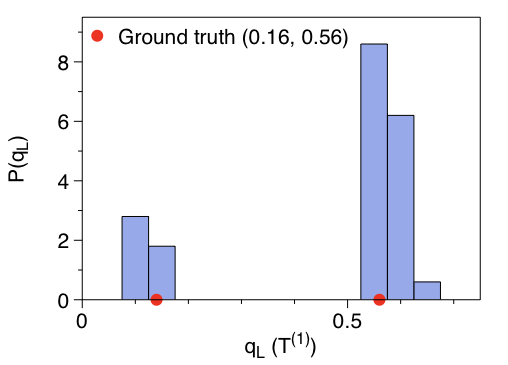}
\includegraphics[width=0.3\columnwidth]{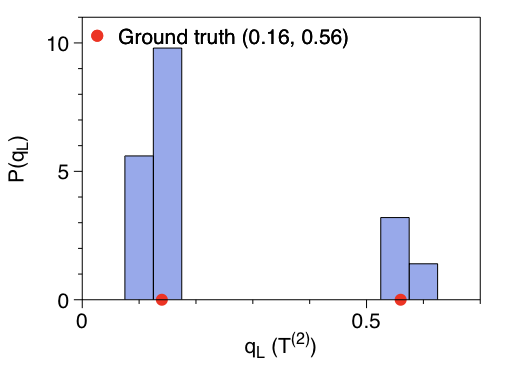}
\includegraphics[width=0.3\columnwidth]{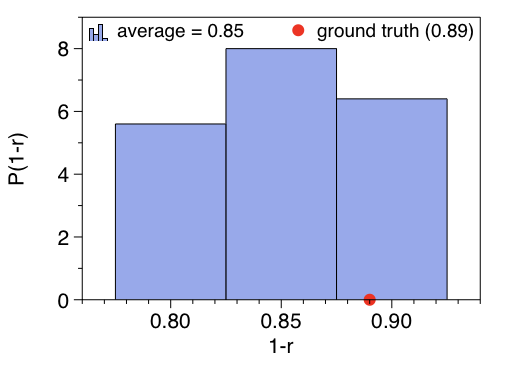}
\caption{{Probability densitiy of each parameter obtained from the 100 Monte Carlo samples of figure \ref{importance_q_r}.}}
\label{parameter_densities}
\end{figure}

\begin{figure}
\centering
\includegraphics[width=0.4\columnwidth]{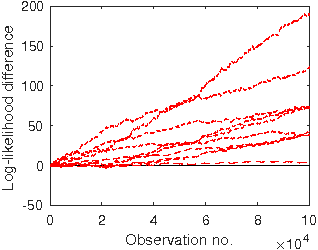}
\caption{Log-likelihood difference between a model with the true values and ten models with slightly perturbed values in the parameters, as a function of the size of the observation sequence. In every case we find that the log-likelihood difference is larger than zero, meaning that the likelihood of the true model is always larger than the likelihoods of the models with perturbed parameters: the model with maximum a posteriori probability is indeed also the model with the correct parameters (not just a model with correct number of layers), and this is an apparent global maximum. This result is robust as it gets more acute as we increase the size of the observed walker sequence.}
\label{perturbation}
\end{figure}

\noindent {\bf Estimated architecture is a likelihood maximum.} As discussed above, we have also confirmed that the likelihood of the parameters, $P(X(1:N)|{\bf T}_L,r)$ (which is the key ingredient in the approximation of $P(L|X(1:N))$), is maximised at the values corresponding to the true model that generates the observations. In order to illustrate this fact further, we again consider the system under study in the main text (a model with $L=3$ layers, with $1-r=0.84$ and ${\bf T}_L=[0.16, 0.76, 0.24]^\top$) together with ten slightly perturbed models with $L=3$ layers and parameters $1-r'=1-r+\Delta$ and ${\bf T}_L'=[0.16+\Delta', 0.76+\Delta'', 0.24+\Delta''']$, where $\Delta, \Delta',\Delta'',\Delta'''$ represent independent Gaussian perturbations of zero mean and variance $5 \times 10^{-4}$. As a function of the time series size (the observation window $X(1:N)$), we have computed the likelihood of each set of parameter values, which in every case we compare with the likelihood obtained for the true parameters. In figure \ref{perturbation} we plot the logarithm of the ratio of these likelihoods, where a positive values indicates that the ratio is larger than one, i.e., that the likelihood of the true parameters is larger than the likelihood of the perturbed parameters. We systematically find that such is the case and, interestingly, we find that any small perturbation grows and generates a monotonically decreasing likelihood-ratio as the size of the observed series increases.\\

\noindent {\bf Scalability.} Finally, we consider the issue of scalability and computational cost. All simulations have been carried out using Matlab R2016a running on an Apple iMac equipped with an Intel Core i7 processor, with 4 cores and 32 GB of RAM. For a first experiment, we have considered the same model and data as in Fig. \ref{additional_estimate}, i.e., a system with $L=2$ layers, parameters $1-r=0.89$, $\sT^{(1)} = 0.14$ and $\sT^{(2)} = 0.56$ and a series of $N=10^4$ observations. As discussed previously, using NPMC instead of deterministic integration reduced by a factor of $\approx 95$ computing time. Therefore, equipped with the NPMC algorithm we can tackle more complex systems. For the last computer experiment we have generated a sequence of $N=20\times 10^3$ observations from a model with $L=5$ layers, and parameters $\sT^{(1)} = 0.1$, $\sT^{(2)} = 0.2$, $\sT^{(3)} = 0.6$, $\sT^{(4)} = 0.7$, $\sT^{(5)} =  0.9$ and $1-r = 0.85$. We have applied the NPMC algorithm, with $J=10$ iterations $M=1,000$ Monte Carlo samples per iteration and $M_c=31$, to approximate the posterior probabilities $P(\tilde L|X(1:N))$ for $\tilde L=1, \ldots, 10$. This simulation took around 370 minutes on the same computer. Fig. \ref{fNPMC2} displays the results, where we observe that the correct model is still clearly identified.

\begin{figure}
\centering
\includegraphics[width=0.4\columnwidth]{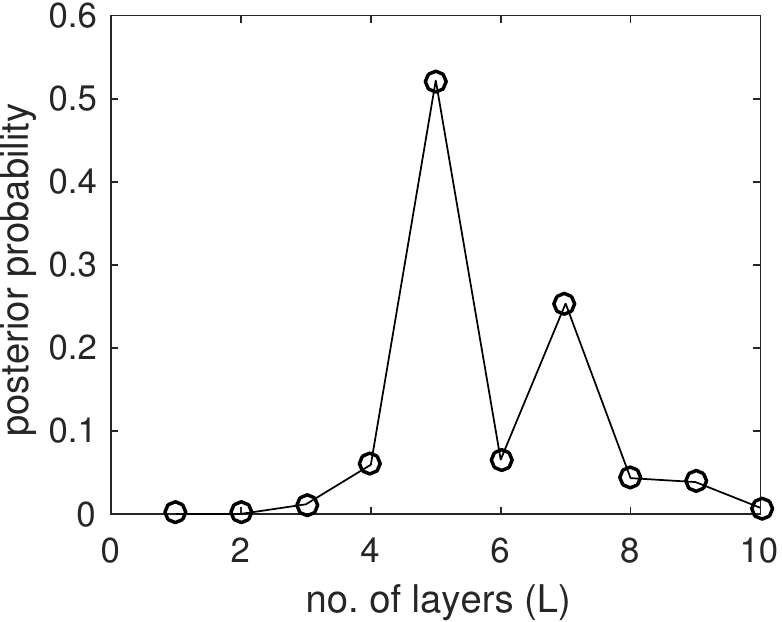}
\caption{Posterior probability $P(L | X(1:N) )$ as a function of the number of layers $L$, computed from a trajectory of $N=20\times 10^3$ time steps generated using a model with $L=5$ and parameters $\sT^{(1)} = 0.1$, $\sT^{(2)} = 0.2$, $\sT^{(3)} = 0.6$, $\sT^{(4)} = 0.7$, $\sT^{(5)} =  0.9$ and $1-r = 0.85$. The algorithm clearly picks out the correct model $L=5$. These results are obtained using the NPMC algorithm.}
\label{fNPMC2}
\end{figure}
 
  \begin{figure}
\centering
\includegraphics[width=0.5\columnwidth]{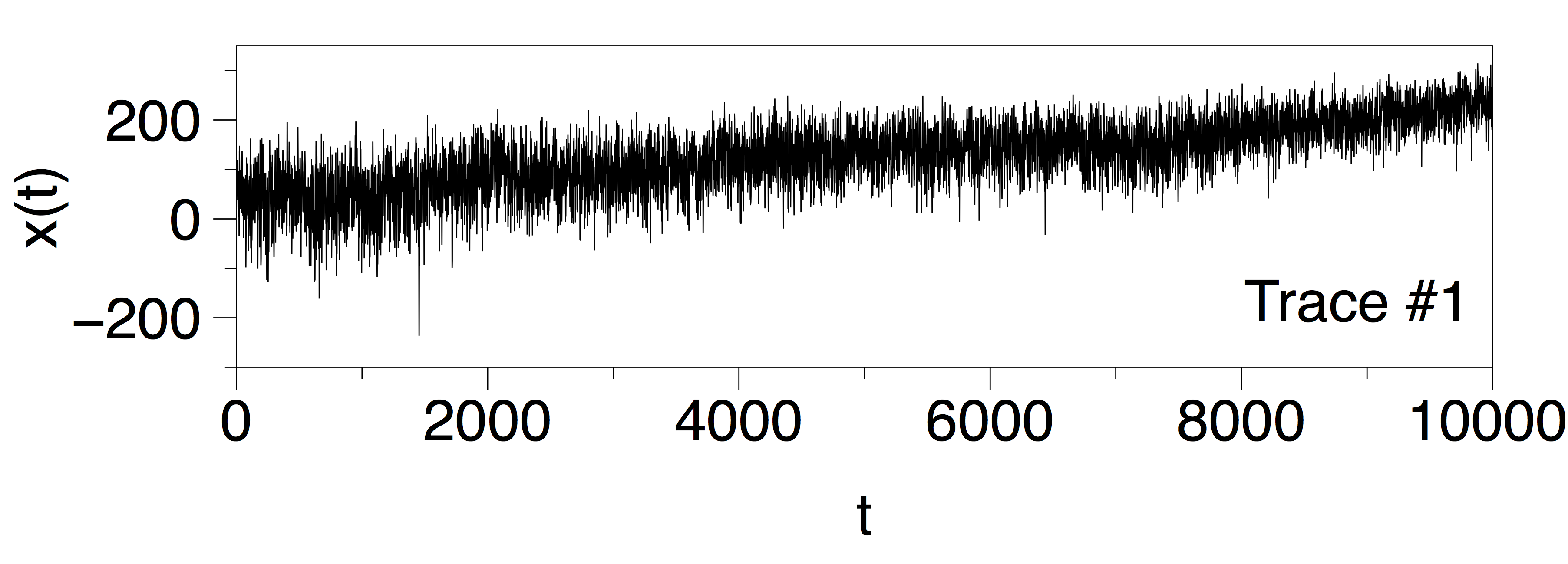}
\includegraphics[width=0.3\columnwidth]{sampleseries1.png}
\includegraphics[width=0.5\columnwidth]{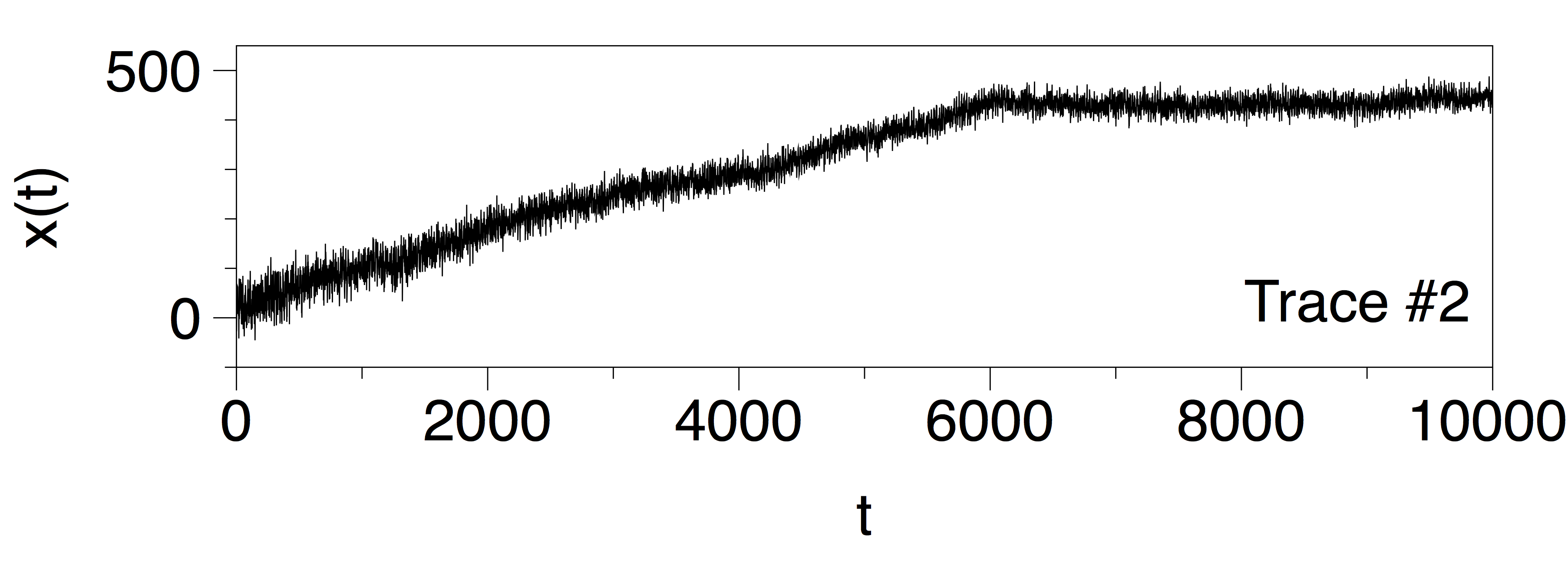}
\includegraphics[width=0.3\columnwidth]{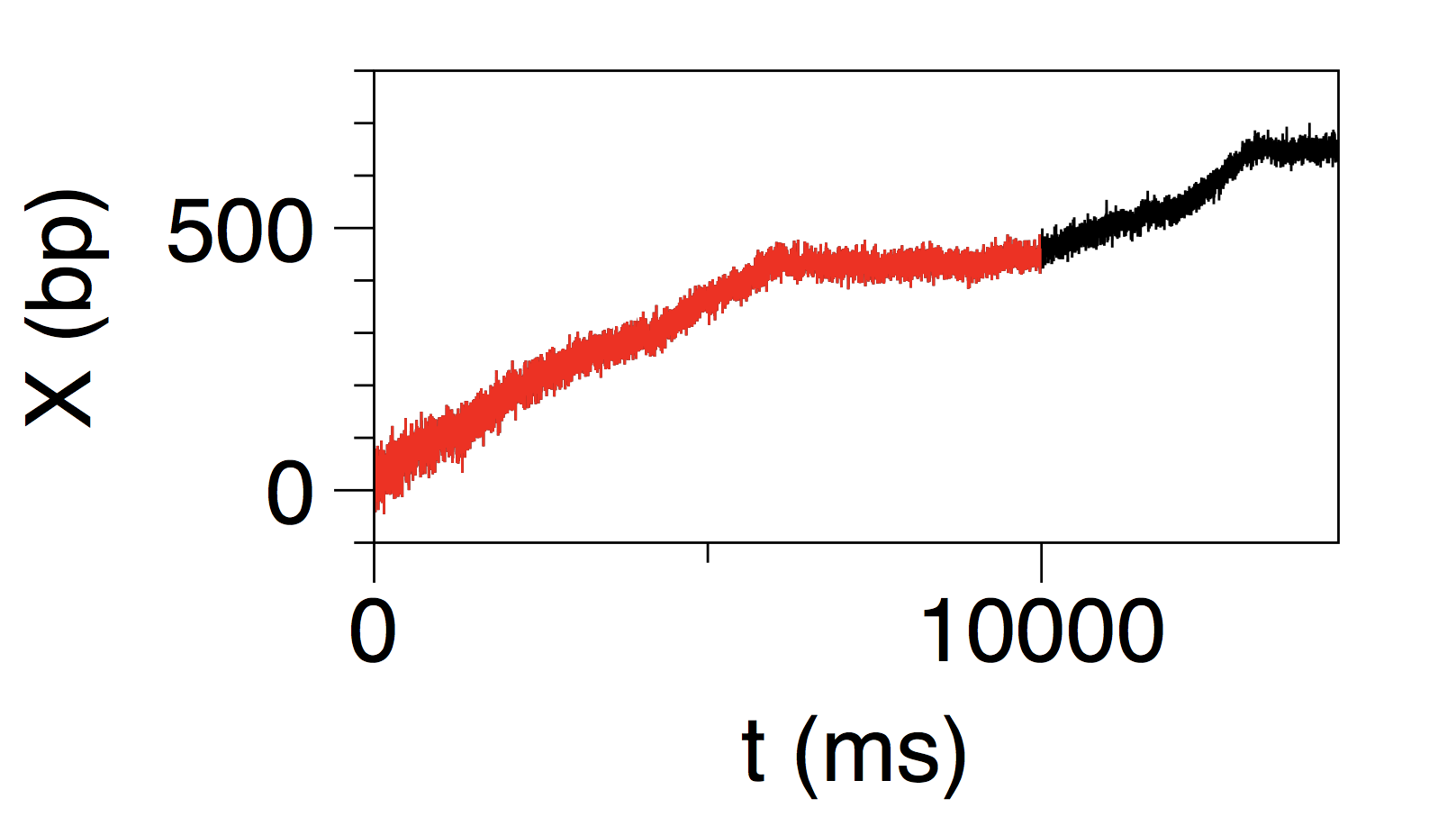}
\includegraphics[width=0.5\columnwidth]{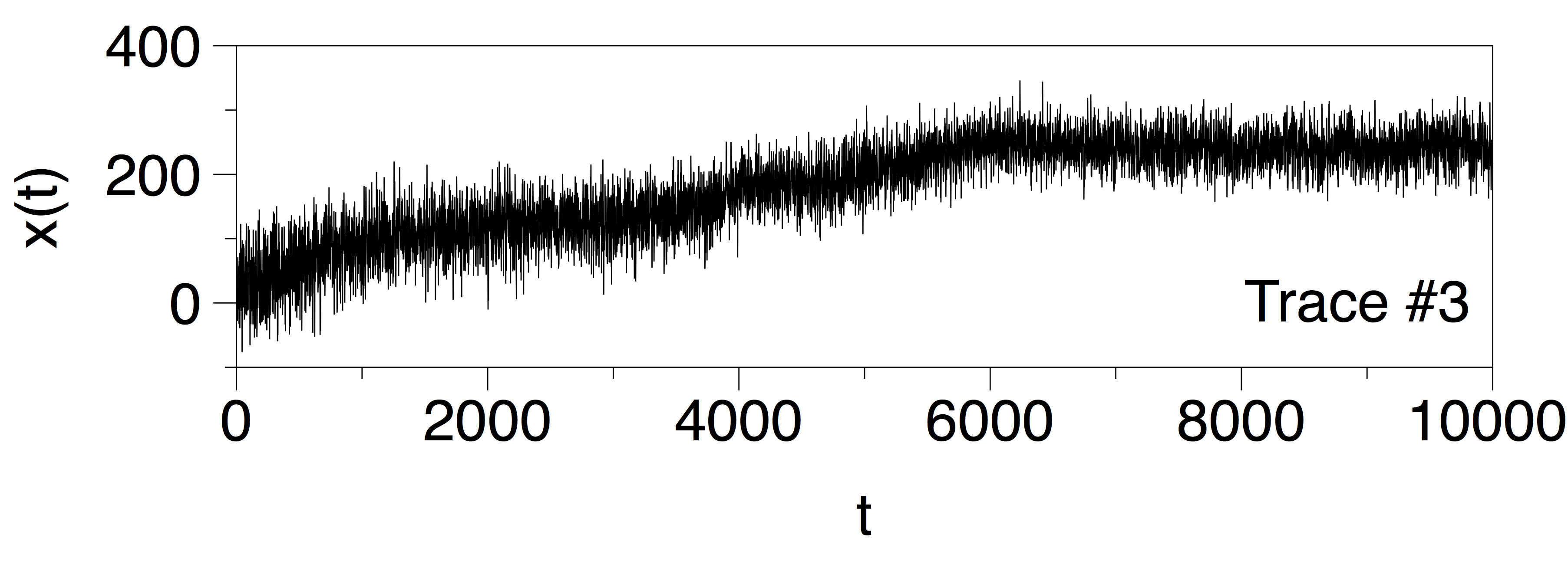}
\includegraphics[width=0.3\columnwidth]{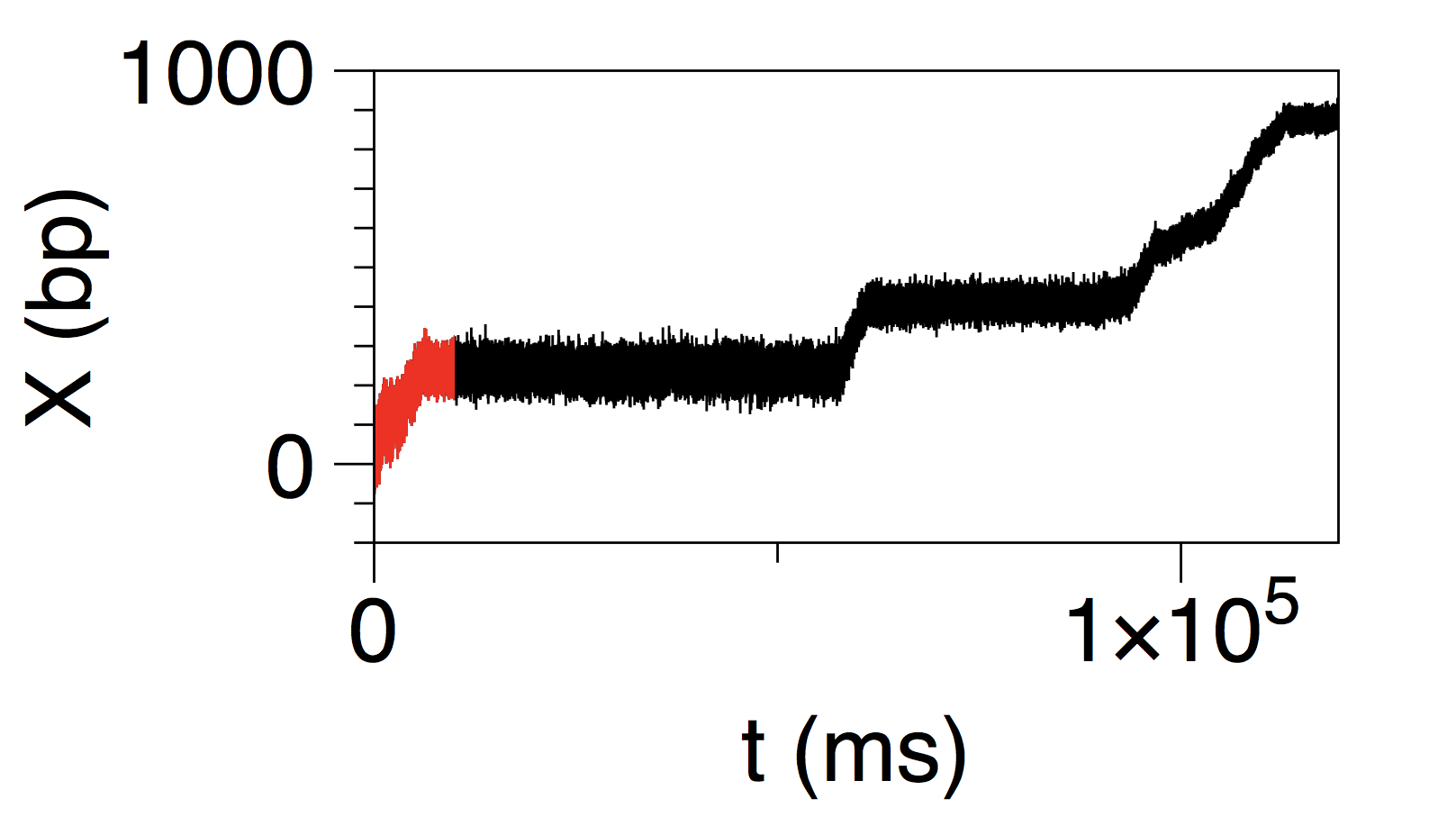}
\includegraphics[width=0.5\columnwidth]{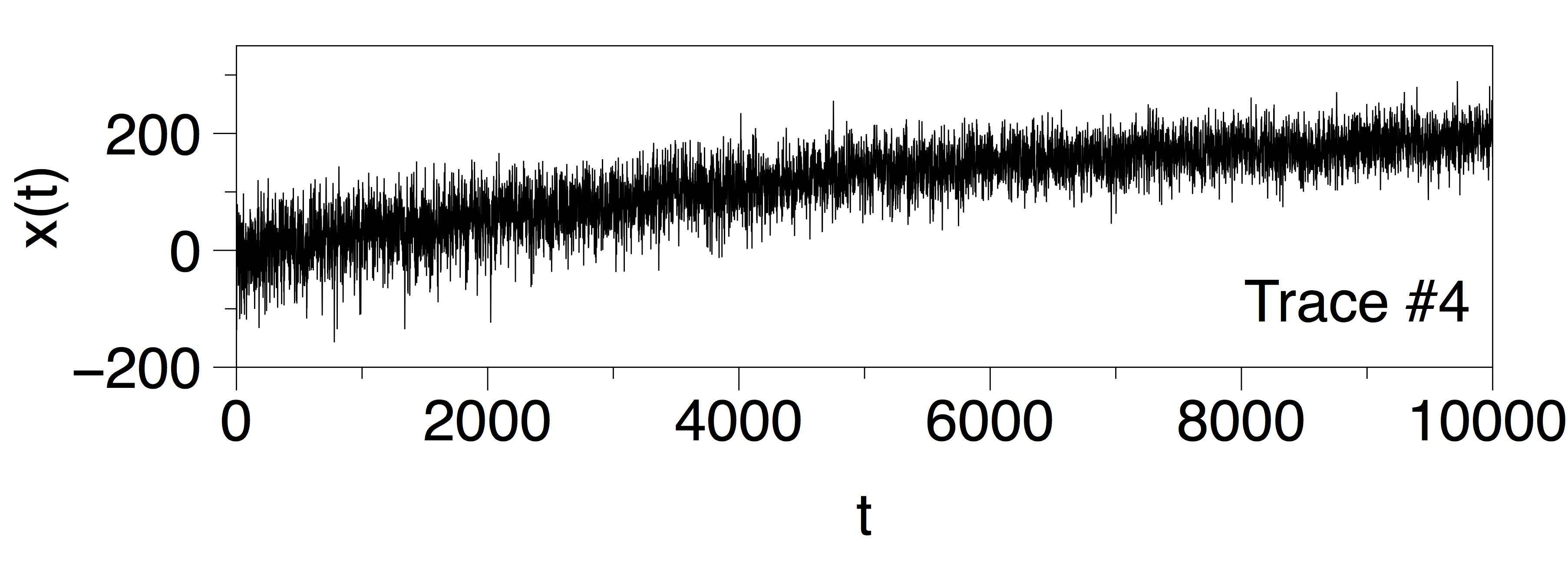}
\includegraphics[width=0.3\columnwidth]{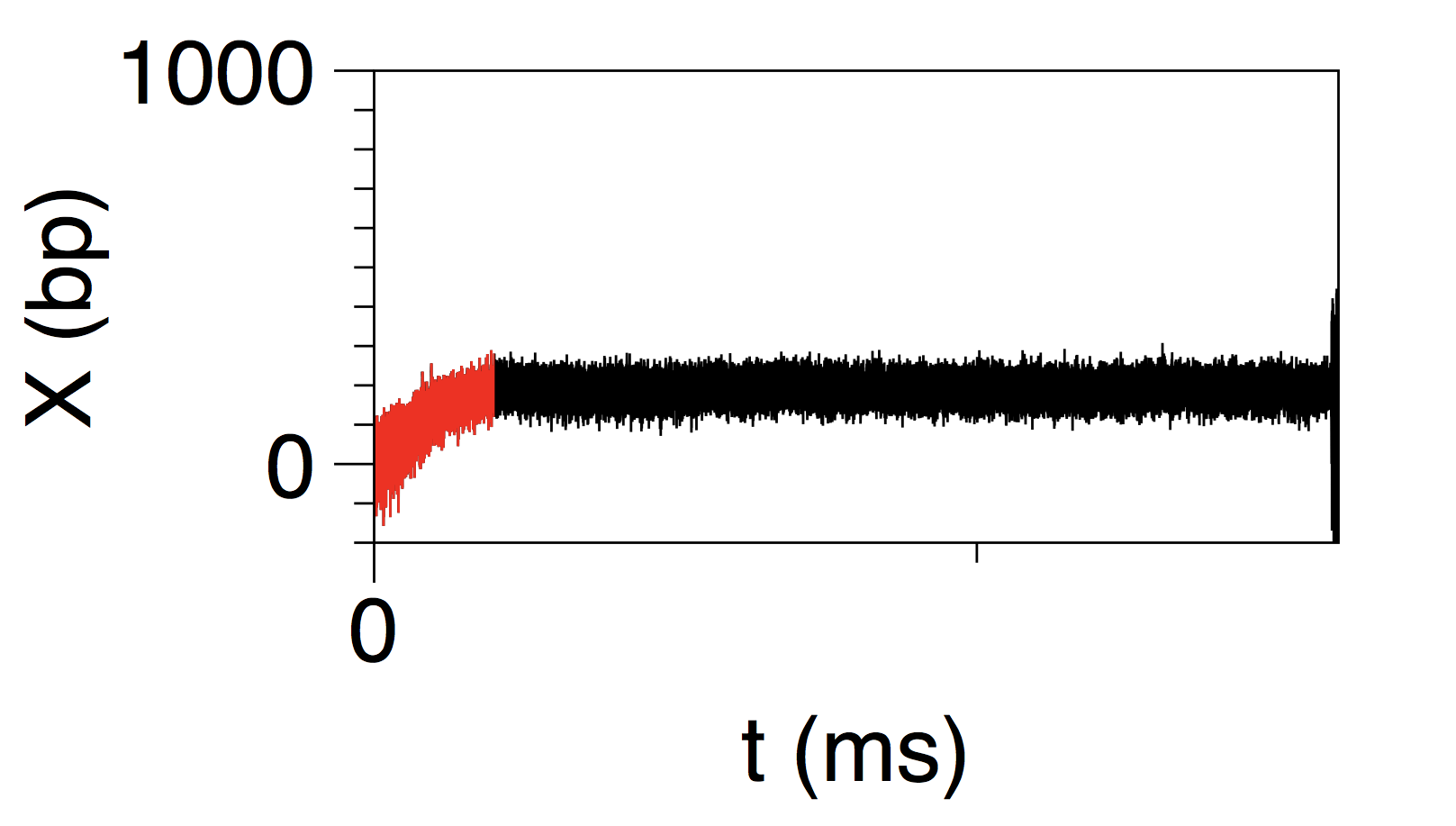}
\includegraphics[width=0.5\columnwidth]{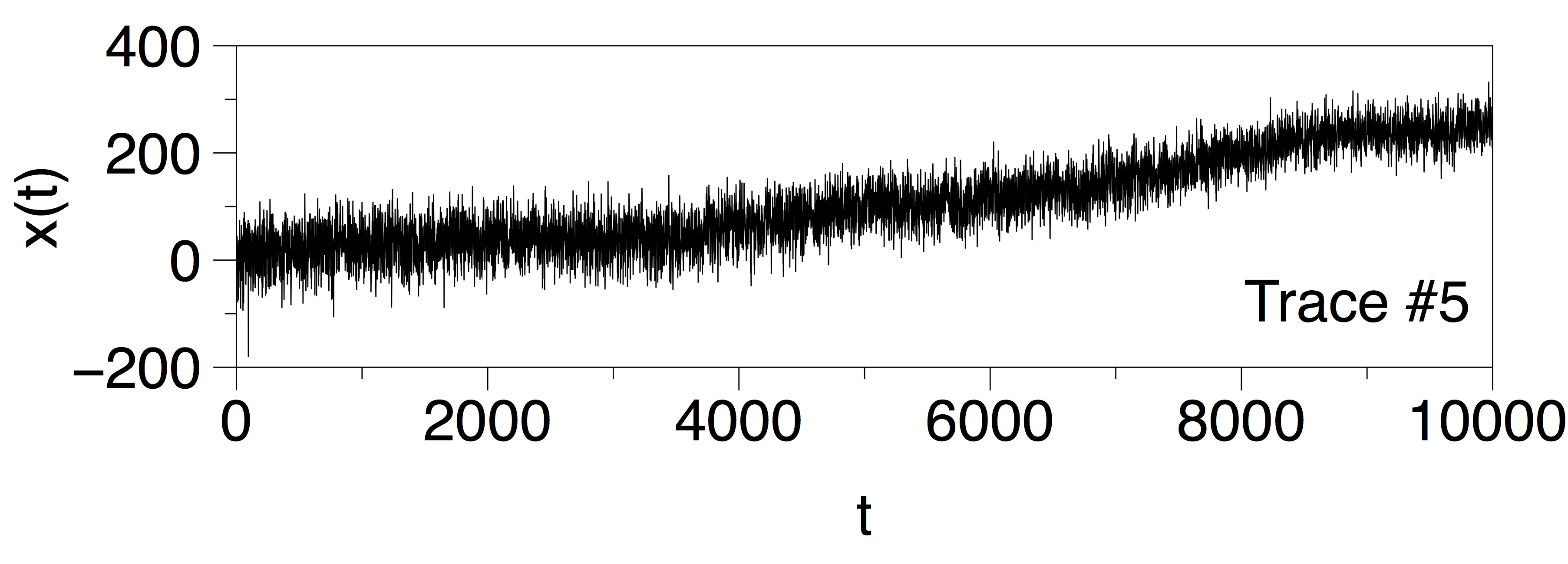}
\includegraphics[width=0.3\columnwidth]{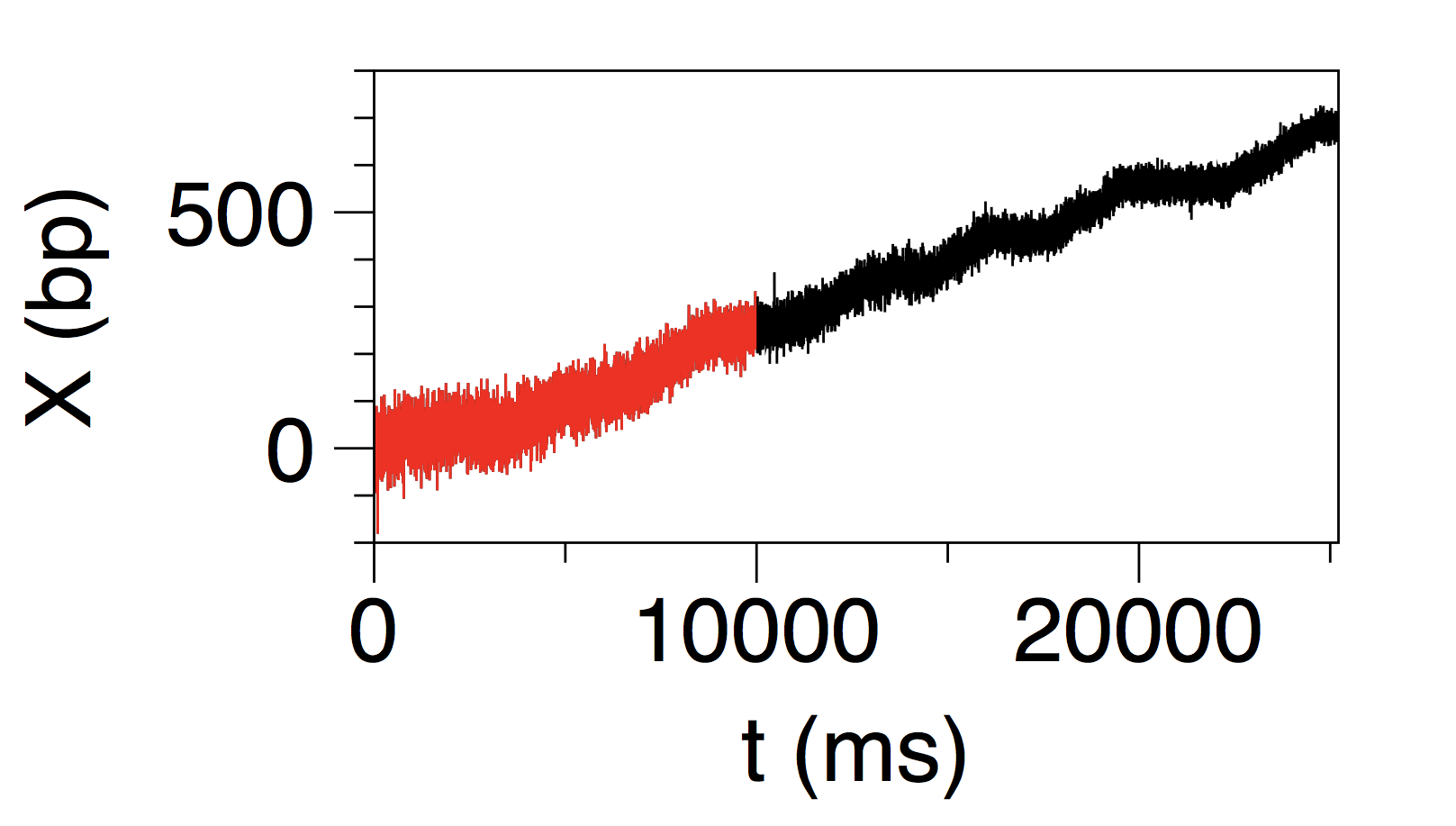}
\caption{Experimental traces of RNA Polymerase I. In particular, and to make the inference problem more challenging, we only select the first $10^4$ first steps (about 10 seconds) for each trace. These parts are highlighted in red the right panels, and plotted in the left panels. Interestingly, only for the traces $\#2$ and $\#3$ the two mechanisms operating --elongation and backtracking-- are observable and clearly discernible, while the rest are hidden below the noise. }
\label{RNA}
\end{figure}
 \section{BIOCHEMICAL NETWORKS: ADDITIONAL DETAILS FOR THE CASE OF RNA POLYMERASE}
 In figure \ref{RNA} we plot (right panels) the five experimental traces of RNA Polymerase I analysed in this work. To make the inference problem more challenging, we systematically select the initial $10^4$ time steps (about 10 seconds, as the signal is sampled at 1kHz), where it is often not evident to visually distinguish the periods where elongation and backtracking are at play.  Due to the inherent noise inside the cell, and as one can see in the figure, only for two cases (traces $\#2$ and $\#3$) the two mechanisms operating --elongation and backtracking-- are observable, while the rest are hidden below the noise. Our methodology (both the layer detection and layer estimation protocols) are succesful at detecting and correctly inferring the presence of these two modes as we find that the more likely number of hidden layers is $L=2$. Intringuingly, if we select a period where apparently only one mode (backtracking) is at play, then our methodology again predicts multiplexity ($D(3)>0$) and $L=2$. This is a surprising result which suggest that, either (i) short elongation times are hidden among a long backtracking pause, or that (ii) while in backtracking, stochastic fluctuations have a non-Markovian character. Either interpretations are in contrast with the state of the art and deserve further investigation.\\

\begin{figure}
\centering
\includegraphics[width=0.6\columnwidth]{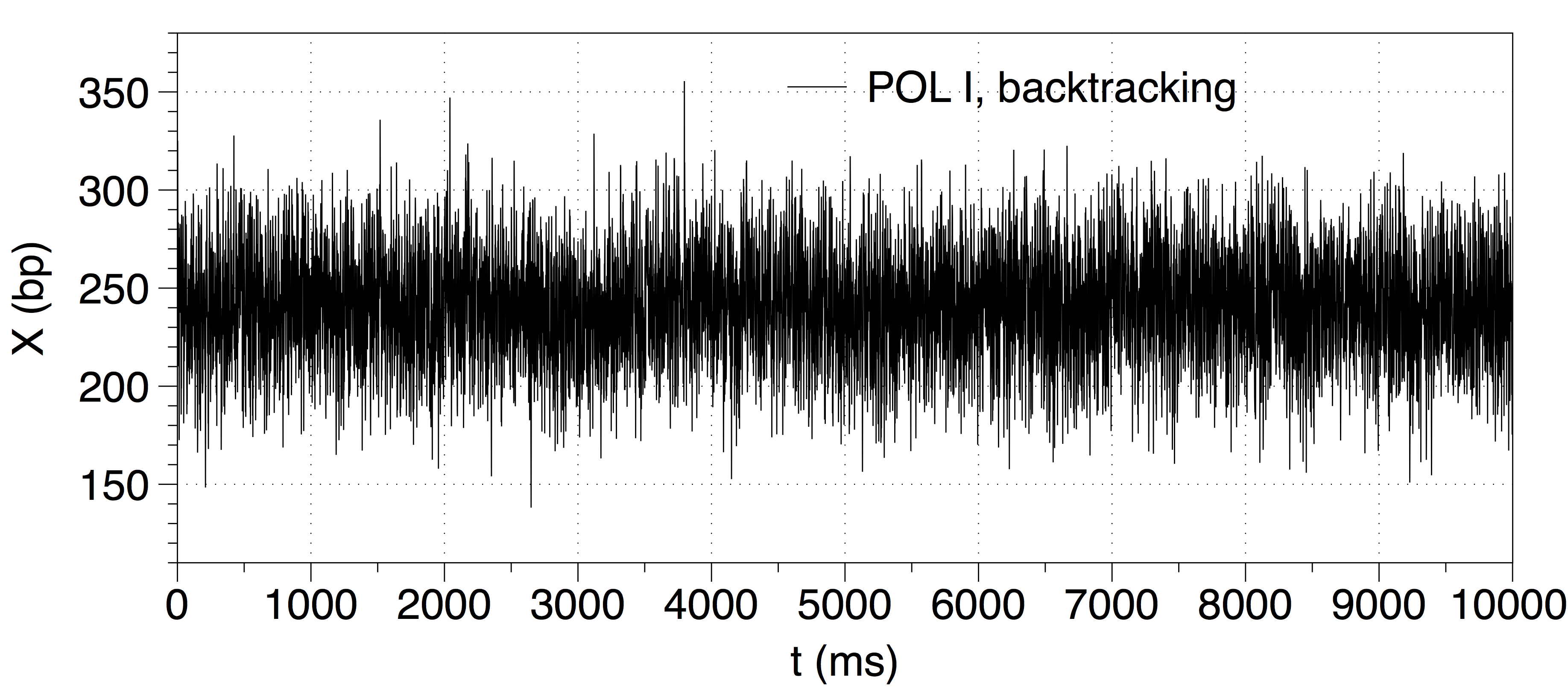}
\includegraphics[width=0.41\columnwidth]{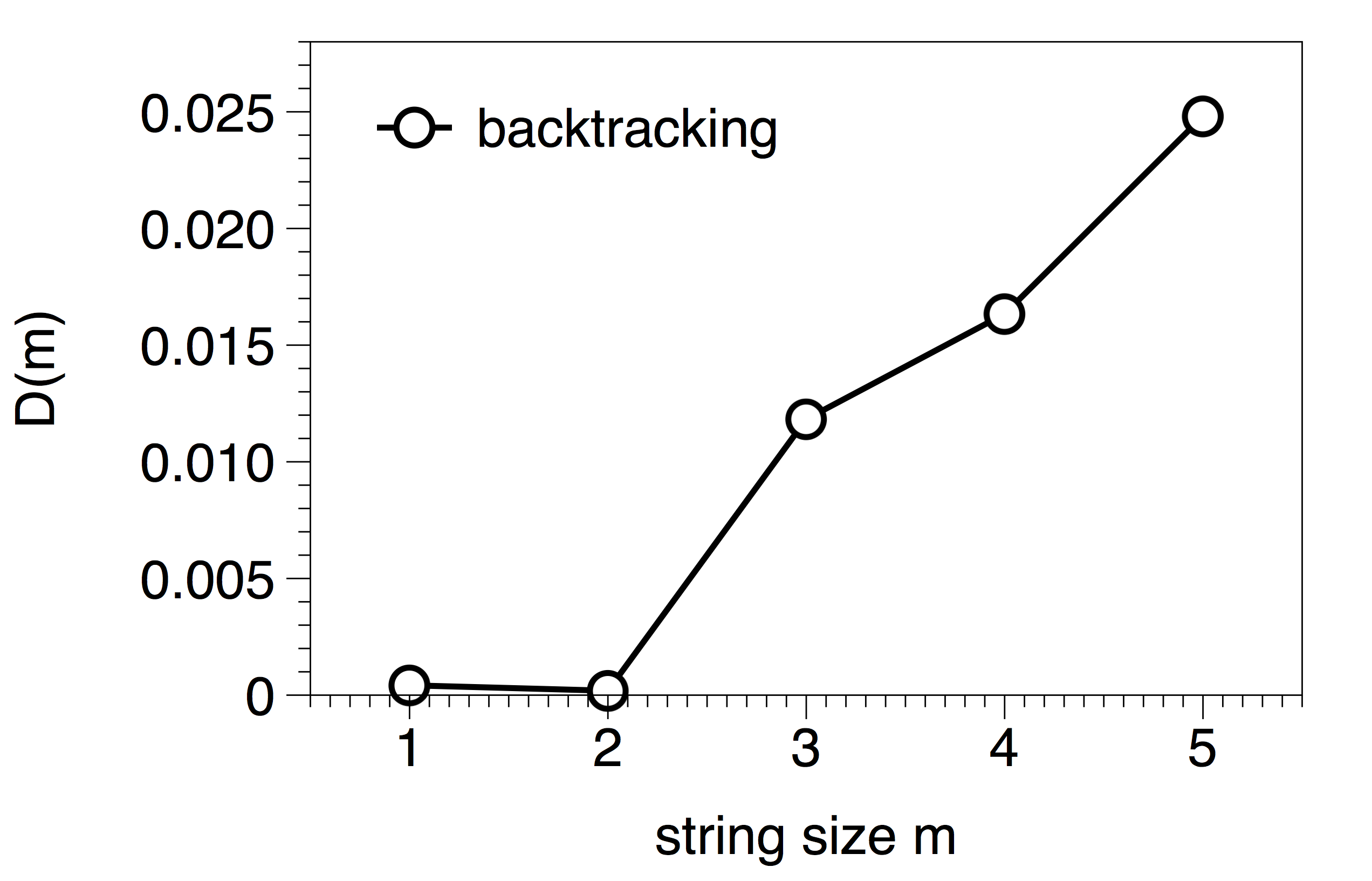}
\includegraphics[width=0.41\columnwidth]{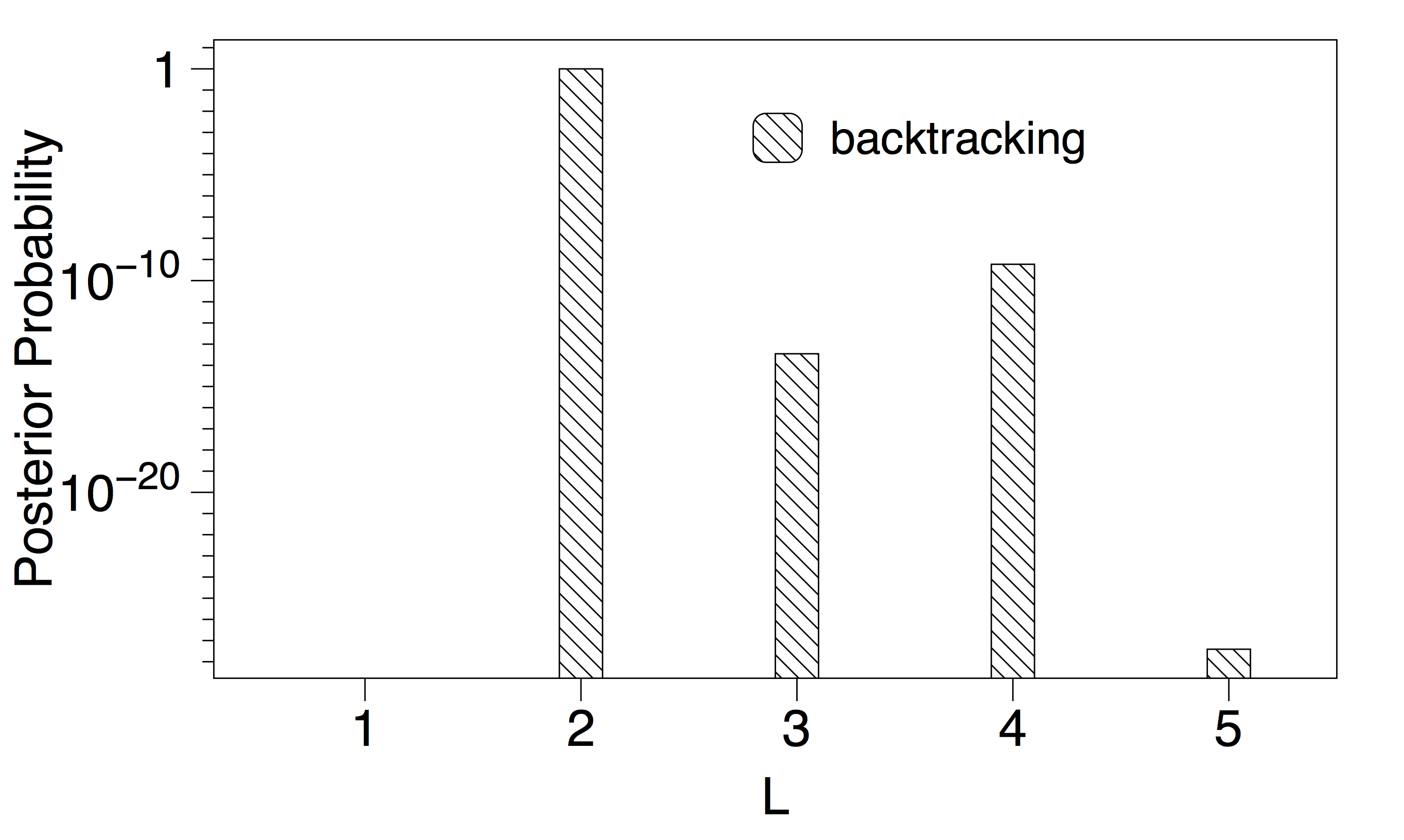}
\caption{(Top panel) Sample of Pol I trajectory of $10^4$ steps in a backtracking pause. (Bottom left panel) $D(m)$ as a function of $m$. We find $D(3)>0$ suggesting that this series is non-Markovian. (Bottom right panel) 
Posterior probability $P(L | X(1:N) )$ as a function of the number of layers $L$, peaking at $L=2$. The probability for $L=1$ is below the $\epsilon$ of the machine, meaning virtually null. This is a surprising result which suggest that, either (i) short elongation times are hidden among a long backtracking pause, or that (ii) while in backtracking, stochastic fluctuations have a non-Markovian character. Either interpretations are in contrast with the state of the art and deserve further investigation.}
\label{backtracking}
\end{figure}

\section{{PROOF OF THEOREM \ref{th_A}}} \label{ap_A}

\cred{It is well known that a higher-order Markov model ($h>1)$ can be converted into a first order Markov system ($h=1$) defined over a higher dimensional state space (see, e.g., \cite{Chen00}). In the same vein, we are going to show that every higher-order Markov model can be converted into an equivalent hidden jump-Markov model of the type discussed in Sec. II (which from now on we call multiplex model) with a sufficiently large number of layers. The proof hence consists of two parts:
\begin{itemize}
\item We first describe a constructive procedure to obtain a multiplex model from the transition probabilities of a generic Markov model of order $h$.
\item We then prove that the probability to generate any sequence of length $h+1$ is the same for the (order $h$) Markov model and the multiplex model.
\end{itemize}
}

\subsection{Construction of the multiplex model} \label{sapConstruction}

\cred{
Recall that the Markov sequence takes values on the space $\mbK = \{ 1, \ldots, K \}$ (this is the node set in the multiplex network) and is defined by the transition probabilities $P_\mbK^h(i_t|i_{t-h:t-1})$, $i_{t-h:t} \in \mbK^{h+1}$. We aim at constructing a model with $L=K^{h-1}$ layers, which implies that the random layer sequence $\ell(t)$ takes values on the set $\mL_{h-1}:={1,\ldots,K^{h-1}}$. As a first step, we establish a bijection between the set of layers $\mL_{h-1}$ and the set of vectors $\mbK^{h-1}$. To be specific, let us introduce the map
$b:\mbK^{h-1}\rightarrow\mL_{h-1}$
which associates every vector of states
 \begin{equation}
  {\bf v}=\left[
  	\begin{array}{c}
	i_{h-1}\\
	i_{h-2}\\
	\vdots\\
	i_1\\
	\end{array}
\right]\in\mbK^{h-1}
\nonumber
\end{equation}
with a unique layer {$b({\bf v}) := 1+ \sum_{s=1}^{h-1} (i_s-1) K^{s-1} \in \mL_{h-1}$.} Intuitively, we interpret the vector
 \begin{equation}
  {\bf v}=\left[
  	\begin{array}{c}
	i_{h-1}\\
	i_{h-2}\\
	\vdots\\
	i_1\\
	\end{array}
\right]\in\mbK^{h-1}
\nonumber
\end{equation}
as a sequence of digits in a base $K$ number system, and the function $b(\cdot)$ symply returns the evaluation of this number (plus one, for notational coherence). Since $b$ is bijective, we can readily construct the inverse function $b^{-1}:\mL_{h-1}\rightarrow\mbK^{h-1}$ that takes a layer number $\ell\in\mL_{h-1}$ and returns its base-$K$ representation (in vector form) $b^{-1}(\ell)$.
}

\cred{
With these ingredients, we can now construct a multiplex model with the form of a Markov chain on the state space $\mbK\times\mL_{h-1}$ described by the transition probabilities
 \begin{equation}
P\left(
\begin{array}{c|c}
\left[
\begin{array}{c}
X(t)\\
\ell(t)
\end{array}
\right]=\left[
\begin{array}{c}
i_t\\
j_t
\end{array}
\right] &
\left[
\begin{array}{c}
X(t-1)\\
\ell(t-1)
\end{array}
\right]=\left[
\begin{array}{c}
i_{t-1}\\
j_{t-1}
\end{array}
\right]
\end{array}
\right)= 
\mbM_*\left(
\begin{array}{c|c}
i_t & i_{t-1}\\
j_t & j_{t-1}
\end{array}
\right)
\nonumber
\end{equation}
where $i_t,i_{t-1}\in\mbK$, $j_t,j_{t-1}\in\mL_{h-1}$ and
\begin{equation}
\mbM_*\left(
\begin{array}{c|c}
i_t & i_{t-1}\\
j_t & j_{t-1}
\end{array}
\right)=\left\{
	\begin{array}{ll}
P_\mbK^h(i_t|i_{t-1:t-h}) &\mbox{if } 
	\begin{array}{ll}
b^{-1}(j_{t-1})=\left(
\begin{array}{c}
i_{t-2}\\
\vdots\\
i_{t-h}
\end{array}
\right) &\mbox{and }
b^{-1}(j_{t})=\left(
\begin{array}{c}
i_{t-1}\\
\vdots\\
i_{t-h+1}
\end{array}
\right),\\
\end{array}\\
0 &\mbox{otherwise.}\\
	\end{array}
\right.
\nonumber
\end{equation}
}

\cred{Note that we have simply encoded the history of states $i_{t-2:t-h}$ into the layer index $j_{t-1}$. For the system to behave ``correctly" we need to check that the transition from layer $j_{t-1}$ to layer $j_t$ is compatible with the sequence of states $i_{t-1:t-h+1}$ that has to be ``recorded" for the next element of {the} chain. Unfortunately, the Markov chain described by the transition probabilities $\mbM_*(\cdot|\cdot)$ is semi-degenerate. To be specific, the layer value $\ell(t)=j_t$ is deterministic given $X(t-1)=i_{t-1}$ and   $\ell(t-1)=j_{t-1}$. Indeed,
\begin{equation}
\mbox{if} \quad 
j_{t-1}=b^{-1}\left(
\begin{array}{c}
i_{h-1}'\\
i_{h-2}'\\
\vdots\\
i_{1}'\\
\end{array}
\right)
\quad \mbox{and $X(t-1)=i_{t-1}$ then} \quad  
 j_{t}=b^{-1}\left(
\begin{array}{c}
i_{t-1}\\
i_{h-1}'\\
\vdots\\
i_{2}'
\end{array}
\right)
\nonumber
\end{equation} 
with probability 1. This determinism is not coherent with our discussion of multiplex models in Sections \ref{sintro} and \ref{smethods}. Fortunately, this difficulty can be easily removed if we further extend the state space of the model.
} 

\cred{
Let us construct a (new but similar) Markov chain on the space $\mbK\times\mL_h$, defined by the transition probabilities
 \begin{equation}
P\left(
\begin{array}{c|c}
\left[
\begin{array}{c}
X(t)\\
\ell(t)
\end{array}
\right]=\left[
\begin{array}{c}
i_t\\
j_t
\end{array}
\right] &
\left[
\begin{array}{c}
X(t-1)\\
\ell(t-1)
\end{array}
\right]=\left[
\begin{array}{c}
i_{t-1}\\
j_{t-1}
\end{array}
\right]
\end{array}
\right)= \mbM\left(
\begin{array}{c|c}
i_t & i_{t-1}\\
j_t & j_{t-1}
\end{array}
\right)
\end{equation}
where
\begin{equation}
\mbM\left(
\begin{array}{c|c}
i_t & i_{t-1}\\
j_t & j_{t-1}
\end{array}
\right):=P_\mbK^h(i_t|i_{t-1:t-h})R_{j_{t-1}j_t}
\label{equ1}
\end{equation}
and 
\begin{equation}
R_{j_{t-1}j_t}:=\left\{
	\begin{array}{ll}
	\frac{1}{K} &\mbox{if } b^{-1}(j_t)=\left(
			\begin{array}{c}
			r\\
			i_{t-1}\\
			\vdots\\
			i_{t-h+1}\\
			\end{array}
		\right) \mbox{ and }
		 b^{-1}(j_{t-1})=\left(
			\begin{array}{c}
			s\\
			i_{t-2}\\
			\vdots\\
			i_{t-h}\\
			\end{array}
		\right) \mbox{ for some $s,r\in\mbK$,}\\
	0 &\mbox{otherwise.}\\
	\end{array}
\right.
\label{equ1.5}
\end{equation}
{(Note that, in this case, the bijection $b$ is a function $\mbK^h \rw \mL_h$, with the same interpretation as before.)}
Intuitively, we have now encoded the system memory (the sequence of states $i_{t-h:t-2}$) into a subset of $K$ layers, instead of a single one. From each layer in this subset there is probability $\frac{1}{K}$ to jump to a new layer belonging to the subset that encodes the sequence $i_{t-h+1:t-1}$. Comparing Eq. \eqref{equ1} with our notation for multiplex models in Sections \ref{sintro} and \ref{smethods} we can readily identify the matrix ${\bf R}_L$ of transition probabilities between layers, while $T_{i_{t-1}i_t}^{(j_{t-1})}=P_\mbK^h(i_t|i_{t-1:t-h})$ (note that $j_{t-1}$ uniquely identifies the sequence $i_{t-2:t-h}$).
}

\subsection{Identity of transition probabilities}

\cred{What remains to be shown is that the conditional probabilities
\begin{equation}
P(X(t)=i_t|X(t-h:t-1)=i_{t-h:t-1}) 
\nonumber
\end{equation}  
generated by the probability function $P_\mbK^h(\cdot|\cdot)$ (for the order-$h$ Markov model) and the Markov kernel $\mbM(\cdot|\cdot)$ in \eqref{equ1} (for the multiplex model) coincide. 
} 

\noindent \cred{
On one hand, if $\{X(t)\}_{t\geq0}$ is an order-$h$ Markov sequence, then 
\begin{equation}
P(X(t)=i_t|X(t-h:t-1)=i_{t-h:t-1}) = P_\mbK^h(i_t|i_{t-h:t-1})
\end{equation}  
by definition. On the other hand, if the sequence $\{X(t)\}_{t\geq0}$ is generated by the multiplex model on $\mbK\times \mL_h$ defined by $\mbM$, we have
\begin{eqnarray}
P(X(t)=i_t &|& X(t-h:t-1)=i_{t-h:t-1}) = \nonumber\\
&=&\sum_{j_t\in\mL_h}\sum_{j_{t-1}\in\mL_h}P(X(t)=i_t,\ell(t)=j_t,\ell(t-1)=j_{t-1}|X(t-h:t-1)=i_{t-h:t-1}) \nonumber\\
&=&\sum_{j_t\in\mL_h}\sum_{j_{t-1}\in\mL_h}P(X(t)=i_t,\ell(t)=j_t|\ell(t-1)=j_{t-1},X(t-h:t-1)=i_{t-h:t-1}) \nonumber\\
&& \times P(\ell(t-1)=j_{t-1}|X(t-h:t-1)=i_{t-h:t-1}) \nonumber\\
&=&\sum_{j_t\in\mL_h}\sum_{j_{t-1}\in\mL_h}\mbM\left(
\begin{array}{c|c}
i_t & i_{t-1}\\
j_t & j_{t-1}
\end{array}
\right)P(\ell(t-1)=j_{t-1}|X(t-h:t-1)=i_{t-h:t-1})
\label{equ2}
\end{eqnarray} 
where the first equality follows from the theorem of total probabilities, the second equality is obtained from the definition of conditional probability and the third identity follows from the definition of the multiplex model on $\mbK\times\mL_h$. Moreover, for the multiplex model constructed in Appendix \ref{sapConstruction},
\begin{eqnarray}
P(\ell(t-1)=j_{t-1}|X(t-h:t-1)=i_{t-h:t-1})= \left\{
	\begin{array}{ll}
	\frac{1}{K}, &\mbox{if } b^{-1}(j_{t-1})=\left(
		\begin{array}{c}
		s\\
		i_{t-2}\\
		\vdots\\
		i_{t-h}\\
		\end{array}
	\right), \quad s\in\mbK\\
	0, &\mbox{otherwise.}\\
	\end{array}
\right.
\label{equ3}
\end{eqnarray}  
Hence substituting \eqref{equ3} and \eqref{equ1} into \eqref{equ2} yields
\begin{equation}
P(X(t)=i_t|X(t-1:t-h)=i_{t-1:t-h})=
\frac{1}{K} \sum_{j_t\in\mL_h} \sum_{j_{t-1} \in S_{b^{-1}}(i_{t-h:t-2})} P_\mbK^h(i_t|i_{t-1:t-h})R_{j_{t-1}j_t},
\label{equ4}
\end{equation}
where
$$
S_{b^{-1}}(i_{t-h:t-2}) := \left\{
	j \in \mL_h: b^{-1}(j) = \left[
		\begin{array}{c}
		s\\
		i_{t-2}\\
		\vdots\\
		i_{t-h}\\
		\end{array}
	\right], \quad s\in\mbK
\right\}.
$$
However, from the definition of $R_{j_{t-1}j_t}$ in Eq. \eqref{equ1.5}  we can further reduce Eq. \eqref{equ4} and obtain
\begin{equation}
P(X(t)=i_t|X(t-h:t-1)=i_{t-h:t-1}) = \frac{1}{K} P_\mbK^h(i_t|i_{t-1:t-h}) 
\sum_{j_t \in S_{b^{-1}}(i_{t-h+1:t-1})} \sum_{\sum_{j_{t-1} \in S_{b^{-1}}(i_{t-h:t-2})}} R_{j_{t-1}j_t}	
\label{equ5}
\end{equation}  
and, for each term in the sum of Eq. \eqref{equ5} we have $R_{j_{t-1}j_t}=\frac{1}{K}$ (from the definition in Eq. \eqref{equ1.5}). 
{Because of the construction of the sets $S_{b^{-1}}(i_{t-h+1:t-1})$ and $S_{b^{-1}}(i_{t-h:t-2})$, there are exactly $K^2$ terms} in the sum of Eq. \eqref{equ5}, which readily yields
\begin{equation}
P(X(t)=i_t|X(t-1:t-h)=i_{t-1:t-h})=P_\mbK^h(i_t|i_{t-h:t-1})
\nonumber
\end{equation}
and completes the proof. $square$
}

%
\section{{PROOF OF THEOREM} \ref{th_B}} \label{ap_B}

\cred{Theorem \ref{th_A} guarantees that any order-$h$ Markov sequence on $\mbK$ can be represented exactly by a multiplex model with $L=K^h$ layers. Hence, for every random sequence $\{X_h(t)\}_{t\geq0}$ we have an equivalent sequence $\{X_L(t)\}_{t\geq0}$ with $L=K^h$ layers and transition probabilities 
\begin{equation}
R_{ij}=P(\ell(t)=j|\ell(t-1)=i),
\nonumber
\end{equation}
 $i,j\in\{1,\ldots,L\}$ for the layers and
\begin{equation} 
T_{is}^{(j)}=P(X_L(t)=s|X_L(t-1)=i,\ell(t-1)=j),
\nonumber
\end{equation}
for $i,s\in\{1,\ldots,K\}$, $j\in\{1,\ldots,L\}$. Therefore, every Markov-$\infty$ sequence $\{X(t)\}_{t\geq0}$ can be {represented} as well by a sequence of random processes $\{X_L(t)\}_{t\geq0}$ described by suitably constructed multiplex models, i.e.,
\begin{equation}
\lim_{L\rightarrow\infty} X_L \stackrel{d}{=}X,
\nonumber
\end{equation}
where $ \stackrel{d}{=}$ denotes equality in distribution.
What we need to prove is that the sequence of multiplex models, described by the pairs $({\bf T}_L,{\bf R}_L)$, $L=K, K^2, K^3, ...$, yields the continuous-layer model described by a proper probability measure $\mM_0$, a Markov kernel $\mM$ on $\mB([0,1))$ and a family of transition matrices ${\bf T}(y)$, $y \in [0,1)$, when $L\rightarrow\infty$.
}

\cred{
Let us note that every transition probability in the multiplex model can be specified by a simple function \cite{Kolmogorov75} defined over subsets of $[0,1)$. To be specific, we construct
\begin{equation}
{\hat T}_{is}^L(y):=\sum_{\ell=1}^L\mathbbm{1}_{[\frac{\ell-1}{L},\frac{\ell}{L})}(y)T_{is}^{(\ell)},
\end{equation}
where
\begin{equation}
\mathbbm{1}_A({y})=\left\{
\begin{array}{ll}
1 &\mbox{if } y \in A\\
0 &\mbox{otherwise}\\
\end{array}
\right.
\end{equation}
is the set-indicator function,
\begin{equation}
\hat\mM^L(y|y^\prime)=\sum_{\ell=1}^L\sum_{j=1}^{L} \mathbbm{1}_{[\frac{j-1}{L},\frac{j}{L})}(y^\prime)\mathbbm{1}_{[\frac{\ell-1}{L},\frac{\ell}{L})}(y)R_{jl}
\end{equation}
and
\begin{equation}
\hat\mM_0^L(y)=\sum_{\ell=1}^L\mathbbm{1}_{[\frac{\ell-1}{L},\frac{\ell}{L})}(y)P(\ell(0)=\ell).
\end{equation}
The limits $\lim_{L\rightarrow\infty}\hat T_{is}^L(y)$, $\lim_{L\rightarrow\infty}\hat\mM^L(y|y^\prime)$ and $\lim_{L\rightarrow\infty}\hat\mM_0^L(y)$ are well defined (because $X_L(t)$ yields the same transition probabilities as $X_h(t)$, for $L=K^h$, and $\lim_{h\rightarrow\infty}X_h\stackrel{d}{=}X$ uniformly over $t$ and $\mbK^{h+1}$). Therefore, from {\cite{Kolmogorov75} (see Theorem 5 in Chapter 8)} the functions
\begin{eqnarray}
T_{is}(y):=\lim_{L\rightarrow\infty}\hat T_{is}^L(y),\nonumber\\
\bar \mM(y|y^\prime):=\lim_{L\rightarrow\infty}\hat \mM^L(y|y^\prime), &\mbox{and }\\
\bar \mM_0(y):=\lim_{L\rightarrow\infty}\hat \mM_0^L(y), \nonumber
\end{eqnarray}
exist and they are measurable w.r.t the Lebesgue measure. In particular, $\bar \mM(y|y^\prime)$ and $\bar \mM_0(y)$ are probability density functions w.r.t. the Lebesgue measure and we obtain $\mM(dy|y^\prime)=\bar \mM(y|y^\prime)dy$, and {$\mM_0(dy)=\bar {\cal M}_0(y)dy$}. $\Box$ 
}

\bibliography{apssamp}

\end{document}